\documentclass[prx,notitlepage,twocolumn,superscriptaddress,amsmath,amssymb,aps,longbibliography]{revtex4-2}

\usepackage{amsmath,amssymb,graphicx,mathtools,dsfont}
\usepackage[dvipsnames]{xcolor}
\usepackage[T1]{fontenc}

\usepackage[normalem]{ulem}
\usepackage{soul}
\usepackage{comment}

\usepackage[most]{tcolorbox}
\usepackage{hyperref}
\hypersetup{colorlinks=true,linkcolor=blue,citecolor=blue,urlcolor=blue}
\usepackage{orcidlink}
\usepackage{bm}
\usepackage{float}
\usepackage{microtype}

\newcommand{\cN}{\mathcal N}
\newcommand{\cC}{\mathcal C}

\newcommand{\av}[1]{\langle #1 \rangle}

\newcommand{\ket}[1]{\mbox{$| #1 \rangle$}}

\newcommand{\NCU}{Institute of Physics, Faculty of Physics, Astronomy and Informatics, Nicolaus Copernicus University in Toru\'n, Grudzi\c{a}dzka 5, 87-100 Toru\'n, Poland}
\newcommand{\IAS}{Institute of Advanced Studies, Nicolaus Copernicus University in Toru\'n, Wile\'nska 4, 87-100 Toru\'n, Poland}
\newcommand{\UIBK}{Institut f\"ur Theoretische Physik, Universit\"at Innsbruck, A-6020 Innsbruck, Austria}
\newcommand{\IQOQI}{Institute for Quantum Optics and Quantum Information, Austrian Academy of Sciences, A-6020 Innsbruck, Austria}
\newcommand{\OIST}{Many-Body Open Quantum Systems Unit, Okinawa Institute of Science and Technology Graduate University, Onna, Okinawa 904-0495, Japan}

\begin{document}

\title{Cavity-induced intertwining of density and pairing order in a degenerate Fermi gas}

\author{Sankalp Sharma\,\orcidlink{0009-0006-2287-1386}}
\affiliation{\NCU}
\affiliation{\IAS}
\author{Farokh Mivehvar\,\orcidlink{0000-0003-4776-1352}}
\affiliation{\UIBK}
\affiliation{\IQOQI}
\affiliation{\OIST}
\author{Helmut Ritsch\,\orcidlink{0000-0001-7013-5208}}
\affiliation{\UIBK}
\author{Tomasz Wasak\,\orcidlink{0000-0002-0958-2276}}
\affiliation{\NCU}
\affiliation{\IAS}


\begin{abstract}
Recent quantum gas cavity-QED experiments demonstrated simultaneous coupling of photons to single atom transitions as well as correlated pairs of ultracold fermions trapped inside optical cavities. This enables simultaneous control over density ordering and pairing. Using extensive numerical simulations, we show that in a transversely driven, two-component degenerate Fermi gas, the interplay of cavity-induced and bare atom--atom interactions controls not only the power threshold for self-organization, but also the type of spatial ordering in the $\mathbb Z_2$-symmetry-broken superradiant state.
In the repulsive interaction regime, unpaired or weakly paired fermions first self-organize through a charge-density-wave instability, and finite-momentum pairing only occurs at much stronger pump strengths. In contrast, an attractive superfluid undergoes a joint density--pairing instability, directly entering into intertwined phase with charge-density-wave and pair-density-wave orders. At strong pumping the photon-enhanced pair interaction generates localized density and pairing order even when the bare contact interaction is repulsive. In this cavity-dominated regime strong spatial localization suppresses the long range superfluid coherence.
Our results identify the role of cavity-induced atomic interactions in supporting intertwined fermionic orders and pave the way for exploring exotic states with multiple orders in highly controlled hybrid light-matter systems.
\end{abstract}

\maketitle


\textit{Introduction}---Ultracold atomic gases in high-finesse optical cavities realize dynamical optical potentials governed by the mutual response of atoms and photons~\cite{RevModPhys.85.553}. Unlike conventional optical lattices imposed by classical laser fields, the intra-cavity optical potential depends on the atomic quantum state. In superradiant states atoms arrange into a spatial pattern that enhances collective coherent scattering into the cavity from the pump beam, while the resulting field stabilizes the spatial order. This feedback mechanism, identified theoretically in Refs.~\cite{PhysRevLett.89.253003,PhysRevA.72.053417}, was first observed in thermal atoms~\cite{PhysRevLett.91.203001} but also is the basis of cavity-mediated self-organization and superradiance in quantum gases~\cite{mivehvar2021cavity}. Its observation with a Bose--Einstein condensate (BEC)~\cite{baumann2010dicke} constituted the first clear real time demonstration of quantum phase transition near $T\approx 0$. More recently, such superradiant self-ordering to a density wave phase was observed in a degenerate Fermi gas~\cite{zhang2021observation}. The realization of coupling strongly interacting Fermi gas to optical cavities~\cite{roux2020strongly,Roux_2021} allowed for independent control of contact and photon-mediated long-range interactions, and the observation of density-wave ordering across the BCS--BEC crossover~\cite{helson2023density}. The emergence of the symmetry breaking spatial order has recently even been resolved in-situ~\cite{h3zm-rnnx}.

For fermions, cavity feedback can act not only on density fluctuations, but also on pairing correlations. Theoretical studies have predicted fermionic (Umklapp-) superradiance, commensuration effects and photon-induced reconstruction of Fermi surface~\cite{PhysRevLett.112.143002,PhysRevLett.112.143003,chen2014superradiance,t4xb-6x3z}. The system exhibits a strong interplay between superradiance and the BCS--BEC crossover, unconventional cavity-mediated pairing and finite-center-of-mass superfluid phases \cite{guo2012cavityBCSBEC,chen2015superradiant,PhysRevLett.123.133601,zheng2020fflo}. Combined with the tunability of short-range interactions throughout the BCS--BEC crossover~\cite{giorgini2008theory,RevModPhys.82.1225}, these effects make cavity-coupled Fermi gases a natural platform for determining when density and pairing orders compete, cooperate, or emerge as intertwined phases. 

Recent experiments enabled collective near-resonant coupling of cavity photons to a molecular photo-association transition, generating pair-polaritons and allowing for photonic access to control short-range pair correlations \cite{konishi2021universal}.
In the dispersive regime, the same microscopic collective coupling induces a dynamical atom--atom interaction, allowing for interaction engineering and strong entanglement generation for quantum metrology~\cite{sharma2025engineering}. The experiment with simultaneous coupling of the cavity field to single atoms and atomic pairs reported coherent interference among superradiance, charge-density-wave (CDW), and pair-density-wave (PDW) orders in a unitary Fermi gas via a Fano-type phase boundary~\cite{zwettler2025cavity}.
However, the question remains open how preexisting superfluidity affects the character and the sequence of self-organized states when photons couple to atomic pairs as well. In particular, it is unclear whether the system develops a density wave before entering finite-momentum pairing state at stronger driving, or if homogeneous pairing allows for direct transition into an intertwined CDW+PDW superradiant phase. Resolving this distinction is essential for determining how superfluidity and photon-induced ordering mechanisms combine to yield finite-momentum pairing.

Here, we show that the presence of superfluidity in a transversely driven two-component Fermi gas with single-atom and atom-pair cavity couplings strongly modifies the types of ordered $\mathbb Z_2$-symmetry-broken~\cite{PhysRevLett.107.140402,mivehvar2021cavity} self-organized states as the pump strength crosses superradiant thresholds. By mapping the phase diagram within the Hartree-Fock-Bogoliubov (HFB) approach~\cite{ring1980nuclear,Dobaczewski_2013, PhysRevA.101.063607,PhysRevA.81.063642,PhysRevA.68.033610} we show the initially unpaired or weakly paired gases first enter the spatially ordered phase through a predominantly density-driven CDW--superradiant instability and acquire finite-momentum pairing only at stronger pumping. In contrast, a sufficiently attractive superfluid undergoes a joint density- and pair-self-organization instability, directly entering an intertwined CDW+PDW superradiant state. At strong pumping strength, the photon-induced pair interaction drives the system toward a cavity-dominated regime of spatially localized density and pairing order with negligible coherence between the separated density regions.
Also, we demonstrate that the atom-pair coupling determines the cooperation or competition of the phases via a Fano-type superradiant phase boundary~\cite{zwettler2025cavity}, also for finite bare atomic interactions. Our findings indicate that preexisting superfluidity might significantly affect the types of allowed self-organized phases. Due to the controlled complexity, this platform offers promising potential for simulating exotic states in light-matter systems~\cite{guo2021optical, lev2025glass} as well as for quantum metrology~\cite{sharma2025engineering}.

\begin{figure}[t]
    \centering
    \includegraphics[width=\columnwidth]{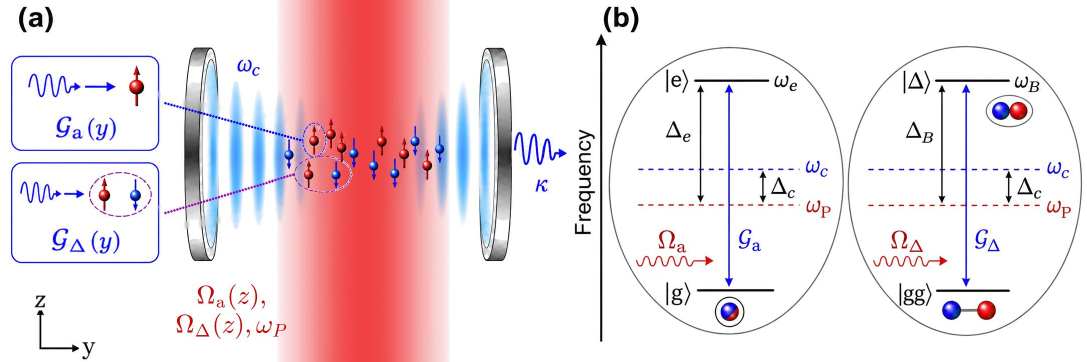}
    \caption{
    (a) Scheme of a two-component Fermi gas inside a planar single-mode optical cavity of width $\kappa$ with transverse laser illumination. The cavity mode of frequency $\omega_c$ is coupled to an internal atomic transition with coupling amplitude $\mathcal G_{\mathrm a}(y)$ and to a molecular resonance with coupling strength $\mathcal G_{\Delta}(y)$. The transverse plane wave pump laser of frequency $\omega_P$ couples to both resonances, the single atom transition with strength $\Omega_\mathrm{a}(z)$ as well as the atom pair resonance with $\Omega_\Delta(z)$. 
    (b) Sketch of relevant optical transitions for single atoms
    $|g\rangle\leftrightarrow|e\rangle$ (left) and atom-pairs
    $|gg\rangle\leftrightarrow|\Delta\rangle$ (right). 
    The pump frequency detunings are denoted by $\Delta_c$, $\Delta_e$, and $\Delta_B$ for the cavity, excited and molecular state, respectively.
    }\label{fig:Transversal}
\end{figure}


\textit{Effective two-channel model}---We consider a uniform two-component Fermi gas inside a single-mode optical cavity with mode function $f(\mathbf{r}) = \cos(k_cy)$, where the wave vector is $k_c = 2\pi/\lambda_c$ and $\lambda_c$ is the photon wavelength. 
The cavity is oriented along the $y$-axis, see Fig.~\ref{fig:Transversal}. 

A transverse pump laser of frequency $\omega_P$ drives the gas and couples single atoms (atom pairs) in the ground state $\ket{g}$ to an excited single-atom $\ket{e}$ (molecular $\ket{\Delta}$) state of energy $\hbar\omega_e$ ($\hbar\omega_B$) with position dependent Rabi frequency $\Omega_\mathrm{a/\Delta}(\mathbf{r}) = \tilde\Omega_\mathrm{a/\Delta} \cos(k_p z)$, where $\tilde\Omega_\mathrm{a/\Delta}$ is the maximum value and $k_p\approx k_c$ is the wave vector of the pump photons; henceforth we set $\hbar\equiv1$.
We assume an elongated trapping geometry where the atomic motion is effectively one-dimensional (1D) along the cavity axis and we approximate $z\approx0$.
The single-atom (atom pair) coupling amplitude to the quantized mode of the cavity is given by $\mathcal{G}_\mathrm{a/\Delta}(\mathbf{r}) = \tilde{\mathcal{G}}_\mathrm{a/\Delta} f(\mathbf{r})$, where $\tilde{\mathcal{G}}_\mathrm{a/\Delta}$ is the maximum value of the cavity QED coupling strength of single photons to single atoms (atom pairs).  At the mean field level, we describe the cavity mode by the coherent amplitude $\alpha$. Throughout the main text, $\alpha$ and $\tilde{\mathcal G}_{\mathrm{a}/\Delta}$ denote the quantities entering the effective theory scaled by the system size, with $|\alpha|^2$ providing the corresponding scaled intra-cavity photon number. Their normalization and scaling in the large-system limit are discussed in the Supplemental Material (SM)~\cite{SM}.
The standard single-atom dipole coupling, described by 
$\tilde{\mathcal{G}}_\mathrm{a}$ and $\tilde\Omega_\mathrm{a}$, together with the photon-pair coupling~\cite{konishi2021universal,zwettler2025cavity}, quantified by $\tilde{\mathcal{G}}_\mathrm{\Delta}$ and $\tilde\Omega_\mathrm{\Delta}$, define the two-channel interaction of the cavity and pump light with the degenerate Fermi gas. 

The pump operates in the far off-resonant dispersive regime, where the detunings --- $\Delta_B = \omega_P - \omega_B$ from the molecular state and $\Delta_e = \omega_P - \omega_e$ from the single atom excited state --- are the largest energy scales compared to the relevant many-body and optical couplings. In such a case, the excited atomic and the molecular states remain only virtually populated and are adiabatically eliminated. 
In the rotating frame of the pump, the effective 1D mean-field Hamiltonian reads
\begin{align} 
&\hat H_{\rm eff} =\; \sum_{\sigma = \uparrow,\downarrow}\!\int\! dy\, \hat\psi_\sigma^\dagger(y) \left[ -\frac{\partial_y^2}{2M} -\mu_\sigma +V_{\mathrm a}(y) \right] \hat\psi_\sigma(y) \nonumber\\ &\; +\int\! dy\; U_{\rm eff}(y)\, \hat\psi_\uparrow^\dagger(y) \hat\psi_\downarrow^\dagger(y) \hat\psi_\downarrow(y) \hat\psi_\uparrow(y) -\Delta_c|\alpha|^2, \label{eq:Heff_main} 
\end{align}
where $\hat\psi_\sigma(y)$ annihilates a fermion of spin $\sigma=\uparrow,\downarrow$ at position $y$, $M$ is the atomic mass, and $\mu_\sigma$ is the corresponding chemical potential, and $\Delta_c = \omega_P - \omega_c$ is the detuning of the pump from the cavity photon frequency. 
The single-particle optical potential and the effective cavity-induced interaction are
\begin{subequations}\label{eq:VU}
\begin{align}
V_{\mathrm a}(y)
&=
g_{\mathrm a}\,
\big|\alpha\cos(k_c y)+\beta_\mathrm{a}\big|^2,
\label{eq:V1_main}\\
U_{\rm eff}(y)
&=
g_0
+
g_{\Delta}\,
\big|\alpha\cos(k_c y)+\beta_\Delta\big|^2,
\label{eq:Ueff_main}
\end{align}
\end{subequations}
where $g_{\mathrm a} \equiv {\widetilde{\mathcal G}_{\mathrm a}^{\,2}}/{\Delta_e}$ and $g_{\Delta} \equiv {\widetilde{\mathcal G}_{\Delta}^{\,2}}/{\Delta_B}$ are the dispersive couplings of the single-atom and atom-pair channels, respectively, while $\beta_\mathrm{a} = {\tilde\Omega_\mathrm{a}}/{\widetilde{\mathcal G}_{\mathrm a}}$ and $\beta_\Delta = {\tilde\Omega_\Delta}/{\widetilde{\mathcal G}_{\Delta}}$ are their respective pumping strengths. 
Since we employ a single pumping laser, we take
$\beta \equiv \beta_{\mathrm a} = \beta_{\Delta}$. 
The parameter $g_0$ is the bare 1D atom-atom contact interaction strength. For details of the derivations, see the SM~\cite{SM}.

Our effective model is defined by Eqs.~(\ref{eq:Heff_main})--(\ref{eq:VU}).
The term $V_{\mathrm a}(y)$ includes the cavity-induced optical potential that couples to the atomic density, and is responsible for self-organization and superradiance~\cite{mivehvar2021cavity}. The important difference, compared to the standard case, is that here the cavity field also enters in the atomic interactions through $U_{\rm eff}(y)$, yielding a spatially dependent pair interaction once the field $\alpha$ is nonzero. As a result, the same intra-cavity amplitude $\alpha$ controls both the single-particle potential and the cavity-induced two-body interactions, leading to a coupling between density order, pairing, and the cavity field. Finally, the effective Hamiltonian is invariant under the simultaneous transformation $y\to y+\lambda_c/2$ and $\alpha\to-\alpha$, corresponding to a $\mathbb Z_2$ symmetry~\cite{PhysRevLett.107.140402,mivehvar2021cavity}. For a solution that breaks this symmetry, the self-consistent density and pairing fields retain the spatial period $\lambda_c$ of the cavity mode.

\textit{Self-consistent HFB framework}---In order to solve the equations, we employ the HFB formalism. The chemical potential $\mu_\sigma$ sets the average atom number in each spin state $\sigma \in\{\uparrow,\downarrow\}$. The two-body interaction in Eq.~(\ref{eq:Heff_main}) 
is decoupled in both Hartree and pairing channels
giving rise to the pairing field $\Delta(y)\equiv U_{\rm eff}(y)F(y)$ with $F(y)\equiv\langle \hat\psi_\downarrow(y)\hat\psi_\uparrow(y)\rangle$. 
We then diagonalize the mean-field Hamiltonian by a Bogoliubov transformation to fermionic quasiparticles $\hat\gamma_n$, whose $n$-th eigenmode is described by particle- and hole-like coherence factors $u_n(y)$ and $v_n(y)$. In our case, we search for the spin-balanced case with $\mu_\uparrow=\mu_\downarrow=\mu$ and $n_\uparrow=n_\downarrow\equiv n/2$, and solve the Bogoliubov--de~Gennes (BdG) equations~\cite{de2018superconductivity, leggett2006quantum, zhu2016bogoliubov, giorgini2008theory}
\begin{align}
\begin{pmatrix}
h(y) & \Delta(y) \\
\Delta^*(y) & -h(y)
\end{pmatrix}
\begin{pmatrix}
u_n(y) \\ v_n(y)
\end{pmatrix}
=
E_n
\begin{pmatrix}
u_n(y) \\ v_n(y)
\end{pmatrix},
\label{eq:BdG_main}
\end{align}
with $h(y)=-\partial_y^2/(2M)-\mu+V_{\mathrm a}(y)+U_{\rm eff}(y)n(y)/2$. The order parameter $\Delta(y)$ and the density $n(y)$ are determined self-consistently from $(u_n,v_n,E_n)$ using the gap equation and the atom number equation. The details on the formalism are in SM~\cite{SM}.
The steady-state cavity amplitude reads as
\begin{align}
    \alpha = \frac{g_{\mathrm a}\beta\,\mathcal{S}_{1}  + g_{\Delta}\beta\,\mathcal{P}_1} 
    {\Delta_c-\big(g_{\mathrm a}\mathcal{S}_2 + g_{\Delta}\mathcal{P}_2\big)+i\kappa},
    \label{eq:alpha_main}
\end{align} 
which closes the set of self-consistent equations in the BdG problem solved in a single unit cell. 
Here, the single-body $\mathcal{S}_k \!\!\!\!\!=\!\!\!\!\!\int_0^{\lambda_c}\! n(y) f^k(y) dy$ and two-body
\mbox{$\mathcal{P}_k =\int_0^{\lambda_c} \av{\hat\psi_\uparrow^\dagger(y) \hat\psi_\downarrow^\dagger(y) \hat\psi_\downarrow(y) \hat\psi_\uparrow(y)} f^k(y)dy$} 
are unit cell contributions describing the photon scattering from the pump to the cavity ($k=1$) and the shift of the cavity resonance ($k=2$); the cavity losses are included through the rate $\kappa$~\cite{mivehvar2021cavity}.
In the mean field approximation, $\av{\hat\psi_\uparrow^\dagger\hat\psi_\downarrow^\dagger\hat\psi_\downarrow\hat\psi_\uparrow}=n_\uparrow(y)n_\downarrow(y)+|F(y)|^2$ showing non-zero pairing $F$ has significant impact on superradiance.
Details on the derivation and properties of $\mathcal{S}_k$ and $\mathcal{P}_k$ across the phase transitions are in SM~\cite{SM}.


\begin{figure*}[!t]
\centering
\includegraphics[width=0.92\textwidth]{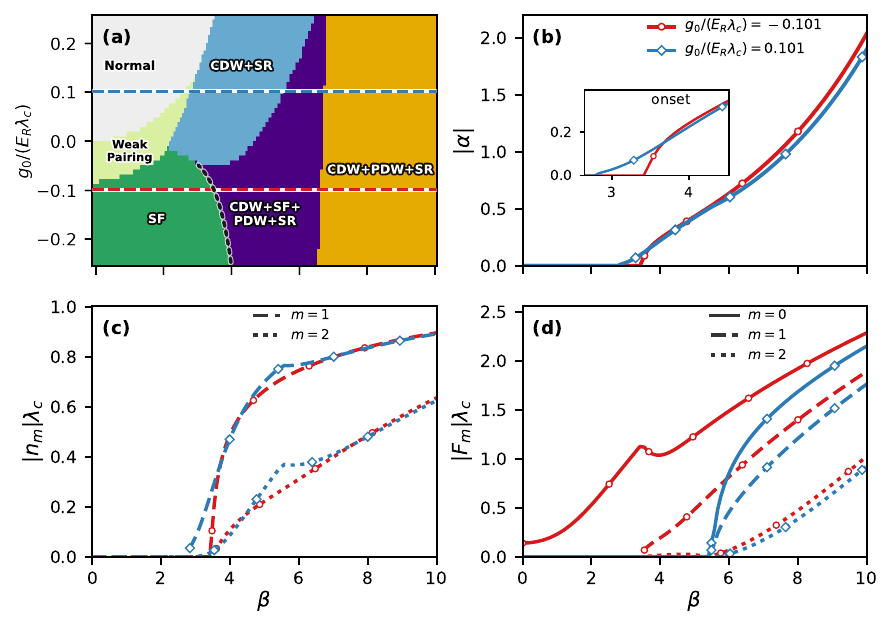}
\caption{
    %
    Mean-field phase diagram and signatures of self-organization.
    \textbf{(a)} Zero temperature phase diagram as a function of the pump strength $\beta$ (dimensionless) and the contact interaction strength $g_0/(E_R \lambda_c)$ (dimensionless). The dashed horizontal lines mark cuts at $g_0/(E_R \lambda_c)=\pm 0.101$ shown in the following plots. The black dotted curve shows the analytical prediction. The labels denote the normal, superfluid (SF), charge-density-wave (CDW), pair-density-wave (PDW), and superradiant (SR) phases. The pale green region ``weak pairing'' corresponds to weak effective attraction, for which the exponentially small pairing amplitude is below our numerical accuracy.
    \textbf{(b)} Intra-cavity field amplitude $|\alpha|$ along the two cuts from
    (a); the inset magnifies the onset of superradiance.
    Fourier components (dimensionless) of \textbf{(c)} the density $|n_m|\lambda_c$ and \textbf{(d)} the pairing field $|F_m|\lambda_c$, where $m\in\{0,1,2\}$.
    In \textbf{(b)}--\textbf{(d)}, the red circles and blue diamonds correspond to $g_0/(E_R \lambda_c)=-0.101$ and $0.101$, respectively. 
    The parameters: $N_{\uparrow}\!=\!N_{\downarrow}\!\!=\!\!1$ atom per unit cell, $\kappa/E_R=0.101$, $\Delta_c/E_R=-3.04$, $g_{\mathrm a}/E_R=-0.152$, and $g_{\Delta}/(E_R \lambda_c)=-0.0152$. 
     }\label{fig:phase_components}
\end{figure*}

\textit{Phase diagram}---Figure~\ref{fig:phase_components}(a) summarizes the interplay of the bare contact interaction and the cavity-induced single-atom and atom-pair couplings. The parameters are such that the single-atom dispersive coupling dominates over the atom-pair coupling, $|g_{\mathrm a} \lambda_c/g_\Delta|=10$.
This hierarchy is motivated by the experimental regime of Ref.~\cite{zwettler2025cavity}, in which the standard single-atom channel provides the dominant background self-organization mechanism, while the weaker pair channel modifies the threshold and the character of the ordered state. We characterize the phases through the intra-cavity photon number, the density inverse participation ratio, the Leggett-type fraction~\cite{PhysRevLett.25.1543} of Cooper pairs, and the Fourier components of the pairing field $F$. 
More details on the phases are presented in SM~\cite{SM}.

At weak pumping $0\leqslant\beta\lesssim2$, repulsive interactions support a normal state, whereas sufficiently attractive interactions produce a spatially uniform superfluid. Close to the non-interacting limit on the attractive side, the magnitude of $\Delta(y)$ is exponentially small and becomes challenging to distinguish from numerical zero~\cite{giorgini2008theory}, which we mark with the pale green region in Fig.~\ref{fig:phase_components}(a).

Upon increasing $\beta$, the repulsive and attractive sides exhibit qualitatively different ordering. For weak or repulsive $g_0$, the system enters the CDW+SR phase, characterized by a lack of pairing, density modulation, and a finite intra-cavity field. For sufficiently attractive $g_0$ and strong enough $\beta$, the anomalous field $F$ develops inhomogeneous spatial pattern quantified by finite momentum pairing, producing an intertwined CDW+SF+PDW+SR regime.
Finite-momentum pairing is the defining characteristic of PDW order, while its coupling to charge order represents an intertwined ordered state~\cite{agterberg2020physics}. At strong pumping $\beta\gtrsim6.5$, the boundary becomes nearly independent of $g_0$, and the CDW+PDW+SR regime extends over the full interaction range considered. In this region, the finite-momentum pairing component persists while the Leggett-type fraction is strongly suppressed, signaling strong localization of the pairing field. This behavior indicates that the state is governed predominantly by the cavity-induced atom-pair interaction, while the bare contact interactions play a secondary role. We next examine the intra-cavity photon number and the leading Fourier components of the density and pairing fields along the representative cuts marked in Fig.~\ref{fig:phase_components}(a).


\textit{Intra-cavity field}---The intra-cavity field amplitude
$|\alpha|$ provides a direct experimentally accessible signature of cavity
self-organization. In Fig.~\ref{fig:phase_components}(b) we show $|\alpha|$
along the attractive and repulsive interaction cuts
$g_0/(E_R\lambda_c)=\pm0.101$ [dashed lines in
Fig.~\ref{fig:phase_components}(a)], where
$E_R=\hbar^2 k_c^2/(2M)$ is the photon recoil energy. At weak pumping, the
intra-cavity field remains negligible, indicating the absence of coherent
scattering into the cavity mode. Above an interaction-dependent threshold,
$|\alpha|$ grows continuously from zero, which signals the onset of the phase with the broken $\mathbb Z_2$ symmetry.

The inset of Fig.~\ref{fig:phase_components}(b) resolves the small difference
between the two thresholds. For the repulsive case, the cavity field appears
at a slightly lower pump strength than for the attractive one. This difference
reflects the distinct initial many-body states: the repulsive interaction
leads to density self-organization from an unpaired state, whereas the
attractive interaction reorganizes an already paired atomic state. At
stronger pumping, the attractive case exhibits a larger intra-cavity field
amplitude due to the additional photon-scattering channel involving the
pairs.


\textit{Ordering patterns}---To characterize the spatial ordering of the atomic density and the pairing field, we resolve them into Fourier components of the cavity wave vector, $n_m=  \frac{1}{\lambda_c}
\int_0^{\lambda_c} dy\, n(y)e^{-imk_cy}$, $F_m =
\frac{1}{\lambda_c} \int_0^{\lambda_c} dy\, F(y)e^{-imk_cy}$.
Here, $m=1,2$ correspond to modulations at $k_c$, $2k_c$, respectively; $F_0$ denotes the uniform pairing component.

In Fig.~\ref{fig:phase_components}(c) we show that the onset of superradiance is accompanied by the appearance of the density harmonics. The fundamental component $|n_1|$ increases rapidly at the self-organization threshold and directly measures the density modulation associated with the broken $\mathbb Z_2$ symmetry. Simultaneously, the second harmonic $|n_2|$ becomes finite and grows with increasing $\beta$, showing that the density pattern develops structure beyond the fundamental cavity-mode modulation. At strong pumping, the density harmonics for $g_0/(E_R\lambda_c)=\pm0.101$ approach similar values, consistent with the nearly vertical CDW+PDW+SR phase boundary in Fig.~\ref{fig:phase_components}(a).

The pairing harmonics in Fig.~\ref{fig:phase_components}(d) reveal how the density ordering becomes intertwined with the anomalous field. For $g_0/(E_R\lambda_c)=-0.101$, the uniform component $|F_0|$ is already finite at $\beta=0$, as expected for the paired superfluid. At the same pump strength at which the density harmonics become finite, the finite-momentum pairing components also emerge. In particular, the simultaneous onset of $|n_1|$ and $|F_1|$ shows that CDW and PDW orders arise from the same self-consistent cavity instability rather than through separate transitions. The weak non-monotonic behavior of $|F_0|$ near the onset indicates a redistribution of pairing weight from the uniform component into finite-momentum harmonics as the paired state adapts to the cavity-induced modulation. The subsequent growth of $|F_2|$ shows that the pairing field also acquires higher spatial Fourier components.

For the repulsive case, the sequence is qualitatively different. The photon number and density harmonics become finite at the transition into the CDW+SR phase, whereas the pairing components remain negligible over an extended range of pump strengths. Only at a larger $\beta$ do $|F_0|$, $|F_1|$, and $|F_2|$ emerge. Thus, on the repulsive side, the cavity first produces density self-organization and subsequently induces pairing once the cavity-induced atomic interaction becomes sufficiently strongly attractive. 
Because the uniform and finite-momentum pairing components emerge within an already formed cavity lattice, the resulting paired state is intrinsically spatially modulated.

At strong pumping, both analyzed $g_0/(E_R\lambda_c)=\pm0.101$ cases display finite Fourier components of the density and pairing fields, explaining why the CDW+PDW+SR phase extends over the full interaction range in Fig.~\ref{fig:phase_components}(a). The increasing values of the higher harmonics indicate stronger density localization and a more pronounced spatial modulation of the pairing field. This interpretation is supported by the analysis of the density inverse participation ratio, the map of the Leggett-type fraction, and representative real-space profiles in SM~\cite{SM}.


\begin{figure}[t]
    \centering
    \includegraphics[width=0.88\columnwidth]{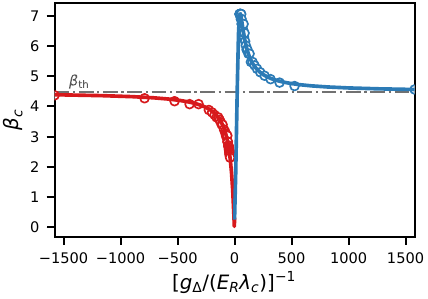}
    \caption{
    Fano-type superradiant phase boundary in the strongly attractive regime. The critical pump amplitude $\beta_c$ to the ordered phase is shown as a function of the dimensionless inverse atom-pair dispersive coupling $x=E_R \lambda_c/g_\Delta$. Red (blue) circles correspond to
    $g_\Delta<0$ ($>0$). The solid curve show the Fano-shape fit to the threshold pump intensity $\beta_c^2(x)$. The horizontal
    dash-dotted line marks the fitted background $\beta_{\rm th}$.
    The parameters: 
    $N_\uparrow=N_\downarrow=1$ per unit cell, $g_0/(E_R\lambda_c)\simeq -3.04$,
    $g_{\mathrm a}/E_R\simeq -0.38$, $\Delta_c/E_R\simeq -4.05$, and
    $\kappa/E_R\simeq 0.10$.
    }
    \label{fig:fano_boundary}
\end{figure}

\textit{Fano-type superradiant phase boundary}---We finally examine how the pairing channel modifies the threshold for self-organization. We choose a strongly attractive bare interaction, $g_0/(E_R\lambda_c)\simeq -3.04$, for which pairing correlations are well developed. This regime facilitates a qualitative comparison with the unitary Fermi gas experiment of Ref.~\cite{zwettler2025cavity}, although the strongly attractive 1D model considered here is not equivalent to a 3D gas at unitarity. In Fig.~\ref{fig:fano_boundary}, we show the critical pump amplitude $\beta_c$ to the ordered phase as a function of the dimensionless inverse atom-pair dispersive coupling $x=[g_\Delta/(E_R\lambda_c)]^{-1}$. This quantity is proportional to the bare molecular detuning, ${x\propto}\Delta_B$. In the limits $x\rightarrow\pm\infty$, corresponding to $g_\Delta\rightarrow 0^\pm$, the contribution of the pairing channel becomes negligible and the threshold approaches the background value $\beta_{\rm th}$ set by the single-atom channel. In contrast, near $x\approx0$, corresponding to large $|g_\Delta|$, the cavity-induced interaction strongly renormalizes the critical value $\beta_c$.

For $g_\Delta<0$, the calculated threshold is lower than $\beta_{\rm th}$, indicating that the single-atom and atom-pair channels act cooperatively with the cavity field that drives self-organization. For $g_\Delta>0$, the two channels instead compete, increasing the threshold above $\beta_{\rm th}$. This change from cooperation to competition across $x=0$ produces the asymmetric dip--peak profile in Fig.~\ref{fig:fano_boundary}.

The asymmetric boundary is characteristic of a Fano-type
resonance~\cite{fano1961effects}, in which a tunable channel interferes
with a nonresonant background. Since $\beta^2$ is proportional to the
pump intensity, we fit the numerical threshold intensity to
$\beta_{\rm fit}^2(x)=\beta_{\rm th}^2
\left[q+(x-x_0)/w\right]^2/
\left\{1+\left[(x-x_0)/w\right]^2\right\}$,
where $\beta_{\rm th}$ is the background threshold, $q$ controls the
asymmetry, and $x_0$ and $w$ denote the resonance position and width,
respectively. In Fig.~\ref{fig:fano_boundary} we demonstrate the fit, with $\beta_{\rm th}\simeq 4.47$ and $q\simeq 1.23$, agrees well with
the numerical results.
In SM, we show other values of $g_0$ yield similar asymmetric profiles~\cite{SM}. Its resemblance to the boundary observed in the unitary Fermi gas experiment~\cite{zwettler2025cavity} suggests the interference of the density and pairing order persists across dimensionalities and interaction regimes.


\textit{Conclusions and perspectives}---We mapped out the interaction--pump phase diagram of an interacting two-component degenerate Fermi gas with simultaneous coupling of cavity photons to internal single atom transitions and molecular type atomic pair states. 
The interplay of cavity-induced and bare contact interactions selects whether the particle self-ordering proceeds first through the buildup of a  density wave or through a joint pair-density instability. 

Our system provides a controlled analogue of
intertwined orders in unconventional superconductors. Such a neutral atomic superfluid offers independently tunable contact and photon-induced interactions that permit direct exploration of when density and pairing phases compete, cooperate, or combine into a finite-momentum paired state \cite{t4xb-6x3z,RevModPhys.87.457, hamidian2016detection, agterberg2020physics}.
Atomic platforms complement the field of cavity quantum materials, where confined electromagnetic fields have been predicted and recently observed to alter superconducting properties \cite{PhysRevLett.123.133601,schlawin2022cavity, kozin2025cavity, Zheng_2026}.

Several directions could explore physics beyond the presented mean-field approach, such as inclusion of quantum fluctuations, population imbalance, extension to higher-dimensions or multimode cavities. Our results identify cavity-coupled Fermi gases as a versatile, programmable platform for generating intertwined orders that are difficult to control in quantum materials~\cite{morales2018coupling}.



\bigskip
\begingroup
\makeatletter
\let\addcontentsline\@gobblethree
\makeatother
\begin{acknowledgments}
This research is part of the project No. 2021/43/P/ST2/02911 co-funded by the National Science Centre and the European Union Framework Programme for Research and Innovation Horizon 2020 under the Marie Sk{\l}odowska-Curie grant agreement No. 945339. 
This research was also funded in part by the Austrian Science Fund (FWF) [grant DOIs: 10.55776/P35891 and 10.55776/PAT2450125]. 
F.\,M.\ acknowledges support from the Joint Excellence in Science and Humanities (JESH) Mobility Programme of the Austrian Academy of Sciences (ÖAW), during which he had numerous fruitful discussions with Santiago F.\ Caballero Benitez. 
For the purpose of Open Access, the author has applied a CC-BY public copyright licence to any Author Accepted Manuscript (AAM) version arising from this submission. 
We gratefully acknowledge Polish high-performance computing infrastructure PLGrid (HPC Centers: ACK Cyfronet AGH, WCSS) for providing computer facilities and support within computational grant no. PLG/2026/019142.
\end{acknowledgments}
\endgroup

\clearpage
\onecolumngrid
\setcounter{section}{0}
\setcounter{equation}{0}
\setcounter{figure}{0}
\renewcommand{\thesection}{S\Alph{section}}
\makeatletter
\@addtoreset{equation}{section}
\makeatother
\renewcommand{\theequation}{S\Alph{section}\arabic{equation}}
\renewcommand{\thefigure}{S\arabic{figure}}
\renewcommand{\theHsection}{SM.\Alph{section}}
\renewcommand{\theHequation}{SM.\Alph{section}.\arabic{equation}}
\renewcommand{\theHfigure}{SM.\arabic{figure}}

\begingroup
\makeatletter
\let\addcontentsline\@gobblethree
\makeatother
\section*{Supplemental Material}
\endgroup
\tableofcontents
\clearpage

\section{Photon--atom-pair coupling in a cavity and adiabatic elimination}
\label{sec:SM_micro}

\begin{figure}[H]
    \centering
    \includegraphics[
        width=\columnwidth,
        trim=5 5 5 5,
        clip
    ]{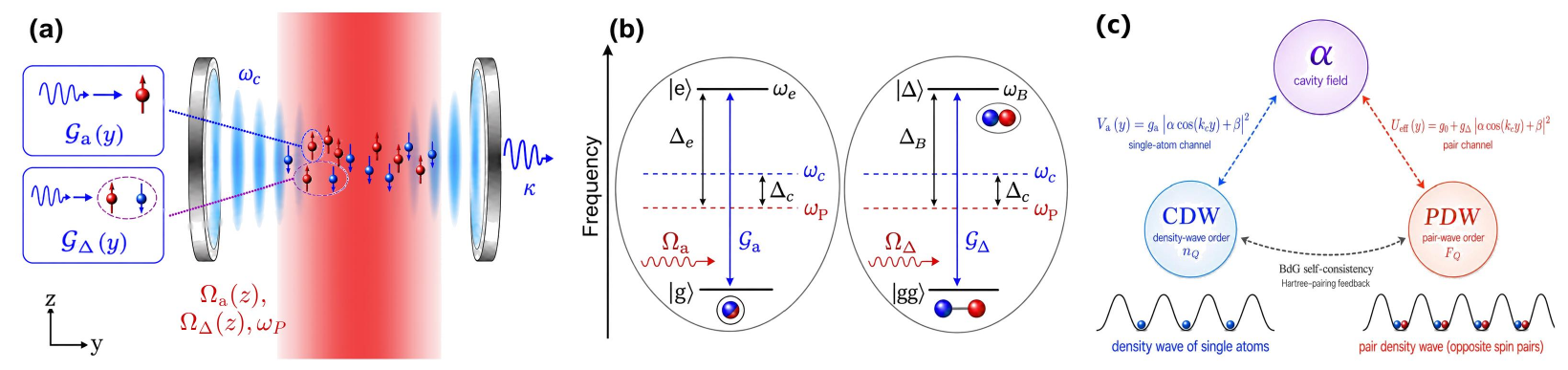}
    \caption{
    \textbf{(a)} Schematic of a quasi-1D two-component Fermi gas transversely driven inside a single-mode optical cavity. The gas is elongated along the cavity axis $y$ and tightly confined transversally. The standing-wave cavity mode of frequency $\omega_c$ couples to a single atom, with position-dependent coupling $\mathcal G_{\rm a}(y)$, and to an opposite-spin atom pair, with coupling $\mathcal G_{\Delta}(y)$. The transverse pump has frequency $\omega_P$ and position-dependent Rabi frequencies $\Omega_{\rm a}(z)$ and $\Omega_{\Delta}(z)$ for the two channels, while photons escape from the cavity at rate $\kappa$.
    \textbf{(b)} Relevant optical transitions for the single-atom channel $\lvert g\rangle\leftrightarrow\lvert e\rangle$ and the molecular channel $\lvert gg\rangle\leftrightarrow\lvert\Delta\rangle$. The corresponding cavity couplings are $\mathcal G_{\rm a}$ and $\mathcal G_{\Delta}$, while the two transitions are driven by the transverse pump with Rabi frequencies $\Omega_{\rm a}$ and $\Omega_{\Delta}$. The blue and red dashed lines indicate the cavity and pump frequencies, $\omega_c$ and $\omega_P$, respectively, with detunings $\Delta_c$, $\Delta_e$, and $\Delta_B$.
    \textbf{(c)} Coupled cavity, density-wave, and pair-density-wave fields. The
    coherent cavity amplitude $\alpha$ couples to the density modulation
    through the single-atom optical potential $V_{\rm a}(y)$ and to the
    pairing modulation through the spatially dependent effective interaction
    $U_{\rm eff}(y)$. The density and pairing fields are mutually coupled
    through the Hartree and pairing channels of the self-consistent Bogoliubov-de Gennes equations, while both depend on the cavity field. The lower sketches
    illustrate the corresponding periodic density and opposite-spin pair density.}
    \label{fig:SM_cavity_setup}
\end{figure}

The cavity configuration, the relevant single-atom and molecular optical transitions, and the resulting coupling among the cavity, density, and pairing channels are summarized in Fig.~\ref{fig:SM_cavity_setup}.

The fermionic atoms in two hyperfine states $\sigma \in \{\uparrow,\downarrow\}$ are described by 1D field operators $\hat\psi_\sigma(y)$. The atomic part of the total Hamiltonian is
\begin{align}
\hat H_A
={}&
\sum_{\sigma=\uparrow,\downarrow}
\int dy\,
\hat\psi_\sigma^\dagger(y)
\left(-\frac{\hbar^2}{2m}\partial_y^2\right)
\hat\psi_\sigma(y)+
 g_0\int dy\,
\hat\psi_\uparrow^\dagger(y)
\hat\psi_\downarrow^\dagger(y)
\hat\psi_\downarrow(y)
\hat\psi_\uparrow(y),
\label{eq:SA_HA}
\end{align}
where $g_0$ is the bare atom--atom coupling describing the strength of the contact interactions. In the effective one-dimensional description, its value is fixed by the three-dimensional $s$-wave scattering length and the transverse confinement.

The cavity supports a single photon mode of frequency $\omega_c$,
\begin{equation}
\hat H_C=\hbar\omega_c\,\hat a^\dagger\hat a,
\label{eq:SA_HC}
\end{equation}
where $\hat a$ annihilates a cavity photon. After projection onto the transverse ground state, the mode function of the cavity is
\begin{equation}
f(y)=\cos(k_cy),
\qquad
k_c=\frac{\omega_c}{c}.
\label{eq:SA_mode}
\end{equation}
The mode function has period $\lambda_c = 2\pi/k_c$, whereas its intensity $\propto f^2(y)$ has period $\pi/k_c = \lambda_c/2$. Because the pump--cavity interference terms are proportional to $\cos(k_cy)$, the full self-consistent mean fields generally have the fundamental period $2\pi/k_c$. The cavity and pump geometry, including the two optical coupling channels and photon loss at rate $\kappa$, is illustrated in Fig.~\ref{fig:SM_cavity_setup}(a).

The electronically excited molecular bound state $\ket{\Delta}$ is represented by a molecular field operator $\hat\psi_\Delta(y)$ with its Hamiltonian
\begin{equation}
\hat H_\Delta
=
\int dy\,
\hat\psi_\Delta^\dagger(y)
\left[-\frac{\hbar^2}{2M}\partial_y^2+E_B\right]
\hat\psi_\Delta(y),
\label{eq:SA_HDelta}
\end{equation}
where $M=2m$ denotes the mass of the molecule. Here $E_B=\hbar\omega_B$ is the internal energy of the excited molecular bound state, measured relative to the threshold energy of two atoms in the ground state.

The coherent conversion between an opposite-spin atom pair and the molecular excitation is driven by a cavity photon and the transverse pump,
\begin{align}
\hat H_{CA}^{(\Delta)}
={}&
\int dy\,
\hat\psi_\Delta^\dagger(y)
\hat\psi_\downarrow(y)
\hat\psi_\uparrow(y)
\left[
\mathcal G_\Delta(y) \hat a
+
\Omega_\Delta(z)e^{-i\omega_Pt}
\right]
+\mathrm{H.c.},
\label{eq:SA_HCA_pair}
\end{align}
where $\mathrm{H.c.}$ is the Hermitian conjugate, and $\mathcal G_{\Delta} (y)$ denotes the atom-pair coupling strength to the quantized cavity mode,
\begin{equation}
\mathcal G_\Delta(y)
=
\widetilde{\mathcal G}_\Delta f(y)
=
\widetilde{\mathcal G}_\Delta\cos(k_cy).
\label{eq:SA_GDelta}
\end{equation}
The constant $\widetilde{\mathcal G}_\Delta$ is the maximum
single atom-pair--cavity-photon coupling strength. The corresponding
atom-pair--molecule transition is also driven by the transverse pump
with the spatially dependent coupling
\begin{equation}
\Omega_\Delta(z)
=
\widetilde{\Omega}_\Delta\cos(k_c z),
\label{eq:SA_pump_pair}
\end{equation}
where $\widetilde{\Omega}_\Delta$ denotes the maximum pump
coupling in the atom-pair channel. The corresponding single-atom pump
coupling is introduced analogously as
$\Omega_{\mathrm a}(z)
=\widetilde{\Omega}_{\mathrm a}\cos(k_c z)$. The parameter $z$ in the Hamiltonian denotes the position of the 1D fermionic cloud. Specifically, we set $z=0$ in the numerical calculations.

We also include a conventional single-atom coupling through electronically excited atomic states $\hat\psi_{e\sigma}(y)$ of internal energy $\hbar\omega_e$,
\begin{equation}
\hat H_e
=
\sum_{\sigma=\uparrow,\downarrow}
\int dy\,
\hat\psi_{e\sigma}^\dagger(y)
\left[-\frac{\hbar^2}{2m}\partial_y^2+\hbar\omega_e\right]
\hat\psi_{e\sigma}(y),
\label{eq:SA_He}
\end{equation}
with the corresponding light--matter coupling
\begin{align}
\hat H_{CA}^{(e)}
={}&
\sum_{\sigma=\uparrow,\downarrow}
\int dy\,
\hat\psi_{e\sigma}^\dagger(y)
\hat\psi_\sigma(y)
\left[
\mathcal G_{\mathrm a}(y) \hat a
+
\Omega_\mathrm{a}(z)e^{-i\omega_Pt}
\right]
+\mathrm{H.c.},
\label{eq:SA_HCA_single}
\end{align}
where $\mathcal G_{\mathrm a}(y)$ denotes the single atom coupling strength to the quantized cavity mode,
\begin{equation}
\mathcal G_{\mathrm a}(y)
=
\widetilde{\mathcal G}_{\mathrm a}f(y)
=
\widetilde{\mathcal G}_{\mathrm a}\cos(k_cy).
\label{eq:SA_Ga}
\end{equation}
Here $\widetilde{\mathcal G}_{\mathrm a}$ is the maximum of the single atom--photon cavity-QED coupling strength. 

The full microscopic Hamiltonian is then
\begin{equation}
\hat H
=
\hat H_A+\hat H_C+\hat H_\Delta+\hat H_e
+\hat H_{CA}^{(\Delta)}+\hat H_{CA}^{(e)}.
\label{eq:SA_Hfull}
\end{equation}

Moving to a frame rotating at the pump frequency $\omega_P$ introduces the detunings
\begin{equation}
\Delta_c=\omega_P-\omega_c,
\qquad
\Delta_B=\omega_P-\omega_B,
\qquad
\Delta_e=\omega_P-\omega_e.
\label{eq:SA_detunings}
\end{equation}
The corresponding single-atom and molecular optical transitions are shown in Fig.~\ref{fig:SM_cavity_setup}(b).
Setting $\hbar=1$, the Hamiltonian in the rotating frame reads
\begin{align}
\hat H
={}&
\sum_{\sigma=\uparrow,\downarrow}
\int dy\,
\hat\psi_\sigma^\dagger(y)
\left(-\frac{\partial_y^2}{2m}\right)
\hat\psi_\sigma(y)
+
 g_0\int dy\,
\hat\psi_\uparrow^\dagger
\hat\psi_\downarrow^\dagger
\hat\psi_\downarrow
\hat\psi_\uparrow -
\Delta_c\hat a^\dagger\hat a
+
\int dy\,
\hat\psi_\Delta^\dagger(y)
\left(-\frac{\partial_y^2}{2M}-\Delta_B\right)
\hat\psi_\Delta(y)
\nonumber\\
&+
\sum_{\sigma=\uparrow,\downarrow}
\int dy\,
\hat\psi_{e\sigma}^\dagger(y)
\left(-\frac{\partial_y^2}{2m}-\Delta_e\right)
\hat\psi_{e\sigma}(y) + \left\{
\int dy\,
\hat\psi_\Delta^\dagger(y)
\hat\psi_\downarrow(y)
\hat\psi_\uparrow(y)
\left[
\mathcal G_\Delta(y) \hat a + \Omega_\Delta(z)
\right]
+\mathrm{H.c.} \right\}
\nonumber\\
&+ \left\{
\sum_{\sigma=\uparrow,\downarrow}
\int dy\,
\hat\psi_{e\sigma}^\dagger(y)
\hat\psi_\sigma(y)
\left[
\mathcal G_{\mathrm a}(y) \hat a +\Omega_{\mathrm a}(z)
\right]
+\mathrm{H.c.} \right\}.
\label{eq:SA_Hrot}
\end{align}

We work in the far-detuned dispersive regime, $|\Delta_B|,|\Delta_e|$ much larger than the relevant many-body and optical coupling scales. The excited molecular and excited atomic fields are then only virtually populated and can be eliminated to leading order in $1/\Delta_B$ and $1/\Delta_e$. Neglecting their time evolution and kinetic energies compared with the corresponding detunings, their Heisenberg equations give
\begin{equation}
\hat\psi_\Delta(y)
\simeq
\frac{1}{\Delta_B}
\hat\psi_\downarrow(y)\hat\psi_\uparrow(y)
\left[
\mathcal G_\Delta(y)\hat a+\Omega_\Delta(z)
\right],
\label{eq:SA_psiDelta_ad}
\end{equation}
and
\begin{equation}
\hat\psi_{e\sigma}(y)
\simeq
\frac{1}{\Delta_e}
\hat\psi_\sigma(y)
\left[
\mathcal G_{\rm a}(y)\hat a+\Omega_{\rm a}(z)
\right].
\label{eq:SA_psie_ad}
\end{equation}

The resulting low-energy Hamiltonian is
\begin{align}
\hat H_{\mathrm{eff}}
={}&
-\Delta_c\hat a^\dagger\hat a
+
\sum_{\sigma=\uparrow,\downarrow}
\int dy\,
\hat\psi_\sigma^\dagger(y)
\left(-\frac{\partial_y^2}{2m}\right)
\hat\psi_\sigma(y)
\nonumber\\
&+
 g_0\int dy\,
\hat\psi_\uparrow^\dagger(y)
\hat\psi_\downarrow^\dagger(y)
\hat\psi_\downarrow(y)
\hat\psi_\uparrow(y)
\nonumber\\
&+
\frac{1}{\Delta_B}
\int dy\,
\hat\psi_\uparrow^\dagger(y)
\hat\psi_\downarrow^\dagger(y)
\hat\psi_\downarrow(y)
\hat\psi_\uparrow(y)
\left[
\mathcal G_\Delta(y) \hat a^\dagger + \Omega_\Delta(z)
\right]
\left[
\mathcal G_\Delta(y) \hat a + \Omega_\Delta(z)
\right]
\nonumber\\
&+
\frac{1}{\Delta_e}
\sum_{\sigma=\uparrow,\downarrow}
\int dy\,
\hat\psi_\sigma^\dagger(y)
\hat\psi_\sigma(y)
\left[
\mathcal G_{\mathrm a}(y) \hat a^\dagger +\Omega_\mathrm{a}(z)
\right]
\left[
\mathcal G_{\mathrm a}(y) \hat a + \Omega_\mathrm{a}(z)
\right].
\label{eq:SA_Heff}
\end{align}
The corresponding Heisenberg equations for the fermionic fields of the ground-state atoms are
\begin{align}
    i\partial_t\hat\psi_\uparrow(y)
    ={}&
    \left[
    -\frac{\partial_y^2}{2m}
    +
    \frac{
    \left[\mathcal G_{\rm a}(y)\hat a^\dagger+\Omega_{\rm a}(z)\right]
    \left[\mathcal G_{\rm a}(y)\hat a+\Omega_{\rm a}(z)\right]
    }{\Delta_e}
    \right]
    \hat\psi_\uparrow(y)
    \nonumber\\
    &+
    \left[
    g_0+
    \frac{
    \left[\mathcal G_\Delta(y)\hat a^\dagger+\Omega_\Delta(z)\right]
    \left[\mathcal G_\Delta(y)\hat a+\Omega_\Delta(z)\right]
    }{\Delta_B}
    \right]
    \hat\psi_\downarrow^\dagger(y)
    \hat\psi_\downarrow(y)
    \hat\psi_\uparrow(y),
\label{eq:SA_EOM_up}
\end{align}
and
\begin{align}
    i\partial_t\hat\psi_\downarrow(y)
    ={}&
    \left[
    -\frac{\partial_y^2}{2m}
    +
    \frac{
    \left[\mathcal G_{\rm a}(y)\hat a^\dagger+\Omega_{\rm a}(z)\right]
    \left[\mathcal G_{\rm a}(y)\hat a+\Omega_{\rm a}(z)\right]
    }{\Delta_e}
    \right]
    \hat\psi_\downarrow(y)
    \nonumber\\
    &+
    \left[
    g_0+
    \frac{
    \left[\mathcal G_\Delta(y)\hat a^\dagger+\Omega_\Delta(z)\right]
    \left[\mathcal G_\Delta(y)\hat a+\Omega_\Delta(z)\right]
    }{\Delta_B}
    \right]
    \hat\psi_\uparrow^\dagger(y)
    \hat\psi_\uparrow(y)
    \hat\psi_\downarrow(y).
\label{eq:SA_EOM_down}
\end{align}

The cavity field obeys
\begin{align}
    i\partial_t\hat a
    ={}&
    -\Delta_c\hat a
    +
    \frac{1}{\Delta_e}
    \sum_{\sigma=\uparrow,\downarrow}
    \int dy\,
    \hat\psi_\sigma^\dagger(y)\hat\psi_\sigma(y)
    \mathcal G_{\rm a}(y)
    \left[
    \mathcal G_{\rm a}(y)\hat a+\Omega_{\rm a}(z)
    \right]
    \nonumber\\
    &+
    \frac{1}{\Delta_B}
    \int dy\,
    \hat\psi_\uparrow^\dagger(y)
    \hat\psi_\downarrow^\dagger(y)
    \hat\psi_\downarrow(y)
    \hat\psi_\uparrow(y)
    \mathcal G_\Delta(y)
    \left[
    \mathcal G_\Delta(y)\hat a+\Omega_\Delta(z)
    \right].
\label{eq:SA_EOM_a}
\end{align}
For a tightly confined cloud centered near an antinode of the transverse pump, i.e., for $z=0$, we have $\Omega_i(z=0) = \tilde\Omega_i$, where $i=\mathrm{a},\Delta$, across the atomic sample. 

Treating the cavity field at the mean field level, we replace $\hat a\rightarrow\alpha=\langle\hat a\rangle$, which leads to
\begin{align}
\hat H_\alpha
={}&
-\Delta_c|\alpha|^2
+
\sum_{\sigma=\uparrow,\downarrow}
\int dy\,
\hat\psi_\sigma^\dagger(y)
\left[
-\frac{\partial_y^2}{2m}
-\mu_\sigma
+V_{\mathrm a}(y)
\right]
\hat\psi_\sigma(y)
\nonumber\\
&+
\int dy\,
U_{\mathrm{eff}}(y)
\hat\psi_\uparrow^\dagger(y)
\hat\psi_\downarrow^\dagger(y)
\hat\psi_\downarrow(y)
\hat\psi_\uparrow(y),
\label{eq:SA_Halpha}
\end{align}
where
\begin{equation}
V_{\mathrm a}(y)
=
\frac{
\left[\alpha^*\mathcal G_{\mathrm a}(y)+\Omega_\mathrm{a}\right]
\left[\alpha\mathcal G_{\mathrm a}(y)+\Omega_\mathrm{a}\right]
}{\Delta_e},
\label{eq:SA_Va}
\end{equation}
and
\begin{equation}
U_{\mathrm{eff}}(y)
=
 g_0
+
\frac{
\left[\alpha^*\mathcal G_\Delta(y)+\Omega_\Delta\right]
\left[\alpha\mathcal G_\Delta(y)+\Omega_\Delta\right]
}{\Delta_B}.
\label{eq:SA_Ueff}
\end{equation}
The resulting coupling of the cavity field to the density and pairing channels is illustrated schematically in Fig.~\ref{fig:SM_cavity_setup}(c). The single-atom channel produces the optical potential $V_{\rm a}(y)$, whereas
the pair channel produces the spatially dependent interaction
related to $U_{\rm eff}(y)$. The density and pairing fields subsequently adjust the cavity amplitude through the self-consistency condition derived below.

\section{Hartree--Fock--Bogoliubov decoupling and self-consistency}
\label{sec:SM_HFB}

Because the atoms are fermions and the interaction involves a contact potential, the two-body interaction in Eq.~\eqref{eq:SA_Halpha} involves only opposite spins. Consequently, we introduce the densities and the anomalous pair amplitude
\begin{equation}
n_\sigma(y)
=
\left\langle
\hat\psi_\sigma^\dagger(y)\hat\psi_\sigma(y)
\right\rangle,
\qquad
F(y)
=
\left\langle
\hat\psi_\downarrow(y)\hat\psi_\uparrow(y)
\right\rangle.
\label{eq:SB_averages}
\end{equation}
Neglecting quadratic fluctuations around these averages gives the Hartree--Fock--Bogoliubov decoupling
\begin{align}
\hat\psi_\uparrow^\dagger \hat\psi_\downarrow^\dagger \hat\psi_\downarrow \hat\psi_\uparrow
\simeq{}& n_\downarrow \hat\psi_\uparrow^\dagger\hat\psi_\uparrow
+ n_\uparrow \hat\psi_\downarrow^\dagger\hat\psi_\downarrow
- n_\uparrow n_\downarrow  + F \hat\psi_\uparrow^\dagger\hat\psi_\downarrow^\dagger
+ F^* \hat\psi_\downarrow\hat\psi_\uparrow
- |F|^2.
\label{eq:SB_decoupling}
\end{align}
All fields in Eq.~\eqref{eq:SB_decoupling} are evaluated at the same position $y$. Here we have omitted the contribution $\propto \langle\hat\psi_\downarrow^\dagger\hat\psi_\uparrow\rangle$, since in our model there are no spin-flip processes and the Hamiltonian conserves the total number of atoms in each spin state, i.e., $N_\uparrow$ and $N_\downarrow$, separately.

We define the pairing field
\begin{equation}
\Delta(y)
\equiv
U_{\mathrm{eff}}(y)F(y).
\label{eq:SB_Delta_def}
\end{equation}
With this definition, the mean-field Hamiltonian is
\begin{align}
\hat H_{\mathrm{MF}}
={}&
-\Delta_c|\alpha|^2+E_0
+
\sum_{\sigma=\uparrow,\downarrow}
\int dy\,
\hat\psi_\sigma^\dagger(y)
 h_\sigma(y)
\hat\psi_\sigma(y)
\nonumber\\
&+
\int dy\,
\left[
\Delta(y)
\hat\psi_\uparrow^\dagger(y)
\hat\psi_\downarrow^\dagger(y)
+
\Delta^*(y)
\hat\psi_\downarrow(y)
\hat\psi_\uparrow(y)
\right],
\label{eq:SB_HMF}
\end{align}
where
\begin{subequations} \label{eq:SB_hsigma}
\begin{align}
h_\uparrow(y)
={}&
-\frac{\partial_y^2}{2m}
-\mu_\uparrow
+V_{\mathrm a}(y)
+U_{\mathrm{eff}}(y)n_\downarrow(y),
\\
h_\downarrow(y)
={}&
-\frac{\partial_y^2}{2m}
-\mu_\downarrow
+V_{\mathrm a}(y)
+U_{\mathrm{eff}}(y)n_\uparrow(y),
\end{align}
\end{subequations}
and
\begin{equation}
E_0
=
-\int dy\,
U_{\mathrm{eff}}(y)
\left[
 n_\uparrow(y)n_\downarrow(y)+|F(y)|^2
\right].
\label{eq:SB_E0}
\end{equation}

For a spin-balanced system, denoting the total density by $n$,
\begin{equation}
\mu_\uparrow=\mu_\downarrow\equiv\mu,
\qquad
n_\uparrow(y)=n_\downarrow(y)\equiv\frac{n(y)}{2},
\label{eq:SB_balance}
\end{equation}
so that $h_\uparrow=h_\downarrow\equiv h$ with
\begin{equation}
h(y)
=
-\frac{\partial_y^2}{2m}
-\mu
+V_{\mathrm a}(y)
+U_{\mathrm{eff}}(y)\frac{n(y)}{2}.
\label{eq:SB_hbalanced}
\end{equation}

Next, we define the standard Nambu spinor
\begin{equation}
\hat\Phi(y)
=
\begin{pmatrix}
\hat\psi_\uparrow(y)\\
\hat\psi_\downarrow^\dagger(y)
\end{pmatrix}.
\label{eq:SB_Nambu}
\end{equation}
Then, diagonalization of Eq.~\eqref{eq:SB_HMF} it equivalent to solving the Hamiltonian
\begin{equation}
\mathcal H_{\mathrm{BdG}}(y)
=
\begin{pmatrix}
 h_\uparrow(y) & \Delta(y)\\
 \Delta^*(y) & -h_\downarrow^*(y)
\end{pmatrix}.
\label{eq:SB_BdGmatrix}
\end{equation}
The corresponding BdG eigenproblem is
\begin{equation}
\mathcal H_{\mathrm{BdG}}(y)
\begin{pmatrix}
u_n(y)\\v_n(y)
\end{pmatrix}
=
E_n
\begin{pmatrix}
u_n(y)\\v_n(y)
\end{pmatrix},
\qquad E_n>0.
\label{eq:SB_BdG}
\end{equation}
We note that the sums below are taken over one representative from each particle-hole pair, chosen as the positive-energy modes.

The Bogoliubov transformation consistent with Eqs.~\eqref{eq:SB_Delta_def}, \eqref{eq:SB_Nambu} and \eqref{eq:SB_BdGmatrix} is
\begin{align}
\hat\psi_\uparrow(y)
={}&
\sum_{E_n>0}
\left[
 u_n(y)\hat\gamma_{n\uparrow}
-
 v_n^*(y)\hat\gamma_{n\downarrow}^\dagger
\right],
\nonumber\\
\hat\psi_\downarrow(y)
={}&
\sum_{E_n>0}
\left[
 u_n(y)\hat\gamma_{n\downarrow}
+
 v_n^*(y)\hat\gamma_{n\uparrow}^\dagger
\right],
\label{eq:SB_Bogoliubov}
\end{align}
where $\hat\gamma_{n\sigma}$, with $\sigma\in \{\uparrow,\downarrow\}$, are the quasiparticle operators.
Indeed, the coefficient of $\hat\gamma_{n\uparrow}$ in the Nambu field $\hat\Phi$ is $(u_n,v_n)^T$, which is the positive-energy eigenvector of Eq.~\eqref{eq:SB_BdG}. The relative signs in Eq.~\eqref{eq:SB_Bogoliubov} are also essential for the anomalous average. Substituting Eq.~\eqref{eq:SB_Bogoliubov} into $F=\langle\hat\psi_\downarrow\hat\psi_\uparrow\rangle$ gives
\begin{align}
F(y)
={}&
\sum_{m,n}
\Bigl[
-u_m(y)v_n^*(y)
\left\langle
\hat\gamma_{m\downarrow}
\hat\gamma_{n\downarrow}^\dagger
\right\rangle
+
 v_m^*(y)u_n(y)
\left\langle
\hat\gamma_{m\uparrow}^\dagger
\hat\gamma_{n\uparrow}
\right\rangle
\Bigr]
\nonumber\\
={}&
-\sum_{E_n>0}
 u_n(y)v_n^*(y)
\left[1-f(E_n)\right]
+
\sum_{E_n>0}
 u_n(y)v_n^*(y)f(E_n)
\nonumber\\
={}&
-\sum_{E_n>0}
 u_n(y)v_n^*(y)
\left[1-2f(E_n)\right],
\label{eq:SB_Fsc}
\end{align}
where
\begin{equation}
f(E)=\frac{1}{e^{E/(k_BT)}+1}.
\label{eq:SB_fermi}
\end{equation}
Combining Eq.~\eqref{eq:SB_Fsc} with the definition $\Delta=U_{\mathrm{eff}}F$ yields
\begin{equation}
\Delta(y)
=
-U_{\mathrm{eff}}(y)
\sum_{E_n>0}
 u_n(y)v_n^*(y)
\left[1-2f(E_n)\right].
\label{eq:SB_gap}
\end{equation}

The spin densities are
\begin{align}
n_\uparrow(y)
={}&
\sum_{E_n>0}
\left[
|u_n(y)|^2f(E_n)
+
|v_n(y)|^2\left(1-f(E_n)\right)
\right],
\label{eq:SB_nup}\\
n_\downarrow(y)
={}&
\sum_{E_n>0}
\left[
|u_n(y)|^2f(E_n)
+
|v_n(y)|^2\left(1-f(E_n)\right)
\right],
\label{eq:SB_ndown}
\end{align}

The cavity amplitude $\alpha$ is obtained from its steady-state mean-field condition; cavity losses enter through the complex detuning $\Delta_c\rightarrow \Delta_c+i\kappa$ in the solution for the steady state.

\section{Bloch formulation and the limit of a large system}
\label{sec:SM_Bloch}

Because the pump--cavity interference terms are proportional to
$\cos(k_cy)$, the system is periodic with the cavity
wavelength and can be represented within a unit cell of length
\begin{equation}
a\equiv\frac{2\pi}{k_c} = \lambda_c,
\qquad
L\equiv N_La,
\label{eq:SC_cell}
\end{equation}
where the integer $N_L$ is the number of periodically repeated cells forming the
total system of length $L$. In the limit of large system size,
$N_L\rightarrow\infty$, the HFB problem can be solved within
one unit cell.

We define the total atomic density and the local
opposite-spin pair density as
\begin{equation}
n(y)
=
n_\uparrow(y)+n_\downarrow(y),
\qquad
p(y)
\equiv \langle \hat\psi_\uparrow^\dagger(y)
\hat\psi_\downarrow^\dagger(y)
\hat\psi_\downarrow(y)
\hat\psi_\uparrow(y) \rangle 
= n_\uparrow(y)n_\downarrow(y)+|F(y)|^2.
\label{eq:SC_densities}
\end{equation}
The corresponding cavity-mode overlaps, that appear in the equation for $\alpha$, see Eq.~(4) in the main text, within one unit cell are
\begin{align}
\mathcal{S}_1 \equiv D_{\mathrm a}
={}&
\int_0^a dy\,
n(y)\cos(k_cy),
&
\mathcal{S}_2 \equiv A_{\mathrm a}
={}&
\int_0^a dy\,
n(y)\cos^2(k_cy),
\label{eq:SC_overlaps_a}\\
\mathcal{P}_1 \equiv D_\Delta
={}&
\int_0^a dy\,
p(y)\cos(k_cy),
&
\mathcal{P}_2 \equiv A_\Delta
={}&
\int_0^a dy\,
p(y)\cos^2(k_cy).
\label{eq:SC_overlaps_Delta}
\end{align}
The corresponding full-system overlaps are
\begin{equation}
D_j^{\mathrm{tot}}
=
N_LD_j,
\qquad
A_j^{\mathrm{tot}}
=
N_LA_j,
\qquad
\mathrm{where\ } j\in\{{\rm a},\Delta\},
\label{eq:SC_total_overlaps}
\end{equation}
and, therefore, increase linearly with the number of cells.

To formulate the cavity equation in the large-system limit, we distinguish
between the single-photon coupling of the finite system and the rescaled
coupling used in the formulation of the problem just within a single unit cell. We denote the former by
$\widetilde{\mathcal G}^{\,\mathrm{phys}}_j$, where the superscript
``phys'' refers to the coupling of the physical finite cavity containing
$N_L$ cells before the large-system rescaling. For the finite system in the cavity, the electric field per cavity photon, and,
hence, the corresponding coupling, scales as
$\widetilde{\mathcal G}^{\,\mathrm{phys}}_j\propto N_L^{-1/2}$.
We define the renormalized couplings used throughout the main text by
\begin{equation}
\widetilde{\mathcal G}_j
\equiv
\sqrt{N_L}\,
\widetilde{\mathcal G}^{\,\mathrm{phys}}_j,
\qquad
j \in \{{\rm a},\Delta\},
\label{eq:SC_G_scaling}
\end{equation}
so that $\widetilde{\mathcal G}_j$ remains finite as
$N_L\rightarrow\infty$.

The effective pump matrix elements
$\widetilde{\Omega}_{\mathrm a}$ and
$\widetilde{\Omega}_\Delta$ are channel dependent but remain
independent of $N_L$. The cavity field amplitude is proportional to
\begin{align}
\mathcal S_{\mathrm{phys}}
={}&
\frac{
\widetilde{\mathcal G}^{\,\mathrm{phys}}_{\mathrm a}
\widetilde{\Omega}_{\mathrm a}
}{\Delta_e}
D_{\mathrm a}^{\mathrm{tot}}
+
\frac{
\widetilde{\mathcal G}^{\,\mathrm{phys}}_\Delta
\widetilde{\Omega}_\Delta
}{\Delta_B}
D_\Delta^{\mathrm{tot}}
\nonumber\\
={}&
\sqrt{N_L}
\left[
\frac{
\widetilde{\mathcal G}_{\mathrm a}
\widetilde{\Omega}_{\mathrm a}
}{\Delta_e}
D_{\mathrm a}
+
\frac{
\widetilde{\mathcal G}_\Delta
\widetilde{\Omega}_\Delta
}{\Delta_B}
D_\Delta
\right],
\label{eq:SC_source_scaling}
\end{align}
and hence scales as $\sqrt{N_L}$. By contrast, the corresponding
dispersive shift of the cavity resonance is
\begin{align}
\mathcal U_{\mathrm{phys}}
={}&
\frac{
\bigl(
\widetilde{\mathcal G}^{\,\mathrm{phys}}_{\mathrm a}
\bigr)^2
}{\Delta_e}
A_{\mathrm a}^{\mathrm{tot}}
+
\frac{
\bigl(
\widetilde{\mathcal G}^{\,\mathrm{phys}}_\Delta
\bigr)^2
}{\Delta_B}
A_\Delta^{\mathrm{tot}}
\nonumber\\
={}&
\frac{
\widetilde{\mathcal G}_{\mathrm a}^{\,2}
}{\Delta_e}
A_{\mathrm a}
+
\frac{
\widetilde{\mathcal G}_\Delta^{\,2}
}{\Delta_B}
A_\Delta,
\label{eq:SC_shift_scaling}
\end{align}
which remains independent of $N_L$. Since $\Delta_c$ and $\kappa$ are
also independent of the system size, the physical coherent cavity
amplitude in the superradiant phase scales as
\begin{equation}
\alpha_{\mathrm{phys}}
\equiv
\langle\hat a\rangle
\propto
\sqrt{N_L}.
\label{eq:SC_alpha_physical_scaling}
\end{equation}
We, therefore, introduce the intensive cavity amplitude
\begin{equation}
\alpha
\equiv
\frac{\alpha_{\mathrm{phys}}}{\sqrt{N_L}},
\label{eq:SC_alpha_intensive}
\end{equation}
which remains finite in the large-system-size limit. Within the
coherent state mean-field approximation, the physical photon number is
\begin{equation}
N_{\mathrm{ph}}
=
|\alpha_{\mathrm{phys}}|^2
=
N_L|\alpha|^2.
\label{eq:SC_photon_scaling}
\end{equation}

The same transverse pump field drives the single-atom and atom-pair
optical processes, but the corresponding effective pump matrix
elements are generally channel dependent. We define
\begin{equation}
\beta_{\mathrm a}
\equiv
\frac{\widetilde{\Omega}_{\mathrm a}}
     {\widetilde{\mathcal G}_{\mathrm a}},
\qquad
\beta_\Delta
\equiv
\frac{\widetilde{\Omega}_\Delta}
     {\widetilde{\mathcal G}_\Delta},
\label{eq:SC_betas}
\end{equation}
and the coupling strengths in the dispersive regime
\begin{equation}
g_{\mathrm a}
\equiv
\frac{\widetilde{\mathcal G}_{\mathrm a}^{\,2}}{\Delta_e},
\qquad
g_\Delta
\equiv
\frac{\widetilde{\mathcal G}_\Delta^{\,2}}{\Delta_B}.
\label{eq:SC_gs}
\end{equation}
Although $g_j$ is quadratic in the corresponding cavity coupling, the
source coefficient $g_j\beta_j$ is linear in it:
\begin{equation}
g_{\mathrm a}\beta_{\mathrm a}
=
\frac{
\widetilde{\mathcal G}_{\mathrm a}
\widetilde{\Omega}_{\mathrm a}
}{\Delta_e},
\qquad
g_\Delta\beta_\Delta
=
\frac{
\widetilde{\mathcal G}_\Delta
\widetilde{\Omega}_\Delta
}{\Delta_B}.
\label{eq:SC_source_coefficients}
\end{equation}

In the numerical calculations, we scan a single dimensionless pump
parameter and impose
\begin{equation}
\beta_{\mathrm a}
=
\beta_\Delta
\equiv
\beta,
\label{eq:SC_beta_common}
\end{equation}
while keeping $g_{\mathrm a}$ and $g_\Delta$ independent. The
single-particle optical potential and effective interaction then take
the compact forms
\begin{align}
V_{\mathrm a}(y)
={}&
g_{\mathrm a}
\left|
\alpha\cos(k_cy)+\beta
\right|^2,
\label{eq:SC_Va}\\
U_{\mathrm{eff}}(y)
={}&
g_0
+
g_\Delta
\left|
\alpha\cos(k_cy)+\beta
\right|^2.
\label{eq:SC_Ueff}
\end{align}

For the common-pump parametrization in
Eq.~\eqref{eq:SC_beta_common}, the steady-state intensive cavity
amplitude is
\begin{equation}
    \alpha = \frac{ \beta ( g_{\mathrm a}D_{\mathrm a} + g_\Delta D_\Delta
    ) }{ \Delta_c - g_{\mathrm a}A_{\mathrm a} - g_\Delta A_\Delta + i\kappa }.
\label{eq:SC_alpha}
\end{equation}
Here $g_{\mathrm a}D_{\mathrm a}$ and
$g_\Delta D_\Delta$ have the same dimensions of energy, although the
two overlaps themselves have different dimensions. In the
one-dimensional continuum normalization,
\begin{equation}
    [D_{\mathrm a}] = [A_{\mathrm a}] = 1,
    \qquad
    [D_\Delta] = [A_\Delta] = L^{-1},
\label{eq:SC_overlap_dimensions}
\end{equation}
while $[g_{\mathrm a}]=E$ and $[g_\Delta]=EL$, where $E$ denotes the dimension of energy and $L$ the dimension of length. Thus, every term in
Eq.~\eqref{eq:SC_alpha} has consistent dimensions. When shown as
dimensionless quantities, the pair-channel overlaps are expressed as
$aD_\Delta$ and $aA_\Delta$.

Equation~\eqref{eq:SC_alpha} closes the self-consistency loop between
the intracavity field and the HFB problem solved within one unit cell:
the density and pairing responses determine $\alpha$, while $\alpha$
influences the atomic system through $V_{\mathrm a}(y)$ and
$U_{\mathrm{eff}}(y)$, as illustrated schematically in
Fig.~\ref{fig:SM_cavity_setup}(c).

Finally, Bloch's theorem is implemented by expanding the BdG amplitudes
in quasimomentum $k$ within the first Brillouin zone,
\begin{equation}
k\in
\left[
-\frac{\pi}{a},
\frac{\pi}{a}
\right).
\label{eq:SC_BZ}
\end{equation}
As $N_L\rightarrow\infty$, the discrete quasimomentum sum becomes
\begin{equation}
\frac{1}{N_L}\sum_k
\longrightarrow
\frac{a}{2\pi}
\int_{-\pi/a}^{\pi/a}dk,
\label{eq:SC_k_integral}
\end{equation}
while extensive observables scale with $N_L$ and all unit-cell
quantities remain finite.

\section{Characterization of the ordered phases}
\label{sec:SM_order_parameters}

In this section we define the observables used to characterize the phases in the $(\beta,\, g_0/(E_R\lambda_c))$ plane. Specifically, we consider the intra-cavity photon number, the Fourier component of the product of the densities of opposite spins at the cavity wave vector $k_c$, the Fourier components of the total density, the inverse participation ratio of the total density, and a measure of the spatial inhomogeneity of the pairing gap inspired by Leggett bound on the superfluid fraction. All spatial integrals are evaluated over one unit cell of length $\lambda_c$.

The cavity order parameter, defining the superradiant phase, is the mean photon number
\begin{equation}
n_{\rm ph}=|\alpha|^2 ,
\label{eq:SM_nph}
\end{equation}
where \(\alpha\) is the steady-state cavity amplitude obtained from
Eq.~\eqref{eq:SC_alpha}. In the normal homogeneous state $D_\mathrm{a}=0$ and $D_\Delta=0$, and the density profile does
not scatter pump photons coherently into the cavity mode, and, therefore,
$n_\mathrm{ph}$ is zero. A finite value of $n_\mathrm{ph}$
signals the onset of the superradiant self-organization. Since the photon number
changes over several orders of magnitude close to the transition, we display
$n_\mathrm{ph}$ on a logarithmic scale.

To characterize the modulation of the densities overlap at the
cavity wave vector, we employ
\begin{equation}
\Theta
=
\int_{0}^{a}dy\,
n_{\uparrow}(y)n_{\downarrow}(y)\cos(k_c y).
\label{eq:SM_theta}
\end{equation}
This quantity enters into the numerator of the expression for the intra-cavity field $\alpha$.
Here, $n_{\uparrow}(y)$ and $n_{\downarrow}(y)$ are the densities of the two
spin components. In our case, i.e., for a spin-balanced gas,
$n_{\uparrow}=n_{\downarrow}=n/2$. The quantity $\Theta$ describes the
component of the local Hartree interaction term 
$n_{\uparrow}(y)n_{\downarrow}(y)$ at the cavity wave vector $k_c$.
Consequently, $\lambda_c|\Theta|$ is a dimensionless characteristic that
signals the atomic density ordering. The absolute
value is reported because the sign of $\Theta$ distinguishes the two spatial configurations related by the symmetry transformation discussed in the main text.

The localization of the atomic density is quantified through the inverse
participation ratio of the total density
\begin{equation}
n(y)=n_{\uparrow}(y)+n_{\downarrow}(y).
\label{eq:SM_total_density}
\end{equation}
We first define the normalized density profile 
\begin{equation}
\mathcal{N}(y)
=
\frac{n(y)}
{\displaystyle\int_{0}^{a}dy'\,n(y')},
\qquad \text{such that }
\int_{0}^{a}dy\,\mathcal{N}(y)=1.
\label{eq:SM_density_weight}
\end{equation}
The continuum inverse participation ratio is then defined by
\begin{equation}
{\rm IPR}
=
\int_{0}^{a}dy\,\mathcal{N}^2(y)
=
\frac{\displaystyle\int_{0}^{a}dy\,n^2(y)}
{\left[\displaystyle\int_{0}^{a}dy\,n(y)\right]^2}.
\label{eq:SM_IPR_dimensional}
\end{equation}
Because this quantity has dimension of inverse length, we report the
dimensionless ratio
\begin{equation}
\widetilde I
=
\lambda_c\,{\rm IPR}
=
\lambda_c\,
\frac{\displaystyle\int_{0}^{a}dy\,n^2(y)}
{\left[\displaystyle\int_{0}^{a}dy\,n(y)\right]^2}.
\label{eq:SM_IPR}
\end{equation}
A perfectly uniform density yields $\widetilde I=1$. Larger values of
$\widetilde I$ indicate a stronger density modulation or localization within
the unit cell. On a discrete grid with $N_y$ points, the limiting range is
$1\leq \widetilde I\leq N_y$, where the upper limit corresponds to all density
being concentrated on a single grid point.

Finally, we introduce a Leggett-type fraction which quantifies the spatial inhomogeneity of the pairing amplitude $\Delta$. 
The original Leggett bound is formulated in terms of the particle density~\cite{PhysRevLett.25.1543} and quantifies the upper bound on the superfluid fraction (see also the discussion around Eq.~\eqref{eq:SM_density_Leggett}). For an interacting Fermi gas, such a fraction with the fermionic density overestimates the actual superfluid fraction~\cite{orso2024superfluid}. Here, we use the same structure only as a diagnostic for the pairing field, treating the density of the condensate of Cooper pairs proportional to the following quantity:
\begin{equation}
n_{\rm pair}(y)=|\Delta(y)|^2 .
\label{eq:SM_pair_weight}
\end{equation}
We then define the Leggett-type fraction
\begin{equation}
f_L^\mathrm{pair}
\equiv
\frac{1}
{\langle n_{\rm pair}\rangle
 \langle n_{\rm pair}^{-1}\rangle},
\qquad
0 \leqslant f_L^{\rm pair} \leqslant 1,
\label{eq:SM_pair_Leggett_like}
\end{equation}
where the average over a unit cell is defined by $ \langle g\rangle = \frac{1}{\lambda_c}\int_{0}^{\lambda_c}dy\,g(y) $ for any function $g(y)$.

This quantity $f_L^{\rm pair}=1$, when $|\Delta(y)|^2$ is spatially uniform, and it is
suppressed $f_L^{\rm pair}<1$ when the pairing field becomes inhomogeneous. It depends
only on the spatial profile of the pairing amplitude. Therefore,
$f_L^{\rm pair}$ should not be interpreted as either the exact superfluid
fraction. In regions where the pairing amplitude is below the numerical threshold, we set
$f_L^{\rm pair}=0$.

\begin{figure}[t!]
    \centering
    \includegraphics[width=\linewidth]{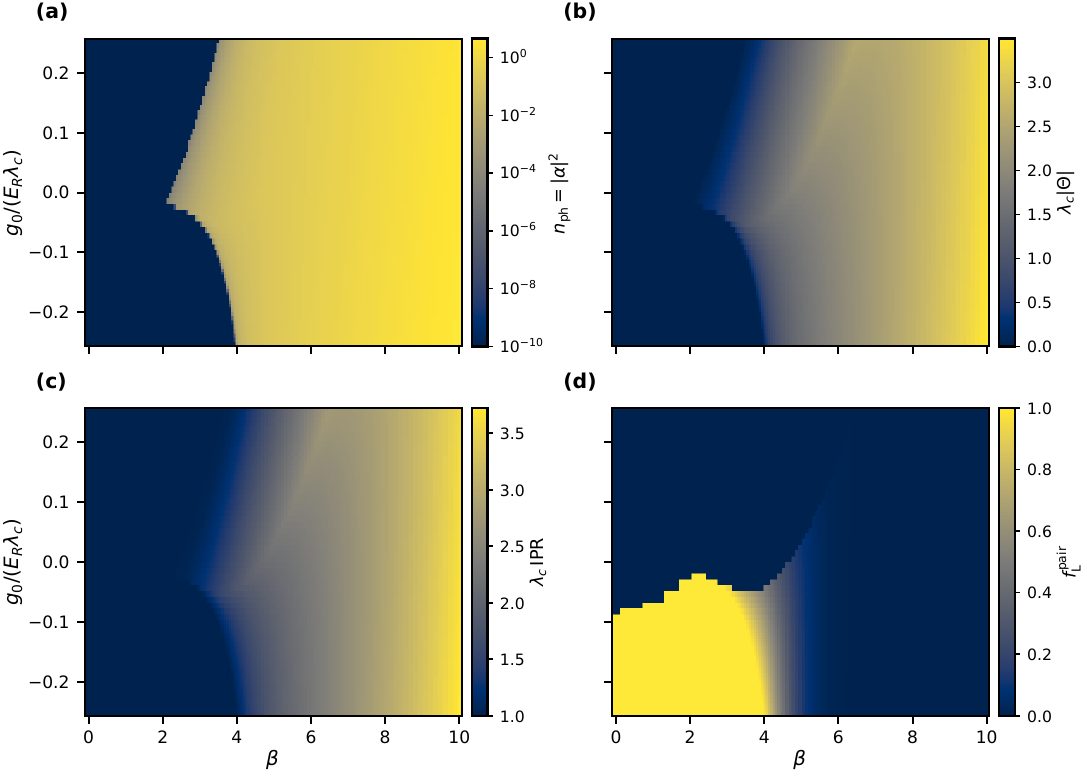}
    \caption{Characterization of the phases in the $(\beta,g_0/(E_R \lambda_c))$ plane. 
    Lengths are expressed in the units of the cavity wavelength $\lambda_c$, and energies in the units of
    the recoil energy $E_R=\hbar^2k_c^2/(2m)$.
    (a) Intra-cavity photon number $n_{\rm ph}=|\alpha|^2$, plotted on a
    logarithmic color scale. The sharp increase of $n_{\rm ph}$ marks the onset
    of superradiant phase.
    (b) Dimensionless $\lambda_c|\Theta|$. A finite $\lambda_c|\Theta|$ indicates a
    modulation of the densitites overlap at wave vector $k_c$.
    (c) Dimensionless inverse participation ratio $\lambda_c\,\mathrm{IPR}$ of the total
    density. The value $\lambda_c \,\mathrm{IPR}=1$ corresponds to a uniform density,
    while larger values indicate stronger density modulation or localization
    inside the unit cell.
    (d) Leggett-type fraction $f_L^{\rm pair}$, computed from $|\Delta(y)|^2$. Within regions where the
    self-consistent pairing field is finite, values close to unity indicate a
    nearly uniform pairing amplitude, whereas smaller values indicate stronger
    spatial inhomogeneity. In regions where the pairing amplitude is below the
    numerical threshold, we set $f_L^{\rm pair}=0$.
    }
    \label{fig:SM_order_parameters}
\end{figure}

Figure~\ref{fig:SM_order_parameters} shows the phase characteristics in the
$(\beta,g_0/(E_R \lambda_c))$ plane. These quantities are used to identify and
characterize the phases discussed in the main text. The photon number in
Fig.~\ref{fig:SM_order_parameters}(a) is negligible below the
self-organization threshold and becomes finite above a critical pump strength $\beta$.
The threshold depends on the contact interaction $g_0$, showing that the
self-organization is modified by the interaction channel of the gas. 

The characteristic~$\Theta$ in Fig.~\ref{fig:SM_order_parameters}(b) follows the buildup of the intra-cavity field amplitude, showing that the superradiant state is accompanied by a modulation of densitties overlap at the cavity wave vector. The inverse participation ratio~$\widetilde I$ in Fig.~\ref{fig:SM_order_parameters}(c) increases with pump strength, indicating that the atomic density becomes progressively more localized within the unit cell.

In the repulsive regime and for weakly attractive $g_0$, a weak curved feature is visible in both $\lambda_c|\Theta|$ and $\widetilde I$, most clearly for intermediate pump strengths. This feature coincides with the appearance of non-zero pairing field $F$. Although $\lambda_c|\Theta|$ and $\widetilde I$ depend only on the density profiles, the emerging pairing field $F$ modifies the effective interaction, which has an impact both on $\widetilde{I}$ and $\Theta$, producing a visible effect. The curve, marking the onset of the feature, therefore, provides an indirect signature of the boundary beyond which density modulation and a finite pairing field coexist. 

Finally, Fig.~\ref{fig:SM_order_parameters}(d) shows the Leggett-type fraction $f_L^\mathrm{pair}$. Within the region where the pairing field is finite, this quantity is close to unity when $|\Delta(y)|^2$ is nearly uniform and decreases as the pairing amplitude $F$ becomes spatially inhomogeneous. Its suppression in the self-organized phase is, therefore, consistent with the development of a spatially modulated pairing component. Importantly, once the density $n_\mathrm{pair}(y)$ drops to zero in some region, the superfluid fraction $f_L^\mathrm{pair}$ drops to zero as well. Physically, in such a case the density of Cooper pairs within a unit cell forms a region that is spatially detached from adjacent cells, which prohibits maintaining global coherence across the entire system.

\begin{figure}[t!]
    \centering
    \includegraphics[width=0.5\linewidth]{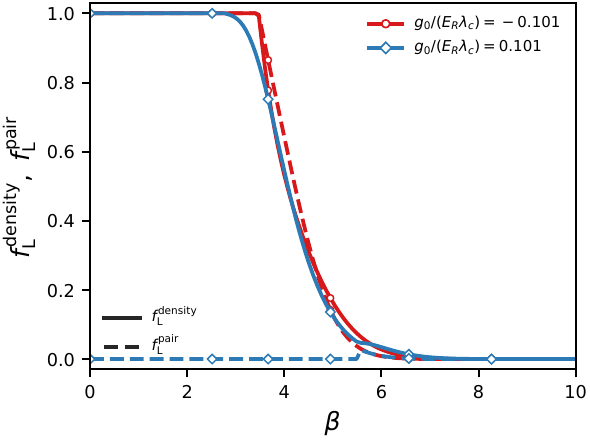}
    \caption{
    Comparison of the Leggett bound $f_L^\mathrm{density}$ (solid) on the superfluid fraction and the Leggett-type fraction $f_L^\mathrm{pair}$ (dashed) for $g_0/(E_R\lambda_c) = +0.101$ (blue diamonds) and $-0.101$ (red circles) as a function of the pump strength $\beta$. The system parameters are the same as in Fig.~\ref{fig:SM_order_parameters} and as Fig.~2 in the main text.}
    \label{fig:SM_fractions}
\end{figure}

In Fig.~\ref{fig:SM_fractions}, we provide also the comparison of the Leggett~\cite{PhysRevLett.25.1543} upper bound on the superfluid fraction
\begin{equation}
    f_L^\mathrm{density} \equiv \frac{1} {\langle n\rangle  \langle n^{-1}\rangle},
\label{eq:SM_density_Leggett}
\end{equation}
where $n(y)$ is the total density of the fermionic gas, with fraction $f_L^\mathrm{pair}$. The plot shows the two cuts for  $g_0/(E_R\lambda_c) = +0.101$ (blue diamonds) and $-0.101$ (red circles), while the solid lines denote the Leggett bound and the dashed ones present the Leggett-type fraction $f_L^\mathrm{pair}$. For the attractive interaction, the fractions $f_L^\mathrm{density}$ and $f_L^\mathrm{pairing}$ provide qualitatively similar results. The latter overestimates slightly the Leggett bound for $1>f_L^\mathrm{pair} \gtrsim f_L^\mathrm{density} \sim 0.3$, while it underestimates it for values $f_L^\mathrm{pair} \lesssim 0.3$. Crucially, for repulsive interactions $g_0$, the value of $f_L^\mathrm{density}$ remains large and finite for the system in the normal phase and in the CDW+SR phase even though the pairing field is absent there and, thus, $f_L^\mathrm{pair}=0$. Only when the system enters the CDW+PDW+SR phase for larger values of the pump $\beta$, $f_L^\mathrm{pair}$ becomes nonzero and small, but on the same order as $f_L^\mathrm{density}$. 
Therefore, in order to characterize the inhomogeneous component of the pairing field, we present the results for $f_L^\mathrm{pair}$.
The actual determination of the superfluid fraction is beyond the scope of the present work and left for future studies~\cite{orso2024superfluid, biagioni2024measurement}.

\section{Density and pairing profiles}
\label{sec:SM_real_space_profiles}

\begin{figure}[t]
    \centering
    \includegraphics[width=\linewidth]{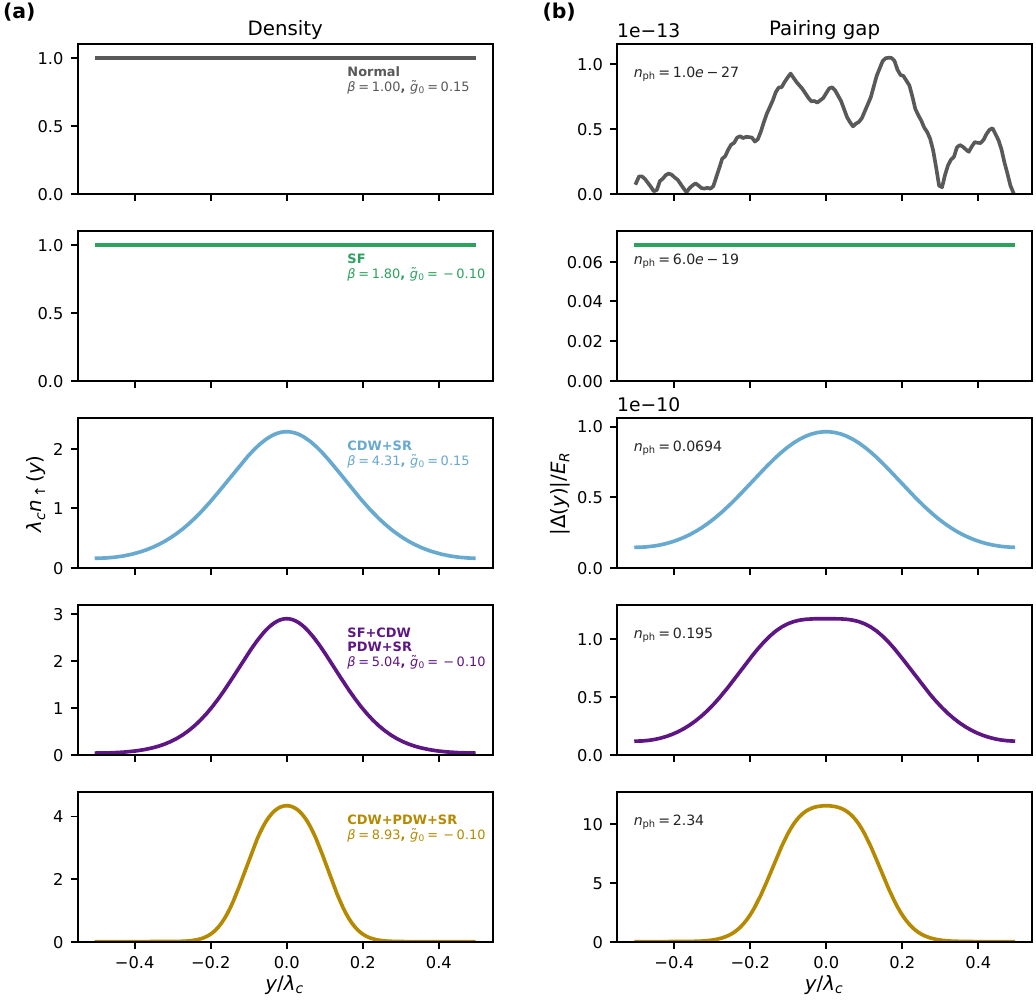}
    \caption{
    Representative position profiles of the density and pairing field in
    different regions of the phase diagram.
    (a) Density $n_{\uparrow}(y)$ (in the units of $\lambda_c^{-1}$). Since the gas is
    spin-balanced, $n_{\uparrow}(y)=n_{\downarrow}(y)$, only one spin component is shown. 
    (b) Dimensionless magnitude of the pairing gap
    $|\Delta(y)|/E_R$. Each row corresponds to the same photon number $n_{\rm ph}=|\alpha|^2$; the labels
    also show $\tilde g_0=g_0/(E_R \lambda_c)$.
    In the normal state the density is uniform and the pairing field is absent (up
    to numerical noise). In the uniform superfluid, $|\Delta(y)|$ is finite and
    constant while the density remains homogeneous. In the superradiant
    density-wave regime, the density becomes strongly modulated by the cavity
    field. For attractive interactions and larger pump strengths, the pairing field develops a pronounced spatial modulation in phase with the density wave, signaling the emergence of a pair-density-wave component. The profiles are shown over one unit cell, $-1/2\leqslant y/\lambda_c\leqslant 1/2$.
    }
    \label{fig:SM_real_space_profiles}
\end{figure}

To complement the order-parameter maps, we examine representative real-space
profiles of the spin density and pairing field in different regions of the
phase diagram. Since we consider a spin-balanced gas,
\begin{equation}
n_{\uparrow}(y)=n_{\downarrow}(y),
\label{eq:SM_spin_balance_profiles}
\end{equation}
it is sufficient to show only one spin component, \(n_{\uparrow}(y)\). The pairing
gap is represented by
\begin{equation}
|\Delta(y)|=|U_{\rm eff}(y)F(y)|,
\label{eq:SM_gap_profile}
\end{equation}
and the photon number $n_{\rm ph}=|\alpha|^2$ is the same in each row of Fig.~\ref{fig:SM_real_space_profiles}.

Figure~\ref{fig:SM_real_space_profiles} illustrates how the phases identified in the $(\beta,\tilde g_0)$ phase diagram appear in real space, where $\tilde g_0=g_0/(E_R \lambda_c)$. The profiles are plotted in dimensionless units: the density as $a n_\uparrow(y)$ and the pairing field as $|\Delta(y)|/E_R$. In the normal state, shown in the first row, the converged photon number is essentially zero and the density is uniform. The pairing field is also absent; the tiny residual structure visible on the scale of $10^{-12}$ is below the numerical threshold and should not be interpreted as a physical pairing order.

For attractive contact interaction and weak pump strength, the system forms a uniform superfluid (denoted by SF in the plot). This is shown in the second row. The density remains spatially uniform and the photon number remains below the numerical threshold used to identify a superradiant state. At the same time, $|\Delta(y)|$ is finite and constant across the unit cell, corresponding to a conventional BCS superfluid.

The third row (CDW+SR) shows a representative superradiant density-wave state for repulsive interactions. Here, the converged photon number becomes finite, and the atomic density develops a strong modulation. This is the origin of the finite order parameter $|\Theta|$ and the increased IPR $\widetilde{I}$. The pairing field remains below the numerical threshold used for phase classification, so this state is marked as a density-ordered superradiant phase rather than a paired phase.

For attractive interactions at intermediate pump strength, shown in the fourth row (labeled as SF+CDW+PDW+SR in the Figure), density order and pairing order coexist. The finite photon number produces a spatial modulation of the density, while the pairing field remains finite but is no longer uniform. The pairing field $|\Delta(y)|$ develops a visible real-space modulation locked to the density structure. This coexistence of a density wave, a superradiant cavity field, and a modulated pairing field corresponds to the regime labeled as $\mathrm{CDW+SF+PDW+SR}$.

At larger pump strength, shown in the fifth row (labeled as $\mathrm{CDW+PDW+SR}$), the atoms are strongly localized near the favorable region of the optical potential. The density peak becomes narrower and the inverse participation ratio increases. The pairing field is also sharply localized. This explains why the Leggett-type fraction $f_L^{\rm pair}$ is suppressed in the strong-pump region.

Each of the position profiles (density and the pairing amplitude), therefore, provide a representative of each class in the phase diagram. In all the cases, the transition into the superradiant phase is accompanied by the formation of a self-organized phase that can be either CDW without pairing or CDW with pairing with position dependent modulation. In particular, in the attractive regime, the
cavity field modifies the pairing field, driving 
a transition from a uniform BCS superfluid to a modulated PDW state.

\section{Experimental observability}
The phases discussed in the main text can, in principle, be probed using a combination of experimental techniques that have already been demonstrated in ultracold Fermi gases, either alone or coupled to high-finesse cavities. 

The cavity field in the superradiant phase is the most direct observable: the onset of a coherent intra-cavity field is measured from the light leaking out through the cavity mirrors and identifies the breaking of the discrete $\mathbb Z_2$ symmetry. In the fermionic density-wave experiment of Ref.~\cite{helson2023density}, this photon signal was used to determine the self-organization threshold while the contact interaction was tuned across the BCS--BEC crossover. This provides the experimental analogue of scanning the bare interaction $g_0$ in the present phase diagram of our work.

The effective 1D regime considered here can be approached by adding strong optical confinement in the two directions transverse to the cavity axis. For example, a 2D optical lattice can partition the gas into an array of elongated tubes oriented along the cavity mode, a geometry already realized with two-component fermionic gases~\cite{moritz2005confinement,liao2010spinimbalance}. When the transverse level spacing $\hbar\omega_\perp$ exceeds the thermal, Fermi, pairing, and cavity-induced energy scales, transverse excitations are suppressed and the low-energy dynamics within each tube is effectively one-dimensional. A single central tube, or a nearly homogeneous group of tubes, can be positioned within the slowly varying transverse envelope of the cavity mode and close to an antinode of the pump field. The pump amplitude is then approximately constant across the occupied transverse wave function, while the cavity standing wave retains the spatial dependence $\cos(k_c y)$ along the tube. If several equivalent tubes are occupied, they can contribute collectively to the same cavity output; the present model then describes a representative tube in the limit of many such tubes. The effective 1D interaction strength $g_0$ is the contact interaction strength renormalized by the confinement and remains tunable through the magnetic Feshbach field~\cite{olshanii1998atomic, bergeman2003atom, RevModPhys.82.1225}.

The superradiant CDW and CDW+PDW phases can be identified by combining the standard measurement of the cavity output with a pair-sensitive optical probe based on cavity-transmission spectroscopy near an atom-pair photoassociation transition. The density Fourier harmonic at the wave vector $k_c$ couples to the usual single-atom dispersive channel and is, therefore, visible through the photon flux.
The CDW order can also be checked more directly by \textit{in-situ} imaging: recent experiments have resolved density-wave patterns in a unitary Fermi gas in an optical resonator and correlated the atomic density modulation with the simultaneously measured cavity field \cite{h3zm-rnnx}. 

The short-range pair correlations are accessible because cavity photons can couple directly to atom pairs through a molecular photoassociation transition. Pair-polariton spectroscopy demonstrated that the optical spectrum provides a direct probe of short-range pair correlations~\cite{konishi2021universal}. This spectroscopy probes the spatially integrated short-range pair correlations sampled by the cavity field, but does not by itself resolve their finite-momentum modulation or directly measure the complex pairing field. 

Therefore, one possible experimental strategy would be to perform repeated pump scans (increasing its value from zero) while varying two independent parameters: the magnetic field, which changes $g_0$, and the molecular detuning, which changes the magnitude and sign of the atom-pair cavity coupling $g_\Delta$. If the system enters into the CDW phase first, the photon flux and CDW signal should appear at the first threshold. However, the onset of PDW order, which occurs for larger pump strengths, does not generate a second emission threshold because the cavity field is already nonzero. Emergence of the PDW order is also not accompanied by any sharp features in the field intensity, as shown in Fig.~2 of the main text. Therefore, in order to reveal the PDW order direct observation of the pairing correlations is required, and one possible direction would be to rapidly project the fermion pairs onto bound molecules and infer their center-of-mass momentum distribution from time-of-flight imaging. Rapid projection and pair-momentum measurements have been demonstrated in Fermi gas experiments without an optical cavity~\cite{regal2004resonance,dyke2021dynamics}. 

Relevant steps have also been implemented in a $^{6}$Li cavity apparatus designed for cavity-QED experiments, where an adiabatic ramp of the magnetic field to the molecular side of the Feshbach resonance, followed by time-of-flight imaging, was used to determine the molecular condensate fraction~\cite{buhler2025direct}. That experiment, however, did not employ rapid projection or resolve finite-momentum pairing. The measurement of molecule momenta near $q=\pm k_c$ would, therefore, constitute only a candidate for the signature of the PDW phase, whose interpretation would require verifying that the projection preserves the initial pair correlations and no additional momenta sidebands are generated due to scattering on the density wave. 

A complementary, longer-term direction is microscopy with atomic resolution of pair correlations, demonstrated in a two-dimensional $^{6}$Li gas without an optical cavity~\cite{yao2025paircorrelations}. Adapting this method to a cavity-QED geometry could probe spatially modulated pair correlations.

If the system enters directly into the CDW+SF+PDW+SR phase, as predicted for an initially attractive superfluid, the onsets of the CDW and the PDW order occur simultaneously. The cooperation or competition between density and pairing channels can be tested at the onset of superradiance by measuring the threshold as a function of molecular detuning at fixed $g_0$. This type of threshold measurement has been performed in a high-finesse cavity-QED experiment with a unitary $^{6}$Li gas~\cite{zwettler2025cavity}, but it cannot establish the presence of PDW order.

Finally, we remark that the theory treats an effective 1D system at the mean-field level, while current experiments involve trapped gases with finite temperature and heating, cavity loss, and atomic cloud inhomogeneity. Nevertheless, two relevant observables have already been demonstrated in cavity-QED experiments with strongly interacting Fermi gases: the emitted photon flux identifies the superradiant threshold, while high-resolution \textit{in-situ} imaging can directly confirm the accompanying density-wave order~\cite{helson2023density,h3zm-rnnx}. Together with an independent measurement of pair correlations, these cavity observables provide one possible direction for testing the predicted change from a CDW+SR phase to an intertwined SF+CDW+PDW+SR phase.

\section{Contact-interaction dependence of the Fano-type phase boundary}
\label{sec:SM_Fano_boundary}

\begin{figure}[!ht]
    \centering
    \includegraphics[width=\columnwidth]
    {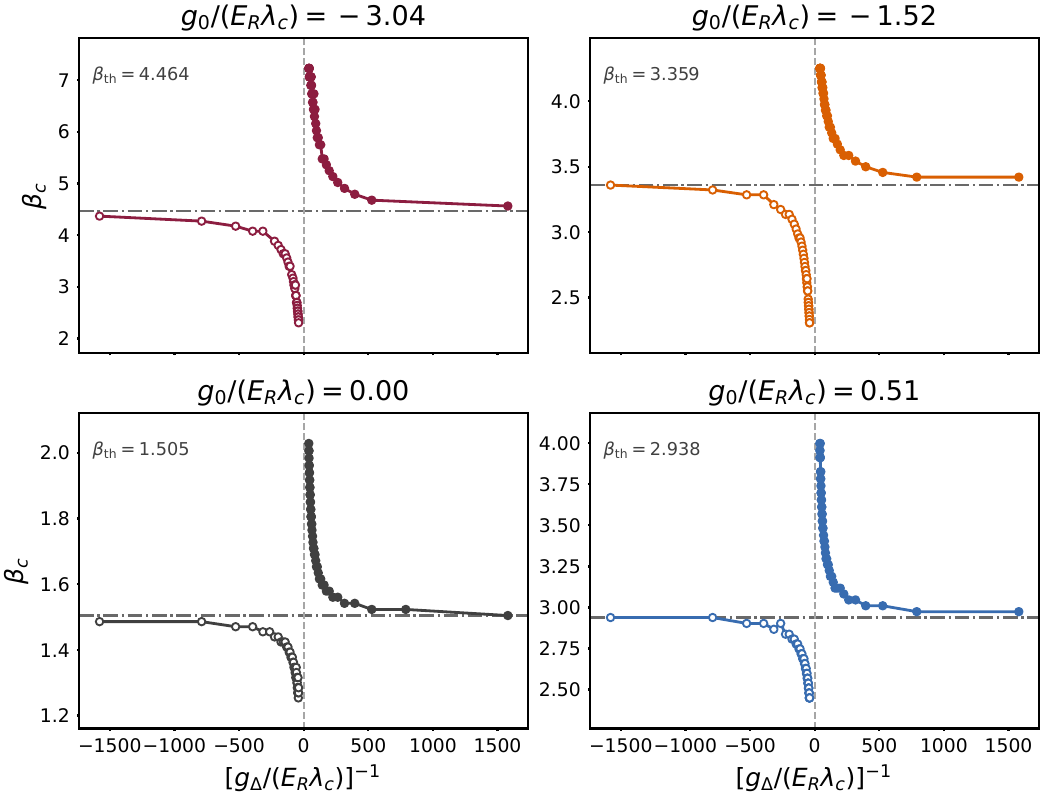}
    \caption{
    Dependence of the Fano-type superradiant phase boundary on the atomic
    interaction strength. The critical pump strength $\beta_c$ is shown as a
    function of the dimensionless inverse cavity-induced interaction
    strength $\left[g_\Delta/(E_R\lambda_c)\right]^{-1}$, which is
    proportional to the molecular detuning. The four panels correspond to
    $g_0/(E_R\lambda_c)=-3.04,-1.52,\,0$, and $0.51$ (from top left to
    bottom right). Open and filled symbols denote $g_\Delta<0$ and
    $g_\Delta>0$, respectively. The horizontal dash-dotted line in each panel
    marks the far-detuned background threshold $\beta_{\rm th}$ obtained in
    the limit $g_\Delta\rightarrow0$.
    The vertical dashed line marks $g_\Delta^{-1}=0$.
    The symbols are the numerically calculated threshold points, while
    the continuous lines connect the displayed points on each side of the
    resonance only as guides to the eye and do not come from a fitting function.
    }
    \label{fig:SM_Fano_boundary}
\end{figure}

Here, we examine how the phase boundary between a homogeneous and a self-organized state changes with $g_0$. For each fixed value of $g_0/(E_R\lambda_c)$, the critical pump strength $\beta_c$ is determined as a function of $g_\Delta$. Since
\begin{equation}
g_\Delta =
\frac{\widetilde{\mathcal G}_\Delta^{\,2}}{\Delta_B},
\end{equation}
$g_\Delta^{-1}$ is proportional to the molecular detuning $\Delta_B$.
In the far-detuned limit, $g_\Delta^{-1}\rightarrow\pm\infty$, the
cavity-induced atom--atom interaction strength vanishes and the phase boundary approaches the background threshold $\beta_{\rm th}$ associated predominantly with the cavity-induced single-atom optical potential.

Closer to the molecular resonance, the pair channel modifies the coherent
cavity source, i.e., in the numerator for the cavity field $\alpha$, we have
\begin{equation}
\mathcal S =
g_{\rm a}D_{\rm a}+g_\Delta D_\Delta .
\label{eq:SM_Fano_source}
\end{equation}
For $g_\Delta<0$, the atomic and pair-channel contributions predominantly
reinforce one another, increasing the cavity response and lowering the pump
strength required for self-organization. For $g_\Delta>0$, the two
contributions compete, reducing $\mathcal S$ and increasing the critical pump
strength. This change between cooperative and competitive coupling produces
the asymmetric Fano-type boundary shown in
Fig.~\ref{fig:SM_Fano_boundary}.

Although $D_{\rm a}$ and $D_\Delta$ enter
Eq.~\eqref{eq:SM_Fano_source} as distinct terms, they are not independent
contributions. The pair-channel term $D_\Delta$ contains both
a density-overlap contribution and a contribution involving the
anomalous amplitude $F$, i.e.,
\begin{equation}
D_\Delta =
D_{\Delta,{\rm density}}
+
D_{\Delta,{\rm pairing}},
\end{equation}
and the density $n_\sigma(y)$ and anomalous amplitude $F(y)$ are obtained from
the same HFB quasiparticle spectrum. The contact interaction enters both the
Hartree and pairing channels and, therefore, controls the density, the
pairing field, and their mutual coupling.

Figure~\ref{fig:SM_Fano_boundary} shows that changing
$g_0$ modifies both the background threshold
$\beta_{\rm th}$ and the resonant contribution to the critical pump strength
$\beta_c$. The background threshold $\beta_{\rm th}$ decreases as
$g_0/(E_R\lambda_c)$ changes from $-3.04$ toward $0$ and then increases again
for $g_0/(E_R\lambda_c)=0.51$, demonstrating a nonmonotonic dependence of
$\beta_{\rm th}$ on $g_0$.

To compare the Fano-type profile for different values of $g_0$, we define the
relative dip and peak values
\begin{equation}
    \delta_- \equiv
    \frac{\beta_{\rm th}-\min_{g_\Delta<0}\beta_c}{\beta_{\rm th}},
    \qquad
    \delta_+ \equiv
    \frac{\max_{g_\Delta>0}\beta_c-\beta_{\rm th}}{\beta_{\rm th}}.
    \label{eq:SM_Fano_relative_changes}
\end{equation}
Here, $\delta_-$ quantifies the maximum relative reduction of the threshold for 
$g_\Delta<0$, while $\delta_+$ measures its maximum relative increase when
$g_\Delta>0$. We define the total normalized dip-to-peak contrast as
\begin{equation}
\mathcal C \equiv
\delta_-+\delta_+
=
\frac{\max_{g_\Delta>0}\beta_c-\min_{g_\Delta<0}\beta_c}
{\beta_{\rm th}}.
\label{eq:SM_Fano_contrast}
\end{equation}
In order to distinguish the asymmetry between the two sides of $g_\Delta=0$, we define
\begin{equation}
    \mathcal A \equiv
    \frac{\delta_+-\delta_-}{\delta_++\delta_-},
\label{eq:SM_Fano_asymmetry}
\end{equation}
where $\mathcal A>0$ means that $\delta_+ > \delta_-$, i.e., the relative value at the peak on the positive side for $g_\Delta>0$ is
larger than the relative value at the dip on the negative side. 
All extrema are evaluated over the sampled range shown in Fig.~\ref{fig:SM_Fano_boundary}.

The largest $\mathcal C$ occurs for the most attractive
interaction analyzed, $g_0/(E_R\lambda_c)=-3.04$, for which $\mathcal C=1.103$ and
$\mathcal A=0.122$. In this regime, the well-developed pairing correlations
enhance the interplay of the density and pairing channels: cooperative
coupling substantially lowers the threshold for $g_\Delta<0$, whereas
competitive coupling strongly raises it for $g_\Delta>0$. The contrast
$\mathcal{C}$ decreases to $\mathcal C=0.579$ for $g_0/(E_R\lambda_c)=-1.52$, where
the asymmetry $\mathcal A=-0.083$. For $g_0/(E_R\lambda_c)=0$ and $0.51$, the contrasts are
instead comparable, i.e., $\mathcal C=0.515$ and $0.528$, while the corresponding
asymmetry measures are $\mathcal A=0.353$ and $0.368$, respectively. Thus, the
result for $g_0>0$ has only a slightly larger $\mathcal{C}$ and $\mathcal{A}$ than the noninteracting result $g_0=0$.

For $g_0/(E_R\lambda_c)=0.51$, the increase of $\beta_c$ remains pronounced
for $g_\Delta>0$. This behavior does not by itself imply a nonzero
anomalous pairing amplitude.
The reason that $g_\Delta$ can still affect the threshold is that $D_\Delta$
also contains the contribution $D_{\Delta,{\rm density}}$.
Thus, even when $F=0$, the presence of cavity-induced interactions with finite $g_\Delta$ can change the
self-organization.

Finally, we note that $g_\Delta^{-1}=0$ corresponds to the molecular
resonance, where the far-detuned elimination of the excited molecular field
is not justified. We, therefore, make no quantitative statement at the
resonance itself. We also remark that we did not employ a fit to a Fano-type profile, since for moderate values of the interaction strength $g_0$, the quality of the fit was insufficient.

\section{Linear response theory of coupled density and pairing modulations}
\label{sec:SM_linear_response}

We now analyze the static linear response of the homogeneous BCS state to a weak perturbation with the spatial periodicity resulting from the coupling to photons. The response involves coupled density and pairing fields and determines both the continuous self-organization threshold and the composition of the critical mode into density and pairing channels. We consider the spin-balanced state defined in Eq.~\eqref{eq:SB_balance} and choose the global phase such that the homogeneous pairing gap $\Delta_0$ is real. Since the interaction with the cavity generates a perturbation that couples to the first Fourier components of the fermionic fields, the response is evaluated at wave vector $k_c$.

\subsection{Homogeneous BCS state and non-self-consistent susceptibility}
\label{sec:SM_lr_bdg}

We expand about the homogeneous solution ($\alpha=0$). From Eqs.~\eqref{eq:SA_Va} and \eqref{eq:SA_Ueff}, the pump then generates the spatially uniform potential and the coupling constant, i.e., 
\begin{equation}
V_{\rm a}^{(0)}
=
g_{\rm a}\beta^2,
\qquad
U_0
=
g_0+g_\Delta\beta^2,
\label{eq:SM_lr_uniform_fields}
\end{equation}
respectively, where we used the effective couplings introduced above. For the balanced system, Eq.~\eqref{eq:SB_hbalanced}, therefore, reduces in momentum space to
\begin{equation}
\xi_k
=
\frac{k^2}{2m}
-\mu
+
V_{\rm a}^{(0)}
+
\frac{U_0 n_0}{2}.
\label{eq:SM_lr_xi}
\end{equation}
Using the same Nambu convention as in Eq.~\eqref{eq:SB_Nambu}, the BdG equations are solved with the eigenvalues $E_{k,s}$ and eigenvectors $\ket{k,s}$ of $H_0$, which we call the homogeneous BdG Hamiltonian:
\begin{equation}
H_0(k)
=
\xi_k\tau_z+\Delta_0\tau_x,
\qquad
E_{k,s}=sE_k,
\text{ where }
E_k\equiv\sqrt{\xi_k^2+\Delta_0^2},
\label{eq:SM_lr_H0}
\end{equation}
where $\tau_z$ and $\tau_x$ are Pauli $z$ and $x$ matrices acting in the Nambu space, while $s=+$ and $s=-$ label the positive- and negative-energy branches, respectively.

We introduce the angle $\theta_k$,
defined by
\begin{equation}
\cos\theta_k \equiv \frac{\xi_k}{E_k},
\qquad
\sin\theta_k \equiv \frac{\Delta_0}{E_k},
\label{eq:SM_lr_theta}
\end{equation}
which allows us to write
\begin{equation}
\frac{H_0(k)}{E_k}
=
\cos\theta_k\,\tau_z
+
\sin\theta_k\,\tau_x.
\label{eq:SM_lr_H_angle}
\end{equation}
The projection operator onto the branch $s$ is
\begin{equation}
P_s(k)
=
|k,s\rangle\langle k,s|
=
\frac12
\left[
\bm 1+s\frac{H_0(k)}{E_k}
\right]
=
\frac12
\left[
\tau_0
+
s\left(
\cos\theta_k\,\tau_z
+
\sin\theta_k\,\tau_x
\right)
\right],
\label{eq:SM_lr_projector}
\end{equation}
where $\bm 1$ is the $2\times2$ identity operator, denoted also by $\tau_0$. The projector representation is convenient because it allows the matrix elements entering the susceptibility to be written in a form that is independent of an arbitrary phase choice of the eigenstates of the BdG Hamiltonian.

In order to proceed, we introduce two independent static perturbations which introduce position-dependent modulations at the cavity wave vector $Q\equiv k_c$,
\begin{equation}
\delta H(y)
=
\cos(Qy)
\left(
\phi_Q\Gamma_n+\eta_Q\Gamma_p
\right),
\qquad \text{ where }
\Gamma_n\equiv\tau_z,
\text{ and }
\Gamma_p\equiv\tau_x.
\label{eq:SM_lr_probe}
\end{equation}
The field $\phi_Q$ perturbs the diagonal part of the BdG Hamiltonian and, therefore, probes the density response, whereas $\eta_Q$ perturbs the real off-diagonal pairing field $\Delta_0$, defined in Eq.~\eqref{eq:SB_Delta_def}. The matrices $\Gamma_n$ and $\Gamma_p$ denote the corresponding coupling matrices in the Nambu space, where the subscripts $n$ and $p$ indicate the density and pairing channels, respectively, that the fields are coupled to.

Consistent with the definitions in Eqs.~\eqref{eq:SB_averages} and \eqref{eq:SB_Delta_def}, the first components of the Fourier transform of the fields take the following approximate form
\begin{equation}
n(y)
=
n_0+\rho_Q\cos(Qy),
\qquad
F(y)
=
F_0+F_Q\cos(Qy)
=
F_0+\frac{p_Q}{2}\cos(Qy),
\text{ where }
p_Q\equiv2F_Q.
\label{eq:SM_lr_harmonics}
\end{equation}
Here, $\rho_Q$ is the amplitude of the density modulation and $F_Q$ is the modulation of the anomalous density (anomalous pairing amplitude)
$
F(y)
=
\left\langle
\hat\psi_\downarrow(y)
\hat\psi_\uparrow(y)
\right\rangle,
$
introduced in Eq.~\eqref{eq:SB_averages}.

To obtain the static susceptibility, let $|\mu\rangle$ denote an eigenstate of
$H_0$, with $H_0|\mu\rangle=E_\mu|\mu\rangle$. Here, $\mu = (k,s)$ collectively denotes the
momentum and the energy branch.
The equilibrium one-body density matrix in Nambu space, i.e., $(R_0)_{ij} = \av{\Phi_j^\dagger \Phi_i}$, is $R_0=f(H_0)$, where $f(E)$ is the Fermi-Dirac distribution function defined in Eq.~\eqref{eq:SB_fermi}. For a weak perturbation $\delta H$, 
we write $R=R_0+\delta R$, where $\delta R$ is the perturbation of the density matrix due to $\delta H$. 
The first order in perturbation theory yields, in the basis of eigenstates of $H_0$,
\begin{equation}
(\delta R)_{\mu\mu'}
=
\delta H_{\mu\mu'}
\frac{f(E_\mu)-f(E_{\mu'})}{E_\mu-E_{\mu'}},
\qquad \text{where }
\delta H_{\mu\mu'}\equiv\langle \mu|\delta H|\mu'\rangle.
\label{eq:SM_lr_deltaR}
\end{equation}

When $E_{\mu}=E_{\mu'}$, the ratio is understood as its limiting value $f'(E_{\mu})$.
For any single-particle operator $O_a$, where $a\in\{n,p\}$ labels the density or pairing channels, its expectation value can be written as
$\langle O_a\rangle=\operatorname{Tr}(O_aR)$, and, therefore, its contribution within the linear response theory is $$\delta\langle O_a\rangle=\operatorname{Tr}(O_a\delta R)$$.

Expanding the trace in the eigenstate basis of $H_0$ and using
Eq.~\eqref{eq:SM_lr_deltaR}, we obtain
\begin{equation}
\delta\langle O_a\rangle
=
\sum_{\mu,\mu'}
\langle\mu'|O_a|\mu\rangle
\delta H_{\mu\mu'}
\frac{f(E_\mu)-f(E_{\mu'})}
{E_\mu-E_{\mu'}}.
\label{eq:SM_lr_deltaO}
\end{equation}

The susceptibility is obtained by differentiating the induced expectation value with respect to the corresponding perturbing field:
\begin{subequations}
\label{SM:chi_def}
    \begin{align}
        \chi_{an}(Q)
        &\equiv
        \left.
        \frac{\partial\,\delta\langle O_a\rangle}{\partial\phi_Q}
        \right|_{\phi_Q=\eta_Q=0}
         =
        \left.
        \operatorname{Tr}
        \left(
        O_a\frac{\partial\delta R}{\partial\phi_Q}
        \right)
        \right|_{\phi_Q=\eta_Q=0}, \\
        \chi_{ap}(Q)
        &\equiv
        \left.
        \frac{\partial\,\delta\langle O_a\rangle}{\partial\eta_Q}
        \right|_{\phi_Q=\eta_Q=0}
        =
        \left.
        \operatorname{Tr}
        \left(
        O_a\frac{\partial\delta R}{\partial\eta_Q}
        \right)
        \right|_{\phi_Q=\eta_Q=0}.      
        \end{align}
\end{subequations}
{Since $\delta\langle O_n\rangle=\rho_Q$ and
$\delta\langle O_p\rangle=p_Q$, these definitions imply}
\begin{equation}
\begin{pmatrix}
\rho_Q\\
p_Q
\end{pmatrix}
=
\begin{pmatrix}
\chi_{nn} & \chi_{np}\\
\chi_{pn} & \chi_{pp}
\end{pmatrix}
\begin{pmatrix}
\phi_Q\\
\eta_Q
\end{pmatrix} \equiv \bm \chi \begin{pmatrix}
\phi_Q\\
\eta_Q
\end{pmatrix}.
\label{eq:SM_lr_chimatrix}
\end{equation}
In this notation, $\chi_{ab}$, where $a,b\in\{n,p\}$, are the matrix elements of the static susceptibility matrix $\bm \chi$, they quantify the susceptibility between the channels $a$ and $b$. The second index $b$ labels the channel of the perturbing field, while the first index $a$ labels the channel in which the response is measured. For example, $\chi_{np}$ describes the density response produced by a perturbation of the pairing field, whereas $\chi_{pn}$ describes the {pairing response} produced by a density perturbation.

Equation~\eqref{eq:SM_lr_deltaO} shows that evaluating the susceptibility
requires the occupation and energy factor together with two matrix
elements: $\langle\mu'|O_a|\mu\rangle$ for the measured observable and
$\delta H_{\mu\mu'}$ for the applied perturbation.

For the perturbation in Eq.~\eqref{eq:SM_lr_probe}, it is useful to describe explicitly which eigenstates are coupled. 
We define
$k_\pm=k\pm Q/2$ and use the notation 
$|k_-,t\rangle$ and $|k_+,s\rangle$ for the eigenstates of $H_0$ at these two momenta. We also define
$\xi_\pm\equiv\xi_{k_\pm}$ and $E_\pm\equiv\sqrt{\xi_\pm^2+\Delta_0^2}$.
As previously, the indices $s,t\in\{+,-\}$ label the energy branches.

The factors $e^{\pm iQy}$ in $\cos(Qy)=(e^{iQy}+e^{-iQy})/2$ change the momentum of the state on which
the perturbation acts by $\pm Q$. The term with $+Q$, therefore, connects
$|k_-,t\rangle$ to $|k_+,s\rangle$. Its matrix element is
\begin{equation}
\delta H_{(k_+,s),(k_-,t)}
=
\frac12
\left[
\phi_Q M_{st}^{n}(k,Q)
+
\eta_Q M_{st}^{p}(k,Q)
\right],
\label{eq:SM_lr_Vmatrix}
\end{equation}
where
$M_{st}^{a}(k,Q)
\equiv
\langle k_+,s|\Gamma_a|k_-,t\rangle$.

The two matrix elements can be interpreted in the following way 
$M_{st}^{n}$ measures how strongly the density perturbation
$\phi_Q\Gamma_n$ couples the states
$|k_-,t\rangle$ and $|k_+,s\rangle$, whereas
$M_{st}^{p}$ measures the coupling of the same two states through the
pairing perturbation $\eta_Q\Gamma_p$. If one of these matrix elements
vanishes, that particular transition cannot be generated through the
corresponding channel. If both are nonzero, their product can contribute to the
susceptibility coupling the density and pairing channels.

For a fixed {pair of states} $(k_-,t)$ and $(k_+,s)$, the product of the matrix element associated with the measured channel $a$ and that associated with the perturbing channel $b$ is
$$C_{ab}^{st}(k,Q)
\equiv
[M_{st}^{a}(k,Q)]^*M_{st}^{b}(k,Q).$$
Equivalently, using the projection operators introduced above,
$$C_{ab}^{st}
=
\mathrm{Tr}
[P_s(k_+)\Gamma_bP_t(k_-)\Gamma_a].$$
Thus, $C_{ab}^{st}$ is the factor of matrix element associated with this pair of states in
the susceptibility $\chi_{ab}$ when the perturbation acts through channel $b$ and the response is evaluated in channel $a$.

For $a=b$, this definition reduces to
$C_{nn}^{st}=|M_{st}^{n}|^2$ and
$C_{pp}^{st}=|M_{st}^{p}|^2$.
These quantities are, therefore, non-negative and measure the strength with
which the transition is accessible through the density or pairing channel.
The off-diagonal elements, contain matrix elements from both channels and enter the mixed
density--pairing susceptibility, which we call the cross-channel susceptibility.

We define $\theta_\pm \equiv \theta_{k_\pm}$ and $\Theta \equiv \theta_++\theta_-$, and substitute the projectors from Eq.~\eqref{eq:SM_lr_projector} into the equation defining
$C_{ab}^{st}(k,Q)$. As a result, we obtain
\begin{subequations}
\label{eq:SM_lr_coherence}
\begin{align}
C_{nn}^{st}
&=
\frac12\left[1+st\cos\Theta\right]
=
\frac12
\left[
1+st\frac{\xi_+\xi_- -\Delta_0^2}{E_+E_-}
\right],
\\
C_{pp}^{st}
&=
\frac12\left[1-st\cos\Theta\right]
=
\frac12
\left[
1-st\frac{\xi_+\xi_- -\Delta_0^2}{E_+E_-}
\right],
\\
C_{np}^{st}=C_{pn}^{st}
&=
\frac{st}{2}\sin\Theta
=
\frac{st}{2}
\frac{\Delta_0(\xi_++\xi_-)}{E_+E_-}.
\label{sm:Cnp}
\end{align}
\end{subequations}
These expressions lead to the following two properties of $C_{ab}^{st}$. First,
$C_{nn}^{st}+C_{pp}^{st}=1$, i.e., for a given pair of states, the relative matrix-element
weight is distributed between the density and pairing channels. Second,
$C_{nn}^{st}C_{pp}^{st}-(C_{np}^{st})^2=0$.
This follows directly from the fact that, for fixed $k,s,t$, the matrix
$C^{st}$ is constructed from the two-component vector
$(M_{st}^{n},M_{st}^{p})$ as an outer product.

Combining these matrix elements with the first-order change of the density
matrix in Eq.~\eqref{eq:SM_lr_deltaR} gives the  static
susceptibility defined in Eq.~\eqref{SM:chi_def}
\begin{equation}
\chi_{ab}(Q)
=
\frac{1}{L}
\sum_{k,s,t}
W_{st}(k,Q)\,
C_{ab}^{st}(k,Q),
\text{ where }
W_{st}(k,Q)
\equiv
\frac{f(sE_+)-f(tE_-)}
{sE_+-tE_-}.
\label{eq:SM_lr_spectral}
\end{equation}
At this point, we emphasize that the above result is a non-self-consistent response function. In the next subsections, we will impose the self-consistency condition, while in the discussion below, we will discuss the properties of the derived formulas.

In Eq.~\eqref{eq:SM_lr_spectral}, the factor $W_{st}$ contains the occupation difference between the two
states and the corresponding energy denominator. The susceptibility,
therefore, receives a large contribution when the two states have different
occupations and $C_{ab}^{st}$ is large.
In our case, for the static equilibrium calculations, $C_{np}^{st}=C_{pn}^{st}$, and, therefore, $$\chi_{np}=\chi_{pn}.$$

An immediate consequence of Eq.~\eqref{eq:SM_lr_coherence} is
\begin{equation}
\Delta_0=0
\quad\Longrightarrow\quad
\chi_{np}(Q)=\chi_{pn}(Q)=0.
\label{eq:SM_lr_selection}
\end{equation}
When $\Delta_0=0$, the eigenstates of $H_0$ have definite particle or hole character. Since $\Gamma_n$ is diagonal in this basis and $\Gamma_p$ connects the particle and hole parts, their matrix elements cannot both be nonzero for the same pair of states, and hence $\chi_{np}=0$. In contrast, $\chi_{pp}$ generally remains finite because $\Gamma_p$ can still connect particle and hole states. 

We now comment on the signs of the static susceptibilities.
The Fermi function decreases continuously with energy:
$f'(E)\leqslant0$. Therefore, for any two energies $E_1$ and $E_2$,
the ratio
$[f(E_1)-f(E_2)]/(E_1-E_2) \leqslant 0$. This remains true when
$E_1=E_2$, where the ratio approaches $f'(E_1)$.
Consequently, $W_{st}(k,Q)\leqslant0$ for every $k,s,t$.

For the density channel,
$C_{nn}^{st}=|M_{st}^{n}|^2\geqslant0$.
Each term contributing to $\chi_{nn}$ is, therefore, the product of a
factor $W_{st} \leqslant 0$ and $|M_{st}^{n}|^2$, which leads to $\chi_{nn}\leqslant0$. 
The same argument gives
$\chi_{pp}\leqslant0$ because
$C_{pp}^{st}=|M_{st}^{p}|^2\geqslant0$.
The off-diagonal factor is different, because
$C_{np}^{st}=(M_{st}^{n})^*M_{st}^{p}$ is a product of two different
matrix elements, and no definite sign follows for $\chi_{np}$.

The representation at hand, however, can be employed for deriving a bound on the cross-channel susceptibility $\chi_{np}$. To this end, we notice that for every vector $\bm z=(z_n,z_p)^T$, we have
$$\bm z^\dagger C^{st}\bm z
=
\left|
z_nM_{st}^{n}+z_pM_{st}^{p}
\right|^2
\geqslant0.$$
Thus, $C^{st}$ is a positive semi-definite matrix. Since $W_{st}\leqslant0$, Eq.~\eqref{eq:SM_lr_spectral} implies that $-\bm\chi$ is also positive semi-definite, which yields
\begin{equation}
|\chi_{np}(Q)|^2
\leqslant
\chi_{nn}(Q)\chi_{pp}(Q),
\text{  and  }
|\mathcal M(Q)|\leqslant 1
\text{  where  }
\mathcal M(Q)
\equiv
\frac{\chi_{np}(Q)}
{\sqrt{\chi_{nn}(Q)\chi_{pp}(Q)}}.
\label{eq:SM_lr_bound}
\end{equation}
The dimensionless quantity $\mathcal M$ measures the density--pairing susceptibility relative to the susceptibilities of each channel.

\subsection{Self-consistent HFB response function of the density and pairing fields}
\label{sec:SM_lr_hfb_response}

The susceptibility $\bm\chi$ introduced above gives the response to prescribed
fields $\phi_Q$ and $\eta_Q.$ 
In the approach within the HFB theory, however, these fields must satisfy the
self-consistency conditions in Eqs.~\eqref{eq:SB_hbalanced} and
\eqref{eq:SB_Delta_def}. We now include these induced changes to obtain the
self-consistent density and pairing response functions.

The cavity modifies both the single-particle potential $V_{\rm a}(y)$ and the
effective interaction $U_{\rm eff}(y)$. For a real pump amplitude $\beta$, we
define the quadrature
$X\equiv\alpha+\alpha^*$.
Close to the continuous self-ordering transition, $\alpha$ and $X$ are small.
Using $f_Q(y)\equiv\cos(Qy)$, the optical potential can be linearized using
$$|\beta+\alpha f_Q(y)|^2
=\beta^2+\beta X f_Q(y)+\mathcal{O}(|\alpha|^2).$$
The bare interaction is, therefore, renormalized to
$$U_0=g_0+g_\Delta\beta^2,$$
while the first-order variations are
$\delta V_{\rm a}(y)=  g_{\rm a} \beta X f_Q(y)$ and
$\delta U_{\rm eff}(y)= g_\Delta \beta  X f_Q(y)$.

The effective single-particle potential of the BdG Hamiltonian
is $V_{\rm a}(y)+U_{\rm eff}(y)n(y)/2$, as follows from
Eq.~\eqref{eq:SB_hbalanced}. Using
$n(y)=n_0+\rho_Q\cos(Qy)$ and keeping only terms linear in the perturbation,
it's first component of the Fourier transform is
$\beta(g_{\rm a}+n_0g_\Delta/2)X+U_0\rho_Q/2$.
Likewise, from the gap equation
$\Delta(y)=U_{\rm eff}(y)F(y)$ and
$F(y)=F_0+p_Q\cos(Qy)/2$, the Fourier transform yields the first component equal to
$\beta g_\Delta F_0X+U_0p_Q/2$. Hence, the two fields needed for evaluating the 
susceptibilities are
\begin{equation}
\phi_Q
=
AX+u\rho_Q,
\qquad
\eta_Q\equiv\Delta_Q
=
BX+up_Q,
\text{\ \ where }
u\equiv\frac{U_0}{2},
\quad
A\equiv
\beta\left(g_{\rm a}+\frac{n_0g_\Delta}{2}\right),
\quad
B\equiv\beta g_\Delta F_0 .
\label{eq:SM_lr_fields}
\end{equation}
The term $u\rho_Q$ is the induced Hartree contribution to the density field.
Similarly, $up_Q=U_0F_Q$ is the change in the pairing field produced by $F_Q$.
 
The terms $AX$ and $BX$ originate directly from the
cavity-induced modulation of $V_{\rm a}$ and $U_{\rm eff}$, respectively.
The same coefficient $u=U_0/2$ appears in both equations because the Hartree contribution
contains $U_0n/2$, while $p_Q=2F_Q$ by definition. 

We collect the perturbed quantities into a vector
\begin{equation}
  \bm x= \begin{pmatrix}\rho_Q \\ p_Q  \end{pmatrix}
\end{equation}
and the coefficients $A$ and $B$ into 
\begin{equation}
    \bm s = 
    \begin{pmatrix}
        A \\ B        
    \end{pmatrix},
\end{equation}
which, as seen from Eq.~\eqref{eq:SM_lr_fields}, are proportional to the pump strength $\beta$. Combining Eq.~\eqref{eq:SM_lr_chimatrix} with
Eq.~\eqref{eq:SM_lr_fields} yields
\begin{equation}
\bm x
=
\bm\chi(\bm sX+u\bm x) \quad
\Longleftrightarrow \quad 
\bm x
=
(\bm K \bm s)\, X,
\text{ where }
\bm K
\equiv
(\bm 1-u\bm\chi)^{-1}\bm\chi .
\label{eq:SM_lr_K}
\end{equation}
Thus, the vector $\bm K\bm s$ has an interpretation of the self-consistent static susceptibility in the density and pairing channel when the perturbing force, in a sense of the linear response theory, is given by the quadrature $X$ of the field $\alpha$, which is, as we assumed, small in the ordered phase. Crucially, the structure of the derivation allows us to identify the contribution to each channel of the response function induced by a self-consistent change in the same or another channel. Therefore, the matrix $K_{ab}$ plays an important role that enables us to quantify the susceptibility of the $a$ channel arising from the $b$ channel, and, thus, the nature of the intertwining of the ordered CDW--PDW phases. The nonlinear dependence of $\bm{K}$ on $\bm \chi$ manifestly encodes the self-consistency condition. 

To write the matrix $\bm K$ explicitly, we define the determinant
\begin{equation}
D
\equiv
\det(\bm 1-u\bm\chi)
=
(1-u\chi_{nn})(1-u\chi_{pp})
-u^2\chi_{np}^2 .
\label{eq:SM_lr_Dm}
\end{equation}
Using $\chi_{np}=\chi_{pn}$, which is valid in the equilibrium formalism 
for the  susceptibility considered here, we obtain
\begin{equation}
\bm K
\equiv
\begin{pmatrix}
K_{nn} & K_{np}\\
K_{pn} & K_{pp}
\end{pmatrix}
=
\frac{1}{D}
\begin{pmatrix}
(1-u\chi_{pp})\chi_{nn}+u\chi_{np}^{\,2}
&
\chi_{np}
\\
\chi_{np}
&
(1-u\chi_{nn})\chi_{pp}+u\chi_{np}^{\,2}
\end{pmatrix},
\label{eq:SM_lr_K_explicit}
\end{equation}
where we used the symmetry of $\bm\chi$.
In particular, this expression shows that $K_{np}=K_{pn}$.

Using the self-consistent susceptibility $\bm K$, the density and the pairing field responses to the cavity quadrature can be written as
\begin{subequations}
\label{eq:SM_lr_ratios}
\begin{align}
\frac{\rho_Q}{X}
&=
K_{nn}A+K_{np}B,
\label{eq:SM_lr_rhoX}
\\
\frac{p_Q}{X}
&=
K_{np}A+K_{pp}B.
\label{eq:SM_lr_pX}
\end{align}
\end{subequations}
The induced response in the gap follows directly from
$\Delta_Q=BX+up_Q$ in Eq.~\eqref{eq:SM_lr_fields}. Dividing this relation
by $X$ and using Eq.~\eqref{eq:SM_lr_pX} gives
\begin{equation}
\frac{\Delta_Q}{X}
=
B+u\frac{p_Q}{X}
=
uK_{np}A+(1+uK_{pp})B.
\label{eq:SM_lr_DeltaX}
\end{equation}

In particular, Eq.~\eqref{eq:SM_lr_DeltaX} separates two contributions to
the modulation of the pairing field $\Delta_Q$. The term $uK_{np}A$ describes the
contribution generated by the cross-coupling to the density channel.
The term $(1+uK_{pp})B$ is associated with the direct cavity modulation of the
pairing field, together with its self-consistent response.

Equations~\eqref{eq:SM_lr_K} also determine the stability of the homogeneous 
solution against a perturbation at finite $Q$ in the absence of a cavity
field. Setting $X=0$ in Eq.~\eqref{eq:SM_lr_K} gives
$(\bm 1-u\bm\chi)\bm x=0$. A nonzero solution for $\bm x$, therefore, requires
\begin{equation}
D=0.
\label{eq:SM_lr_matter_instability}
\end{equation}
At this point, the susceptibility matrix $\bm K$ is singular, indicating an instability of the homogeneous state. Away from such an instability, $D\neq0$, and $X=0$ implies $\rho_Q=p_Q=\Delta_Q=0$ at linear order.
We remark that in our numerical solutions, we find $D\neq0$ along the boundary between SF and SF+CDW+PDW+SR phases of the phase diagram from the main text. The finite-$Q$ density and pairing modulations at this boundary are, therefore, induced by the cavity field rather than by an instability of the homogeneous state in the absence of the cavity field.

The presented reasoning demonstrates why the system cannot enter directly into a phase with PDW order from the normal state. Indeed, when $\Delta_0=0$, the anomalous field vanishes, i.e., $F_0=0$, and, hence, the coefficient $B=0$, as seen from Eq.~\eqref{eq:SM_lr_fields}.
Then, Eq.~\eqref{eq:SM_lr_selection} gives $\chi_{np}=0$. Provided
$D\neq0$, Eqs.~\eqref{eq:SM_lr_ratios} imply
$p_Q=\Delta_Q=0$ at linear order, while a finite density component 
$\rho_Q$ remains possible. On the other hand, when $\Delta_0 \neq 0$, the system can develop finite density modulation, entering into the CDW phase, without supporting modulation in the pairing field, i.e., without PDW order, as observed in the phase diagram in the main text.

\subsection{Self-consistency for cavity field and the pump threshold}
\label{sec:SM_lr_cavity_loop}

We now combine the density and pairing response functions with the steady-state
condition for the intra-cavity field. As derived in Eq.~\eqref{eq:SC_alpha}, the
cavity amplitude is
\begin{equation}
\alpha
=
\frac{
\beta\left(
g_{\rm a}D_{\rm a}
+
g_\Delta D_\Delta
\right)
}{
\Delta_c
-
g_{\rm a}A_{\rm a}
-
g_\Delta A_\Delta
+i\kappa
}.
\label{eq:SM_lr_cavity_equation}
\end{equation}

For the state considered here, the Hamiltonian does not mix the two
spin components, and the spin coherence is absent.
The mean field decoupling via the Wick's theorem yields
\begin{equation}
\cC(y)
\equiv
\left\langle
\hat n_\uparrow(y)\hat n_\downarrow(y)
\right\rangle
=
\frac{n^2(y)}{4}+|F(y)|^2 .
\label{eq:SM_lr_contact}
\end{equation}

Exploiting the description in terms of the lowest order of the Fourier components, introduced in
Eq.~\eqref{eq:SM_lr_harmonics}, and keeping terms only to linear order in
$\rho_Q$ and $p_Q$, $C(y) = \cC_0 + \cC_Q \cos(Q y)$, where 
\begin{equation}
\cC_Q
=
\frac{n_0}{2}\rho_Q
+
F_0p_Q .
\label{eq:SM_lr_contactQ}
\end{equation}
The first term arises from the density modulation, whereas the second
contains the modulation of the anomalous field. 

For one unit cell $\lambda_c=2\pi/Q$ , we introduce
$\cN_Q\equiv\int_0^a dy\,f_Q^2(y)=a/2$. With the same normalization in Eq.~\eqref{eq:SC_alpha}, the overlaps are
$D_{\rm a}=\cN_Q\rho_Q$ and
$D_\Delta=\cN_Q\cC_Q$. The numerator in Eq.~\eqref{eq:SM_lr_cavity_equation}
is  
\begin{equation}
J
\equiv
\beta
\left(
g_{\rm a}D_{\rm a}
+
g_\Delta D_\Delta
\right)
=
\cN_Q
\left(
A\rho_Q+Bp_Q
\right)
=
\cN_Q\,\bm s^T\bm x .
\label{eq:SM_lr_J}
\end{equation}
Thus, $J$ describes coherent scattering of photons from the pump beam into the cavity due to the density and modulations of the anomalous field. Notice that the same coefficients $A$ and $B$ appear here and in Eq.~\eqref{eq:SM_lr_fields} because both follow from the same linear expansion of $V_{\rm a}(y)$ and $U_{\rm eff}(y)$ in the cavity quadrature~$X$.

The denominator in Eq.~\eqref{eq:SM_lr_cavity_equation} is evaluated using
the homogeneous solution. We, therefore, define
\begin{equation}
\widetilde\Delta_c^{(0)}
\equiv 
\Delta_c
-
g_{\rm a}A_{\rm a}^{(0)}
-
g_\Delta A_\Delta^{(0)} ,
\label{eq:SM_lr_Deltac0}
\end{equation}
where the superscript $(0)$ denotes evaluation at
$\rho_Q=p_Q=0$. Indeed, near the continuous transition, $J$ is already first order
in the infinitesimal fields $\rho_Q$ and $p_Q$. A first-order change $\delta\widetilde\Delta_c$ of the denominator would, therefore, enter the cavity amplitude through the product
$J\,\delta\widetilde\Delta_c$. This product is of second order in the
small fields and can be neglected in the linear analysis.

Consequently, the linearized steady-state cavity field becomes
\begin{equation}
\alpha
=
\frac{J}
{\widetilde\Delta_c^{(0)}+i\kappa},
\text{ and }
X
=
\chi_c J,
\text{ where  }
\chi_c
\equiv
\frac{
2\widetilde\Delta_c^{(0)}
}{
[\widetilde\Delta_c^{(0)}]^2+\kappa^2
}.
\label{eq:SM_lr_Rc}
\end{equation}
Here, $\chi_c$ is the static cavity susceptibility relating the quadrature $X$ to $J$.

The self-consistent fermionic response obtained in the previous subsection
is $\bm x=\bm K\bm sX$. Substituting this into
Eq.~\eqref{eq:SM_lr_J}, and then using $X=\chi_cJ$, results in
\begin{equation}
(1- \cN_Q \chi_c \bm s^T\bm K\bm s) X  = 0 .
\label{eq:SM_lr_loop}
\end{equation}
The solution without an intra-cavity field, i.e., $X=0$, always exists. However, a nonzero  
solution is possible when
\begin{equation}
1 = \cN_Q\chi_c\, \bm s^T\bm K\bm s 
\label{eq:SM_lr_threshold}
\end{equation}
This condition determines the threshold pump strength for the self-organization.

It is useful to make this self-consistency condition more explicit. Let
$\lambda_\nu$ and $\bm e_\nu$, denote the eigenvalues and
normalized eigenvectors of the susceptibility, i.e.,
$$\bm\chi\bm e_\nu=\lambda_\nu\bm e_\nu,$$ 
with $\nu=\pm$.
Since $-\bm\chi$ is positive
semi-definite, as shown above, both eigenvalues satisfy
$\lambda_\nu\leqslant0$. Therefore, the eigenvalues of $\bm K$ are
\begin{equation}
k_\nu
=
\frac{\lambda_\nu}
{1-u\lambda_\nu}.
\label{eq:SM_lr_k_eigen}
\end{equation}
The response along $\bm e_\nu$ remains finite as long as
$1-u\lambda_\nu\neq0$. If this quantity vanishes, the response diverges,
indicating the finite-$Q$ instability discussed in the previous
subsection. For the stable homogeneous solutions considered here,
$1-u\lambda_\nu>0$.
Since $\lambda_\nu\leqslant0$, it follows that
$k_\nu\leqslant0$.
Thus, the eigenvalues of $\bm K$ are non-positive for the stable
homogeneous solution.

Due to the specific form of the threshold condition in Eq.~\eqref{eq:SM_lr_threshold}, which contains the factor $\bm s^T \bm K \bm s$, it can now be split into the two eigenmodes of $\bm K$. Indeed, expanding  $\bm s$ in the orthonormal basis $\bm e_\pm$, gives
\begin{equation}
\bm s^T\bm K\bm s
=
\sum_{\nu=\pm}
k_\nu
\left|
\bm s\cdot\bm e_\nu
\right|^2 .
\label{eq:SM_lr_mode_decomposition}
\end{equation}
Therefore, we can rewrite Eq.~\eqref{eq:SM_lr_threshold}, for the threshold pump strength $\beta_c$, as
\begin{equation}
G_+(\beta_c) + G_-(\beta_c) = 1,
\label{eq:SM_lr_Gthreshold}
\end{equation}
where
\begin{equation}
G_\nu \equiv \cN_Q\chi_c\, k_\nu \left| \bm s\cdot\bm e_\nu \right|^2.
\label{eq:SM_lr_Gnu}
\end{equation}
Here, $k_\nu$ measures the strength of the response matrix $\bm K$ along the mode  $\bm e_\nu$, whereas n$|\bm s\cdot\bm e_\nu|^2$ measures how strongly the density and pairing channels contribute to that response mode.

Finally, we conclude that the self-organization condition in Eq.~\eqref{eq:SM_lr_Gthreshold} can be satisfied when the pump beam is red detuned from the effective cavity frequency, which is shifted by the interaction with the atomic cloud. Indeed, since, as we have established, $k_\nu\leqslant0$, from Eq.~\eqref{eq:SM_lr_Gnu} we need $\chi_c < 0$, which, from Eq.~\eqref{eq:SM_lr_Rc}, is possible only for negative detunings, $\widetilde\Delta_c^{(0)}<0$.

\subsection{Critical pump and Fano-type phase boundary}
\label{sec:SM_lr_fano}

The quadratic form in Eq.~\eqref{eq:SM_lr_Gthreshold} can explicitly be written as
\begin{equation}
\bm s^T\bm K\bm s
=
A^2K_{nn}
+
2ABK_{np}
+
B^2K_{pp}.
\label{eq:SM_lr_threepaths}
\end{equation}
The first and third terms are the contributions obtained when the cavity
couples through the density and pairing channels separately. The middle term
contains the cross-channel response $K_{np}$ 
and therefore it can be interpreted as a contribution quantifying the interference between the density and pairing channels.

Factoring out explicitly the pump field amplitude, i.e, using $A=\beta\mathcal A$ and $B=\beta\mathcal B$, with
$\mathcal A \equiv g_{\rm a}+n_0g_\Delta/2$,
$\mathcal B \equiv g_\Delta F_0$, and
$$r_{\rm eff} \equiv \frac{\mathcal B}{\mathcal A} = \frac{F_0}{g_{\rm a}g_\Delta^{-1}+\frac{n_0}{2}},$$ Eq.~\eqref{eq:SM_lr_threepaths}
becomes
\begin{equation}
\bm s^T\bm K\bm s
=
\beta^2\mathcal A^2
\left(
K_{nn}
+
2r_{\rm eff}K_{np}
+
r_{\rm eff}^2K_{pp}
\right).
\label{eq:SM_lr_Gbright}
\end{equation}
The threshold intensity $\beta_c^2$ from Eq.~\eqref{eq:SM_lr_threshold}, therefore, satisfies
\begin{equation}
\beta_c^2
=
\left.
\frac{1}{
\cN_Q \chi_c\mathcal A^2
\left(
K_{nn}
+
2r_{\rm eff}K_{np}
+
r_{\rm eff}^2K_{pp}
\right)
}
\right|_{\beta=\beta_c}.
\label{eq:SM_lr_betac}
\end{equation}
We notice that this equation is implicit because, on the right-hand side, the homogeneous quantities
$U_0$, $n_0$, $F_0$, the response matrix $\bm K$, and the effective cavity
detuning entering in $R_c$ all depend on the pump strength $\beta$. The threshold has to be obtained self-consistently from the condition above.

Nevertheless, the role of the cross-channel response $K_{np}$ can be isolated by comparing
the three contributions in  Eq.~\eqref{eq:SM_lr_betac}.  The terms $K_{nn}$ and $K_{pp}$ are non-positive, since $\bm K$ is negative semi-definite.
If $r_\mathrm{eff} K_{np}<0$, this off-diagonal term makes $\bm s^T\bm K\bm s$ more negative and 
the threshold pump is, consequently, reduced. 
On the other hand, if $r_\mathrm{eff} K_{np}>0$, it partially
cancels the diagonal contributions and raises the threshold intensity $\beta_c^2$. 
In this operational sense the two cases correspond, respectively, to cooperation and competition of the CDW and PDW orders. We choose the phase of the real anomalous field such that $F_0>0$. Since $\mathcal A<0$ throughout the strongly attractive interaction range used for the Fano-boundary calculation, $r_{\rm eff}$ has the opposite sign to $g_\Delta$. This phase choice does not affect the result, i.e., changing $F_0\rightarrow-F_0$ reverses the signs of both $r_{\rm eff}$ and $K_{np}$, leaving their product unchanged.

The connection with a Fano-type superradiant phase boundary becomes transparent when the these background quantities, such as $n_0$, $F_0$, $\bm K$, vary slowly across the region of interest. 
To this end, we define $z\equiv1/r_{\rm eff}$ and
$\overline{\bm K}\equiv-\bm K$, which is positive semi-definite in the stable regime. 
If $\mathcal A$, $\overline{\bm K}$, and $\chi_c$ are treated as
approximately constant, Eq.~\eqref{eq:SM_lr_betac} becomes
\begin{equation}
\beta_c^2(z)
\simeq
\frac{
z^2
}{
\cN_Q(-\chi_c)\mathcal A^2
\left(
\overline K_{nn}z^2
+
2\overline K_{np}z
+
\overline K_{pp}
\right)
}.
\label{eq:SM_lr_fano_proxy}
\end{equation}
For $\det\overline{\bm K}>0$, the term in the denominator can be written as
$\overline K_{nn}[(z-z_0)^2+\gamma^2]$, where
$z_0=-\overline K_{np}/\overline K_{nn}$ and
$\gamma=\sqrt{\det\overline{\bm K}}/\overline K_{nn}$.
Introducing $\epsilon=(z-z_0)/\gamma$ and
$q_F=z_0/\gamma$, Eq.~\eqref{eq:SM_lr_fano_proxy} takes the asymmetric profile which is characteristic of the Fano resonance,
\begin{equation}
\beta_c^2 \simeq
\beta_{\rm bg}^2
\frac{(q_F + \epsilon)^2}{1+\epsilon^2},
\label{eq:SM_lr_fano_form}
\end{equation}
where $\beta_{\rm bg}^{-2}=\cN_Q(-\chi_c)\mathcal A^2\overline K_{nn}$. The asymmetry parameter $q_F$ is determined by the off-diagonal response, i.e., if $\overline K_{np}=0$, then $z_0=0$ and $q_F$ vanishes. 

The experimentally and numerically used control variable is the molecular detuning $\Delta_B$, which, in our studies, is probed by the inverse
of the atom-pair coupling strength
$x=[g_\Delta/(E_Ra)]^{-1}$. Rewriting the $z$ variable in terms of $x$, we obtain
\begin{equation}
z
=
\frac{g_{\rm a}}{E_RaF_0}\,x  + \frac{n_0}{2F_0}
\text{ and }
\epsilon = 
\frac{g_{\rm a}}{\gamma E_RaF_0}\,x  + \bigg( \frac{n_0}{2\gamma  F_0} - \frac{K_{np}}{\sqrt{\mathrm{det}\bm K}} \bigg)
\label{eq:SM_lr_inverseg}
\end{equation}
Thus, if $n_0$ and $F_0$ vary slowly with $g_\Delta$, $z$ is a linear function of  $x$. Moreover, when the single-atom contribution to $\mathcal A$ dominates, i.e, 
$|g_{\rm a}|\gg|n_0g_\Delta/2|$, one has $\mathcal A\simeq g_{\rm a}$, so the prefactor in Eq.~\eqref{eq:SM_lr_fano_form} also varies weakly. In this regime, the
threshold intensity $\beta_c^2$ as a function of $x$ is expected to be given by the Fano-type profile.
We employ a fitting function of this structure in the main text, where we analyze the numerical data for the strongly attractive fermionic gas. Outside this strongly interacting regime, all the quantities $n_0$, $F_0$, $\bm K$, $\mathcal A$, and $\chi_c$ depend on $g_\Delta$, and, consequently, deviations from the Fano-type profile are expected.

\subsection{Composition and interpretation of the critical mode}
\label{sec:SM_lr_composition}
Here, we identify the contribution of the density and pairing channels to the instability of the homogeneous state at the transition to the ordered state.
Since both $\rho_Q$ and $p_Q$ are proportional to
the small cavity quadrature $X$, their ratio approaches a finite
value at the transition. Using Eqs.~\eqref{eq:SM_lr_rhoX} and \eqref{eq:SM_lr_pX}, we obtain
\begin{equation}
\frac{p_Q}{\rho_Q}
=
\frac{
\chi_{np}A
+
(\chi_{pp}-ud_\chi)B
}{
(\chi_{nn}-ud_\chi)A
+
\chi_{np}B
},
\label{eq:SM_lr_composition}
\end{equation}
where we introduced $d_\chi\equiv\det\bm\chi =\chi_{nn}\chi_{pp}-\chi_{np}^2$. 

This quantity is complementary to the decomposition of the condition for the threshold pump strength into $G_\pm$. Indeed,  $G_\nu$ describes how strongly each response
eigenmode, encoded via $\bm K$, contributes to the self-organization transition, whereas $p_Q/\rho_Q$ describes the composition of the resulting fermionic perturbation induced by the emergence of the ordering.

Regarding the pairing field, given in Eq.~\eqref{eq:SM_lr_DeltaX}, it has two components,
\begin{equation}
\frac{\Delta_Q}{X} \sim u\chi_{np}A + (1-u\chi_{nn})B.
\end{equation}
Interestingly, the susceptibility $\chi_{pp}$ does not appear here. The term $u\chi_{np}A$ is present when the perturbation in the density impacts the gap equation via the presence of the cross-channel susceptibility. In contrast,
the term $(1-u\chi_{nn})B$ is associated with the direct perturbation of the pairing field through the spatial dependence of $U_{\rm eff}(y)$. 

The pairing field perturbation $\Delta_Q$ is dominated by the density channel only when
\begin{equation}
|u\chi_{np}A|
\gg
|(1-u\chi_{nn})B|.
\label{eq:SM_lr_density_induced}
\end{equation}
When the two contributions are comparable, the complete expression in
Eq.~\eqref{eq:SM_lr_DeltaX} must be used. These analytical observations are discussed together with the numerical results below.

\subsection{Results in the static linear response theory and the structure of the critical mode}
\label{sec:SM_lr_results}

Here, we apply the formalism, developed in the previous subsections, at the boundary of CDW+SF+PDW+SR and the homogeneous SF phase, where $\Delta_0\neq0$.

\begin{figure*}[t]
\centering
\includegraphics[width=0.94\textwidth]{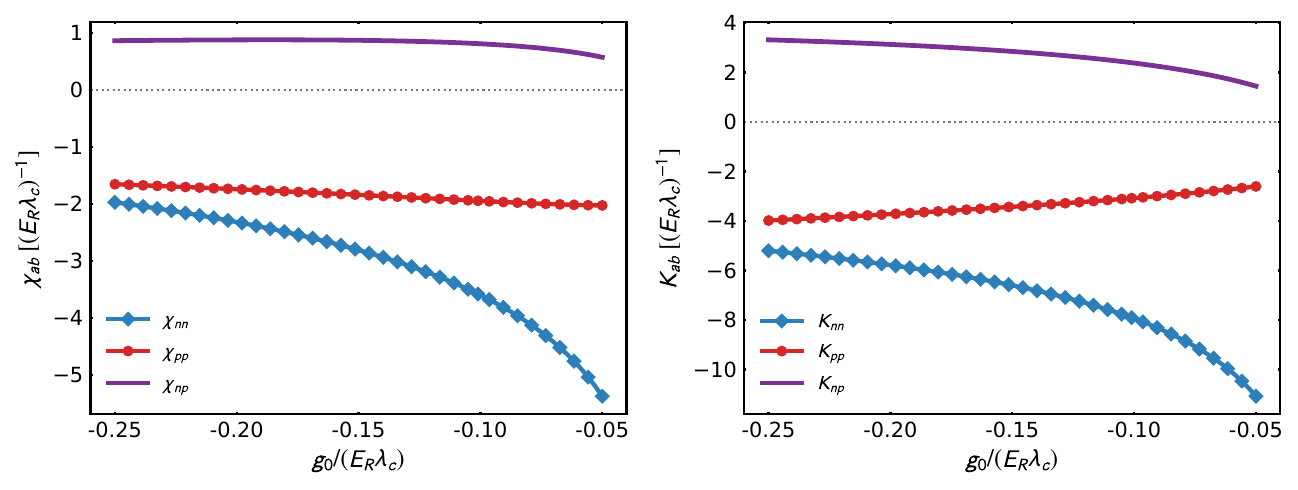}
\caption{
    The susceptibilities $\chi_{ab}$ (left panel) and $K_{ab}$ (right panel), both in the units of $1/(E_R\lambda_c)$, with $a,b\in\{n,p\}$ evaluated at the boundary of the SF and CDW+SF+PDW+SR phases, shown in Fig.~2 from the main text. Left: the matrix element of the non-self-consistent susceptibility $\bm\chi$. Right: the elements of the self-consistent susceptibility matrix $\bm K$. As the attractive bare atom-atom interaction strength  is reduced, the density response in the density channel becomes increasingly dominant while the cross-channel 
response function remains finite. The gray dotted lines mark zero.
}
\label{fig:SM_lr_bare_dressed}
\end{figure*}

We scan the contact interaction $g_0$, in analogy to the phase diagram presented in the main text.
To this end, for each $g_0$, we determine the threshold $\beta_c(g_0)$ 
Eq.~\eqref{eq:SM_lr_Gthreshold}. 
The parameters are the same as in Fig.~2 of the main text, but for convenience we repeat them here, i.e.,
\begin{equation}
\frac{\kappa}{E_R}=0.101,
\qquad
\frac{\Delta_c}{E_R}=-3.04,
\qquad
\frac{g_{\rm a}}{E_R}=-0.152,
\qquad
\frac{g_\Delta}{E_R \lambda_c}=-0.0152,
\label{eq:SM_lr_result_parameters}
\end{equation}
with $N_\uparrow=N_\downarrow=1$ is the number of atoms per unit cell.

We first examine how the density and pairing field susceptibilities change as the bare atom-atom interaction coupling strength $g_0$ is reduced. Figure~\ref{fig:SM_lr_bare_dressed} compares the susceptibility $\bm\chi$ with the self-consistent $\bm K=(\bm 1-u\bm\chi)^{-1}\bm\chi$ from Eq.~\eqref{eq:SM_lr_K}. 
From the figure, we see that $\chi_{np}>0$, while, as expected, $\chi_{pp}<0$ and $\chi_{nn}<0$.

A clear change in the character of the response occurs as $g_0$ is varied.
Toward weaker attraction, $|\chi_{nn}|$ increases substantially, while
$\chi_{np}$ decreases. The response of the system, therefore, is dominated by the density channel.
The susceptibility in the pairing channel, $\chi_{pp}$, changes more moderately. The increase of $|\chi_{nn}|$ is consistent with the proximity of the CDW+SR phase toward weaker attraction. 
As can be seen from the right panel for the susceptibility $\bm K$, the self-consistency modifies values of the susceptibilities, but the qualitative features are similar. Interestingly, $|\chi_{pp}|$ decreases while $|K_{pp}|$ increases when $g_0$ becomes more negative. This shows that the self-consistency condition is important for the pairing channel response. In general, however, for $u<0$, the eigenvalues $\lambda_\nu$ of $\bm\chi$ are non-positive. Therefore, the eigenvalues of $\bm{K}$, given by $k_\nu=\lambda_\nu/(1-u\lambda_\nu)$ satisfy $|k_\nu|>|\lambda_\nu|$. This shows that the self-consistently induced density and pairing fields enhance the response of the system.

\begin{figure*}[t]
\centering
\includegraphics[width=0.32\textwidth]{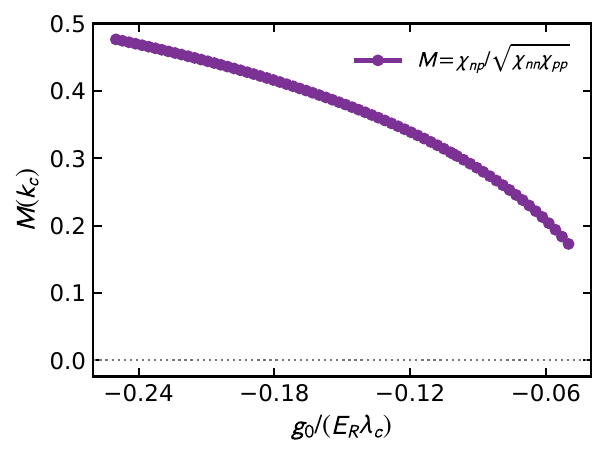}
\hfill
\includegraphics[width=0.32\textwidth]{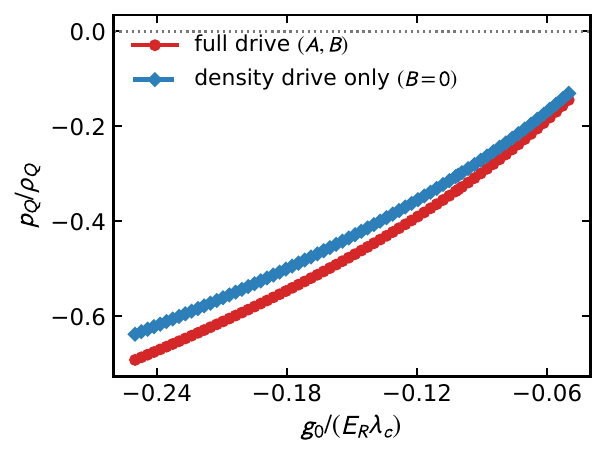}
\hfill
\includegraphics[width=0.32\textwidth]{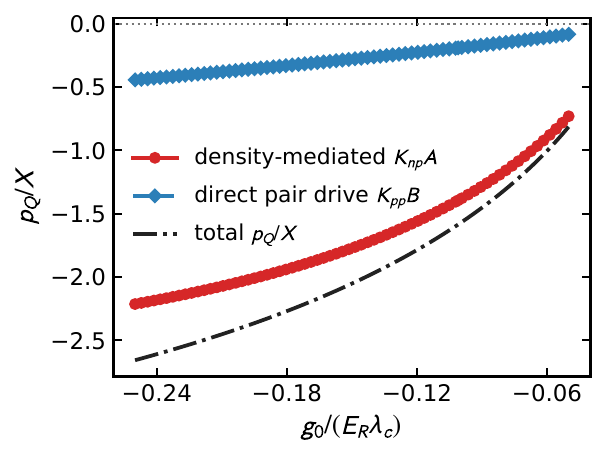}
\caption{
    Analysis of the density and pairing channels coupling along the boundary between SF and CDW+SF+PDW+SR phases.
    From left to right: 
    normalized cross-channel susceptibility $\mathcal M$ (left panel), 
    ratio $p_Q/\rho_Q$ of the orders (red) and the contribution with $B=0$ (blue) (middle panel), 
    the decomposition of the pairing field $p_Q/X=K_{np}A+K_{pp}B$ (right panel). The gray dotted lines mark zero.
}
\label{fig:SM_lr_intertwining}
\end{figure*}

As $g_0$ is reduced, the density response becomes
progressively stronger --- both $|\chi_{nn}|$ and $|K_{nn}|$ increase.
At the same time, the cross-channel components $\chi_{np}$ and $K_{np}$ vary more moderately than the corresponding density components and remain nonzero throughout the considered range of $g_0$, but become progressively smaller relative to the response in the density channel.
Thus, due to the vicinity of the CDW+SR phase, the coupling of the two channels persists, but the response along the superradiant phase boundary becomes dominated by the density channel as $g_0$ tends to zero.

In Fig.~\ref{fig:SM_lr_intertwining} (left panel), we present the normalized cross-channel susceptibility $\mathcal M$, defined in Eq.~\eqref{eq:SM_lr_bound}.
The value of $\mathcal M$, which is bounded by 1, is already appreciably different from zero, and its values are increasing from $\approx 0.2$ for weak to $\approx 0.5$ for stronger attractions.
Such behavior can be explained as follows. A large
gap $\Delta_0$ leads to strong particle--hole mixing in the 
eigenstates of the BdG Hamiltonian. Consequently, the same pair of eigenstates can have sizeable matrix elements of both $\Gamma_n$ and $\Gamma_p$, leading to a larger mixed susceptibility. As the pairing $\Delta_0$ weakens, the response function $\chi_{np}$ decreases due to $\Delta_0$ scaling in Eq.~\eqref{sm:Cnp}. 

To determine the contribution of the density channel to the pairing
field $p_Q$, we first calculate the threshold $\beta_c$ using both $A$
and $B$. We then evaluate Eqs.~\eqref{eq:SM_lr_rhoX} and
\eqref{eq:SM_lr_pX} at this threshold while keeping $U_0$, $\bm\chi$,
$\bm K$, and $A$ fixed, but setting $B=0$ only in these equations.
The remaining terms are $K_{nn}A$ in $\rho_Q/X$ and $K_{pn}A$ in
$p_Q/X$.

Specifically, from Eqs.~\eqref{eq:SM_lr_rhoX} and
\eqref{eq:SM_lr_pX}, we obtain
\begin{equation}
\left.
\frac{p_Q}{\rho_Q}
\right|_{B=0}
=
\frac{K_{pn}}{K_{nn}}
=
\frac{K_{np}}{K_{nn}},
\label{eq:SM_lr_ratio_B0}
\end{equation}
where the last equality follows from $K_{pn}=K_{np}$.

The middle panel in Fig.~\ref{fig:SM_lr_intertwining} shows that the
ratios obtained with $B=0$ and with both $A$ and $B$ differ only slightly
over the considered range of $g_0$. For $B=0$, the ratio
$K_{pn}/K_{nn}$ compares the contributions of the density channel to the
pairing and density fields, respectively. Its magnitude is largest for
stronger attraction and decreases as the attraction is reduced.
Including $B$ slightly increases the magnitude of $p_Q/\rho_Q$ for the
parameters considered here.

\begin{figure*}[t]
\centering
\includegraphics[width=0.94\textwidth]{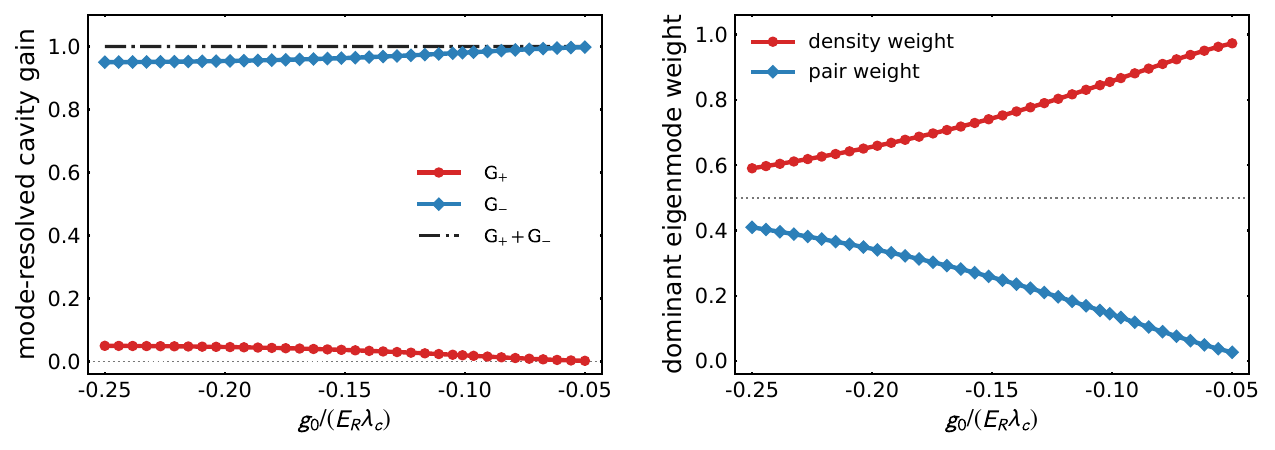}
\caption{
    Left: contributions $G_\pm$ of the two density--pairing response eigenmodes
    to the cavity self-consistency condition, cf. Eq.~\eqref{eq:SM_lr_Gthreshold}; 
    $G_+ + G_- = 1$ at the threshold $\beta_c$. 
    Right: density and pairing channel weights of the
    dominant eigenvector $\bm e_-$ of $\bm K$. 
    Strong attraction leads to strong mode hybridization, whereas toward weaker attraction, the density channel dominates
    The dotted line at $1/2$ marks equal density and anomalous-field weight.
}
\label{fig:SM_lr_mode_character}
\end{figure*}  

This conclusion becomes particularly clear by resolving the pairing field into 
\begin{equation*}
\frac{p_Q}{X}
=
K_{np}A+K_{pp}B .
\label{eq:SM_lr_pair_decomp_result}
\end{equation*}
The right panel of Fig.~\ref{fig:SM_lr_intertwining} shows that
$|K_{np}A|>|K_{pp}B|$. The dominant contribution to $p_Q$, therefore, comes from the cross-channel response $K_{np}$. The direct contribution $K_{pp}B$ is smaller and has the same sign as $K_{np}A$, so the two
terms enhance one another. Their magnitudes both decrease toward
weaker attraction, with the direct contribution becoming particularly small.
This last observation is due to $B \propto F_0$, which tends to zero, when the interaction strength $|g_0|$ decreases.

\begin{figure}[t]
\centering
\includegraphics[width=0.6\linewidth]{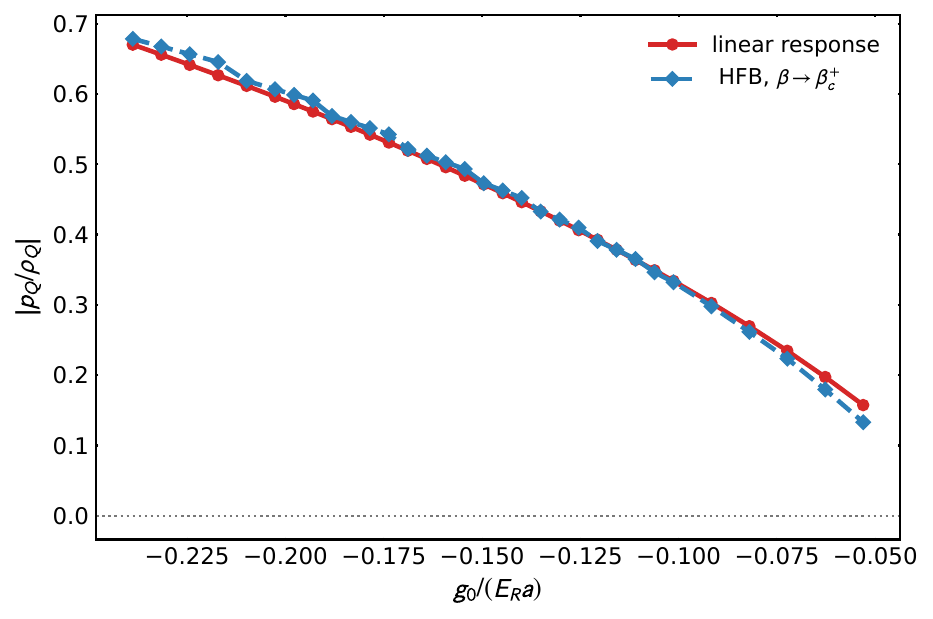}
\caption{
    The ratio $p_Q/\rho_Q$ of the density and pairing orders along the boundary of the SF and CDW+SF+PDW+SR phases. The prediction within the linear response theory
    $p_Q/\rho_Q$ is evaluated at its threshold pump $\beta_c$ and is compared with
    $r_c^{\rm HFB}$ extracted from the full nonlinear HFB solutions for the numerically determined  $\beta_c$. The good agreement of the numerics with the analytical predictions shows that the linear response theory reproduces the density--pairing composition of the ordered state. The gray dotted line marks zero.
}
\label{fig:SM_lr_HFB_comparison}
\end{figure}

We next investigate the mode participation for the threshold pump strength, see Eq.~\eqref{eq:SM_lr_Gthreshold}. 

In Fig.~\ref{fig:SM_lr_mode_character} (left panel), we shows that $G_-$, related to eigenvalue $\lambda_- \leqslant \lambda_+$ of $\bm K$, already accounts for
most of the contribution for strong attractions, and approaches the saturation of the condition as the attraction is reduced. In contrast, $G_+$ is small and vanishes when the bare atom-atom interaction strength $g_0\to0$.  The cavity predominantly amplifies one eigenmode of the coupled density--pairing response matrix $\bm K$.

Explicitly, the eigenvector corresponding to the eigenvalue $\lambda_-$ is written as $$\bm e_-= \begin{pmatrix}
    e_{-,n} \\ e_{-,p}
\end{pmatrix},$$ and, thus, we define $w_n=|e_{-,n}|^2$ and $w_p=|e_{-,p}|^2$.
The right panel of Fig.~\ref{fig:SM_lr_mode_character} shows that for strong attractions the hybridization of the density and pairing channels is significant, as $w_n$ is similar in magnitude to $w_p$. As the attraction is weaker, $w_n$ increases continuously toward unity and $w_p = 1 - w_n$ decreases toward zero, where $G_-$ approaches unity. The self-organization, therefore, evolves smoothly from a regime with strong mixing between the density and pairing channels toward a regime in which the system's response is dominated solely by the density channel.

In Fig.~\ref{fig:SM_lr_HFB_comparison}, we plot $|p_Q/\rho_Q|$ from Eq.~\eqref{eq:SM_lr_composition}, and observe that this magnitude decreases from a sizable value on the strongly attractive side to a much smaller value at weaker
attraction. This trend is consistent with both the decrease of
$\mathcal M$ and the change of $\bm e _-$ toward the density channel, discussed in the previous paragraph.

Finally, we compare the linear response theory with the fully numerical solutions within the self-consistent HFB theory used to calculate the phase diagram in the main text. 
The procedure is as follows.
For each $g_0$, let $\beta_c^{\rm HFB}(g_0)$ denote the corresponding threshold value for the transition from the SF phase to the self-organized CDW+SF+PDW+SR phase, obtained from the full HFB calculations. The fields $\rho_Q$ and $p_Q$ are extracted from the solutions in the ordered phase with $\beta>\beta_c^{\rm HFB}$, see Eq.~\eqref{eq:SM_lr_harmonics}. Close to the transition, we construct the fit
\begin{equation}
    p_Q=r_c^{\rm HFB}\rho_Q,
    \label{eq:SM_lr_HFB_ratio_fit}
\end{equation}
using a linear function
This procedure is more stable numerically than taking the ratio of two
independently vanishing fields.

In Fig.~\ref{fig:SM_lr_HFB_comparison}, we compare the ratio $r_c^{\rm HFB}$ of the fields at the numerically evaluated phase boundary with the prediction $p_Q/\rho_Q$ evaluated at $\beta = \beta_c$ obtained within the linear response theory.
The two calculations are in very good agreement over the full interaction range shown.

In summary, the static linear response theory predicts the intertwining character of the PDW and CDW phases, the coupling of the density and pairing fields, their relative amplitudes, and how they emerge from the nonlinear self-consistent HFB solution at the superradiant phase transition. The dependence on the interaction strength $g_0$ shows similar behaviour over different quantities characterizing the system, namely, stronger attraction produces larger cross-channel coupling, and a larger
$|p_Q/\rho_Q|$, while toward weaker attraction the critical response of the system continuously transforms to predominantly density-wave in character. 
Also, the finite $B=0$ response and the decomposition, presented in Fig.~\ref{fig:SM_lr_intertwining} (middle panel), show that the pairing field is formed predominantly through the cross-channel response function to intra-cavity field.

\bibliography{refs_arxiv}

@misc{SM,
  title = {Supplemental {M}aterial},
  note  = {additional material containing microscopic Hamiltonian, adiabatic elimination of the excited states,  Hartree-Fock-Bogoliubov theory, Bloch wave function formulation, density and pairing profiles, Fano-type profiles, linear response theory and analytical predictions of a phase boundary.}
}

@article{konishi2021universal,
  title={Universal pair polaritons in a strongly interacting {Fermi} gas},
  author={Konishi, Hideki and Roux, Kevin and Helson, Victor and Brantut, Jean-Philippe},
  journal={Nature},
  volume={596},
  number={7873},
  pages={509--513},
  year={2021},
  publisher={Nature Publishing Group UK London},
  doi = {https://doi.org/10.1038/s41586-021-03731-9}
}

@article{sharma2025engineering,
  title = {Engineering interactions by collective coupling of atom pairs
           to cavity photons for entanglement generation},
  author = {Sharma, Sankalp and Chwede{\'n}czuk, Jan and Wasak, Tomasz},
  journal = {Phys. Rev. Research},
  volume = {7},
  pages = {L012038},
  year = {2025},
  doi = {10.1103/PhysRevResearch.7.L012038}
}

@article{zhang2021observation,
  title={Observation of a superradiant quantum phase transition in an intracavity degenerate {Fermi} gas},
  author={Zhang, Xiaotian and Chen, Yu and Wu, Zemao and Wang, Juan and Fan, Jijie and Deng, Shujin and Wu, Haibin},
  journal={Science},
  volume={373},
  number={6561},
  pages={1359--1362},
  year={2021},
  publisher={American Association for the Advancement of Science},
  doi = {10.1126/science.abd4385}
}

@article{helson2023density,
  title={Density-wave ordering in a unitary {Fermi} gas with photon-mediated interactions},
  author={Helson, Victor and Zwettler, Timo and Mivehvar, Farokh and Colella, Elvia and Roux, Kevin and Konishi, Hideki and Ritsch, Helmut and Brantut, Jean-Philippe},
  journal={Nature},
  volume={618},
  number={7966},
  pages={716--720},
  year={2023},
  publisher={Nature Publishing Group UK London},
  doi = {https://doi.org/10.1038/s41586-023-06018-3}
}

@article{RevModPhys.87.457,
  title = {Colloquium: Theory of intertwined orders in high temperature superconductors},
  author = {Fradkin, Eduardo and Kivelson, Steven A. and Tranquada, John M.},
  journal = {Rev. Mod. Phys.},
  volume = {87},
  issue = {2},
  pages = {457--482},
  numpages = {26},
  year = {2015},
  month = {May},
  publisher = {American Physical Society},
  doi = {10.1103/RevModPhys.87.457},
  url = {https://link.aps.org/doi/10.1103/RevModPhys.87.457}
}

@article{guo2012cavityBCSBEC,
  title = {Ultracold {Fermi} gas in a single-mode cavity:
           Cavity-mediated interaction and {BCS-BEC} evolution},
  author = {Guo, Xiaoyong and Ren, Zhongzhou and Guo, Guangjie
            and Peng, Jie},
  journal = {Phys. Rev. A},
  volume = {86},
  pages = {053605},
  year = {2012},
  doi = {10.1103/PhysRevA.86.053605}
}

@article{zheng2020fflo,
  title = {Cavity-induced {F}ulde--{F}errell--{L}arkin--{O}vchinnikov
           superfluids of ultracold {F}ermi gases},
  author = {Zheng, Zhen and Wang, Z. D.},
  journal = {Phys. Rev. A},
  volume = {101},
  pages = {023612},
  year = {2020},
  doi = {10.1103/PhysRevA.101.023612}
}

@article{giorgini2008theory,
  title = {Theory of ultracold atomic {Fermi} gases},
  author = {Giorgini, Stefano and Pitaevskii, Lev P. and Stringari, Sandro},
  journal = {Rev. Mod. Phys.},
  volume = {80},
  issue = {4},
  pages = {1215--1274},
  numpages = {0},
  year = {2008},
  month = {Oct},
  publisher = {American Physical Society},
  doi = {10.1103/RevModPhys.80.1215},
  url = {https://link.aps.org/doi/10.1103/RevModPhys.80.1215}
}

@article{agterberg2020physics,
  title={The physics of pair-density waves: cuprate superconductors and beyond},
  author={Agterberg, Daniel F and Davis, JC S{\'e}amus and Edkins, Stephen D and Fradkin, Eduardo and Van Harlingen, Dale J and Kivelson, Steven A and Lee, Patrick A and Radzihovsky, Leo and Tranquada, John M and Wang, Yuxuan},
  journal={Annual Review of Condensed Matter Physics},
  volume={11},
  number={1},
  pages={231--270},
  year={2020},
  publisher={Annual Reviews},
  doi ={https://doi.org/10.1146/annurev-conmatphys-031119-050711}
}

@article{RevModPhys.85.553,
  title = {Cold atoms in cavity-generated dynamical optical potentials},
  author = {Ritsch, Helmut and Domokos, Peter and Brennecke, Ferdinand and Esslinger, Tilman},
  journal = {Rev. Mod. Phys.},
  volume = {85},
  issue = {2},
  pages = {553--601},
  numpages = {0},
  year = {2013},
  month = {Apr},
  publisher = {American Physical Society},
  doi = {10.1103/RevModPhys.85.553},
  url = {https://link.aps.org/doi/10.1103/RevModPhys.85.553}
}

@article{PhysRevLett.89.253003,
  title = {Collective Cooling and Self-Organization of Atoms in a Cavity},
  author = {Domokos, Peter and Ritsch, Helmut},
  journal = {Phys. Rev. Lett.},
  volume = {89},
  issue = {25},
  pages = {253003},
  numpages = {4},
  year = {2002},
  month = {Dec},
  publisher = {American Physical Society},
  doi = {10.1103/PhysRevLett.89.253003},
  url = {https://link.aps.org/doi/10.1103/PhysRevLett.89.253003}
}

@article{PhysRevLett.112.143002,
  title = {Fermionic Superradiance in a Transversely Pumped Optical Cavity},
  author = {Keeling, J. and Bhaseen, M. J. and Simons, B. D.},
  journal = {Phys. Rev. Lett.},
  volume = {112},
  issue = {14},
  pages = {143002},
  numpages = {5},
  year = {2014},
  month = {Apr},
  publisher = {American Physical Society},
  doi = {10.1103/PhysRevLett.112.143002},
  url = {https://link.aps.org/doi/10.1103/PhysRevLett.112.143002}
}

@article{baumann2010dicke,
  title={Dicke quantum phase transition with a superfluid gas in an optical cavity},
  author={Baumann, Kristian and Guerlin, Christine and Brennecke, Ferdinand and Esslinger, Tilman},
  journal={nature},
  volume={464},
  number={7293},
  pages={1301--1306},
  year={2010},
  publisher={Nature Publishing Group UK London},
  doi ={https://doi.org/10.1038/nature09009}
}

@article{mivehvar2021cavity,
  title={Cavity QED with quantum gases: New paradigms in many-body physics},
  author={Mivehvar, Farokh and Piazza, Francesco and Donner, Tobias and Ritsch, Helmut},
  journal={Advances in Physics},
  volume={70},
  number={1},
  pages={1--153},
  year={2021},
  publisher={Taylor \& Francis},
  doi = {https://doi.org/10.1080/00018732.2021.1969727}
}

@article{PhysRevLett.123.133601,
  title = {Cavity-Mediated Unconventional Pairing in Ultracold Fermionic Atoms},
  author = {Schlawin, Frank and Jaksch, Dieter},
  journal = {Phys. Rev. Lett.},
  volume = {123},
  issue = {13},
  pages = {133601},
  numpages = {6},
  year = {2019},
  month = {Sep},
  publisher = {American Physical Society},
  doi = {10.1103/PhysRevLett.123.133601},
  url = {https://link.aps.org/doi/10.1103/PhysRevLett.123.133601}
}

@article{PhysRevLett.112.143003,
  title = {Umklapp Superradiance with a Collisionless Quantum Degenerate {Fermi} Gas},
  author = {Piazza, Francesco and Strack, Philipp},
  journal = {Phys. Rev. Lett.},
  volume = {112},
  issue = {14},
  pages = {143003},
  numpages = {5},
  year = {2014},
  month = {Apr},
  publisher = {American Physical Society},
  doi = {10.1103/PhysRevLett.112.143003},
  url = {https://link.aps.org/doi/10.1103/PhysRevLett.112.143003}
}

@article{RevModPhys.82.1225,
  title = {Feshbach resonances in ultracold gases},
  author = {Chin, Cheng and Grimm, Rudolf and Julienne, Paul and Tiesinga, Eite},
  journal = {Rev. Mod. Phys.},
  volume = {82},
  issue = {2},
  pages = {1225--1286},
  numpages = {0},
  year = {2010},
  month = {Apr},
  publisher = {American Physical Society},
  doi = {10.1103/RevModPhys.82.1225},
  url = {https://link.aps.org/doi/10.1103/RevModPhys.82.1225}
}

@article{Zheng_2026,
doi = {10.1088/1367-2630/ae868d},
url = {https://doi.org/10.1088/1367-2630/ae868d},
year = {2026},
month = {jul},
publisher = {IOP Publishing},
volume = {28},
number = {7},
pages = {073204},
author = {Zheng, Zhen and Zhu, Shi-Liang and Wang, Z D},
title = {Cavity-induced multispin interactions and phase transitions in ultracold {Fermi} gases},
journal = {New Journal of Physics}
}

@article{t4xb-6x3z,
  title = {Fate of the {Fermi} Surface Coupled to a Single-Wave-Vector Cavity Mode},
  author = {Frank, Bernhard and Pini, Michele and Lang, Johannes and Piazza, Francesco},
  journal = {Phys. Rev. Lett.},
  volume = {136},
  issue = {14},
  pages = {143403},
  numpages = {8},
  year = {2026},
  month = {Apr},
  publisher = {American Physical Society},
  doi = {10.1103/t4xb-6x3z},
  url = {https://link.aps.org/doi/10.1103/t4xb-6x3z}
}

@article{PhysRevA.72.053417,
  title = {Self-organization of atoms in a cavity field: Threshold, bistability, and scaling laws},
  author = {Asb\'oth, J. K. and Domokos, P. and Ritsch, H. and Vukics, A.},
  journal = {Phys. Rev. A},
  volume = {72},
  issue = {5},
  pages = {053417},
  numpages = {12},
  year = {2005},
  month = {Nov},
  publisher = {American Physical Society},
  doi = {10.1103/PhysRevA.72.053417},
  url = {https://link.aps.org/doi/10.1103/PhysRevA.72.053417}
}

@article{kozin2025cavity,
  title = {Cavity-enhanced superconductivity via band engineering},
  author = {Kozin, Valerii K. and Thingstad, Even and Loss, Daniel
            and Klinovaja, Jelena},
  journal = {Phys. Rev. B},
  volume = {111},
  pages = {035410},
  year = {2025},
  doi = {10.1103/PhysRevB.111.035410}
}

@article{schlawin2022cavity,
  title = {Cavity quantum materials},
  author = {Schlawin, Frank and Kennes, Dante M. and Sentef, Michael A.},
  journal = {Appl. Phys. Rev.},
  volume = {9},
  pages = {011312},
  year = {2022},
  doi = {10.1063/5.0083825}
}

@article{moritz2005confinement,
  title   = {Confinement Induced Molecules in a 1D {Fermi} Gas},
  author  = {Moritz, Henning and St{\"o}ferle, Thilo and G{\"u}nter, Kenneth
             and K{\"o}hl, Michael and Esslinger, Tilman},
  journal = {Phys. Rev. Lett.},
  volume  = {94},
  pages   = {210401},
  year    = {2005},
  doi     = {10.1103/PhysRevLett.94.210401}
}

@article{liao2010spinimbalance,
  title   = {Spin-imbalance in a one-dimensional {Fermi} gas},
  author  = {Liao, Yean-an and Rittner, Ann Sophie C. and Paprotta, Tobias
             and Li, Wenhui and Partridge, Guthrie B. and Hulet, Randall G.
             and Baur, Stefan K. and Mueller, Erich J.},
  journal = {Nature},
  volume  = {467},
  pages   = {567--569},
  year    = {2010},
  doi     = {10.1038/nature09393}
}

@article{h3zm-rnnx,
  title = {Microscopy of Cavity-Induced Density-Wave Ordering in Ultracold Gases},
  author = {B\"uhler, Tabea and Fabre, Aur\'elien and Bolognini, Gaia and Xue, Zeyang and Zwettler, Timo and Del Pace, Giulia and Brantut, Jean-Philippe},
  journal = {Phys. Rev. Lett.},
  volume = {136},
  issue = {14},
  pages = {143401},
  numpages = {7},
  year = {2026},
  month = {Apr},
  publisher = {American Physical Society},
  doi = {10.1103/h3zm-rnnx},
  url = {https://link.aps.org/doi/10.1103/h3zm-rnnx}
}

@article{guo2021optical,
  title={An optical lattice with sound},
  author={Guo, Yudan and Kroeze, Ronen M and Marsh, Brendan P and Gopalakrishnan, Sarang and Keeling, Jonathan and Lev, Benjamin L},
  journal={Nature},
  volume={599},
  number={7884},
  pages={211--215},
  year={2021},
  publisher={Nature Publishing Group UK London},
  doi ={https://doi.org/10.1038/s41586-021-03945-x}
}

@article{lev2025glass,
  title = {Multimode Cavity QED Ising Spin Glass},
  author = {Marsh, Brendan P. and Schuller, David Atri and Ji, Yunpeng and Hunt, Henry S. and Socolof, Giulia Z. and Bowman, Deven P. and Keeling, Jonathan and Lev, Benjamin L.},
  journal = {Phys. Rev. Lett.},
  volume = {135},
  issue = {16},
  pages = {160403},
  numpages = {7},
  year = {2025},
  month = {Oct},
  publisher = {American Physical Society},
  doi = {10.1103/x19r-pzyb},
  url = {https://link.aps.org/doi/10.1103/x19r-pzyb}
}

@article{PhysRevLett.25.1543,
  title = {Can a Solid Be "Superfluid"?},
  author = {Leggett, A. J.},
  journal = {Phys. Rev. Lett.},
  volume = {25},
  issue = {22},
  pages = {1543--1546},
  numpages = {0},
  year = {1970},
  month = {Nov},
  publisher = {American Physical Society},
  doi = {10.1103/PhysRevLett.25.1543},
  url = {https://link.aps.org/doi/10.1103/PhysRevLett.25.1543}
}

@article{fano1961effects,
  title = {Effects of Configuration Interaction on Intensities and Phase Shifts},
  author = {Fano, U.},
  journal = {Phys. Rev.},
  volume = {124},
  issue = {6},
  pages = {1866--1878},
  numpages = {0},
  year = {1961},
  month = {Dec},
  publisher = {American Physical Society},
  doi = {10.1103/PhysRev.124.1866},
  url = {https://link.aps.org/doi/10.1103/PhysRev.124.1866}
}

@book{ring1980nuclear,
  title     = {The Nuclear Many-Body Problem},
  author    = {Ring, Peter and Schuck, Peter},
  year      = {1980},
  publisher = {Springer-Verlag},
  address   = {Berlin, Heidelberg},
  series    = {Theoretical and Mathematical Physics},
  url       = {https://link.springer.com/book/9783540212065}
}

@inbook{Dobaczewski_2013,
   title={Hartree—Fock—Bogoliubov Solution of the Pairing Hamiltonian in Finite Nuclei},
   ISBN={9789814412490},
   url={http://dx.doi.org/10.1142/9789814412490_0004},
   DOI={10.1142/9789814412490_0004},
   booktitle={Fifty Years of Nuclear BCS},
   publisher={WORLD SCIENTIFIC},
   author={Dobaczewski, J. and Nazarewicz, W.},
   year={2013},
   month=Mar, pages={40–60} }

@article{PhysRevA.101.063607,
  title = {Hartree-Fock-Bogoliubov theory of trapped one-dimensional imbalanced {Fermi} systems},
  author = {Patton, Kelly R. and Sheehy, Daniel E.},
  journal = {Phys. Rev. A},
  volume = {101},
  issue = {6},
  pages = {063607},
  numpages = {10},
  year = {2020},
  month = {Jun},
  publisher = {American Physical Society},
  doi = {10.1103/PhysRevA.101.063607},
  url = {https://link.aps.org/doi/10.1103/PhysRevA.101.063607}
}

@article{PhysRevA.81.063642,
  title = {Hartree-Fock-Bogoliubov theory of dipolar {Fermi} gases},
  author = {Zhao, Cheng and Jiang, Lei and Liu, Xunxu and Liu, W. M. and Zou, Xubo and Pu, Han},
  journal = {Phys. Rev. A},
  volume = {81},
  issue = {6},
  pages = {063642},
  numpages = {5},
  year = {2010},
  month = {Jun},
  publisher = {American Physical Society},
  doi = {10.1103/PhysRevA.81.063642},
  url = {https://link.aps.org/doi/10.1103/PhysRevA.81.063642}
}

@article{PhysRevA.68.033610,
  title = {Hartree-Fock-Bogoliubov theory versus local-density approximation for superfluid trapped fermionic atoms},
  author = {Grasso, Marcella and Urban, Michael},
  journal = {Phys. Rev. A},
  volume = {68},
  issue = {3},
  pages = {033610},
  numpages = {10},
  year = {2003},
  month = {Sep},
  publisher = {American Physical Society},
  doi = {10.1103/PhysRevA.68.033610},
  url = {https://link.aps.org/doi/10.1103/PhysRevA.68.033610}
}

@book{de2018superconductivity,
  title     = {Superconductivity of Metals and Alloys},
  author    = {de Gennes, Pierre-Gilles},
  publisher = {CRC Press},
  year      = {2018},
  doi       = {10.1201/9780429497032},
  url       = {https://doi.org/10.1201/9780429497032}
}

@book{leggett2006quantum,
  title     = {Quantum Liquids: Bose Condensation and Cooper Pairing in Condensed-Matter Systems},
  author    = {Leggett, Anthony J.},
  year      = {2006},
  publisher = {Oxford University Press},
  doi       = {10.1093/acprof:oso/9780198526438.001.0001},
  url       = {https://doi.org/10.1093/acprof:oso/9780198526438.001.0001}
}

@book{zhu2016bogoliubov,
  title     = {Bogoliubov-de Gennes Method and Its Applications},
  author    = {Zhu, Jian-Xin},
  year      = {2016},
  publisher = {Springer},
  address   = {Cham},
  series    = {Lecture Notes in Physics},
  volume    = {924},
  doi       = {10.1007/978-3-319-31314-6},
  url       = {https://doi.org/10.1007/978-3-319-31314-6}
}

@article{chen2014superradiance,
  title = {Superradiance of Degenerate {Fermi} Gases in a Cavity},
  author = {Chen, Yu and Yu, Zhenhua and Zhai, Hui},
  journal = {Phys. Rev. Lett.},
  volume = {112},
  issue = {14},
  pages = {143004},
  numpages = {5},
  year = {2014},
  month = {Apr},
  publisher = {American Physical Society},
  doi = {10.1103/PhysRevLett.112.143004},
  url = {https://link.aps.org/doi/10.1103/PhysRevLett.112.143004}
}

@article{chen2015superradiant,
  title = {Superradiant phase transition of {Fermi} gases in a cavity across a Feshbach resonance},
  author = {Chen, Yu and Zhai, Hui and Yu, Zhenhua},
  journal = {Phys. Rev. A},
  volume = {91},
  issue = {2},
  pages = {021602(R)},
  numpages = {5},
  year = {2015},
  month = {Feb},
  publisher = {American Physical Society},
  doi = {10.1103/PhysRevA.91.021602},
  url = {https://link.aps.org/doi/10.1103/PhysRevA.91.021602}
}

@article{roux2020strongly,
  title={Strongly correlated Fermions strongly coupled to light},
  author={Roux, Kevin and Konishi, Hideki and Helson, Victor and Brantut, Jean-Philippe},
  journal={Nature Communications},
  volume={11},
  number={1},
  pages={2974},
  year={2020},
  publisher={Nature Publishing Group UK London},
  doi ={https://doi.org/10.1038/s41467-020-16767-8}
}

@article{Roux_2021,
doi = {10.1088/1367-2630/abeb91},
url = {https://doi.org/10.1088/1367-2630/abeb91},
year = {2021},
month = {apr},
publisher = {IOP Publishing},
volume = {23},
number = {4},
pages = {043029},
author = {Roux, K and Helson, V and Konishi, H and Brantut, J P},
title = {Cavity-assisted preparation and detection of a unitary {Fermi} gas},
journal = {New Journal of Physics}
}

@article{morales2018coupling,
  title={Coupling two order parameters in a quantum gas},
  author={Morales, Andrea and Zupancic, Philip and L{\'e}onard, Julian and Esslinger, Tilman and Donner, Tobias},
  journal={Nature materials},
  volume={17},
  number={8},
  pages={686--690},
  year={2018},
  publisher={Nature Publishing Group UK London},
  doi = {https://doi.org/10.1038/s41563-018-0118-1}
}

@article{PhysRevLett.91.203001,
  title = {Observation of Collective Friction Forces due to Spatial Self-Organization of Atoms: From Rayleigh to Bragg Scattering},
  author = {Black, Adam T. and Chan, Hilton W. and Vuleti\ifmmode \acute{c}\else \'{c}\fi{}, Vladan},
  journal = {Phys. Rev. Lett.},
  volume = {91},
  issue = {20},
  pages = {203001},
  numpages = {4},
  year = {2003},
  month = {Nov},
  publisher = {American Physical Society},
  doi = {10.1103/PhysRevLett.91.203001},
  url = {https://link.aps.org/doi/10.1103/PhysRevLett.91.203001}
}

@article{PhysRevLett.107.140402,
  title = {Exploring Symmetry Breaking at the Dicke Quantum Phase Transition},
  author = {Baumann, K. and Mottl, R. and Brennecke, F. and Esslinger, T.},
  journal = {Phys. Rev. Lett.},
  volume = {107},
  issue = {14},
  pages = {140402},
  numpages = {5},
  year = {2011},
  month = {Sep},
  publisher = {American Physical Society},
  doi = {10.1103/PhysRevLett.107.140402},
  url = {https://link.aps.org/doi/10.1103/PhysRevLett.107.140402}
}

@article{hamidian2016detection,
  title={Detection of a Cooper-pair density wave in Bi2Sr2CaCu2O8+ x},
  author={Hamidian, MH and Edkins, Stephen David and Joo, Sang Hyun and Kostin, A and Eisaki, H and Uchida, S and Lawler, MJ and Kim, E-A and Mackenzie, Andrew P and Fujita, K and others},
  journal={Nature},
  volume={532},
  number={7599},
  pages={343--347},
  year={2016},
  publisher={Nature Publishing Group UK London},
  doi = {https://doi.org/10.1038/nature17411}
}

@article{olshanii1998atomic,
  title = {Atomic Scattering in the Presence of an External Confinement and a Gas of Impenetrable Bosons},
  author = {Olshanii, M.},
  journal = {Phys. Rev. Lett.},
  volume = {81},
  issue = {5},
  pages = {938--941},
  numpages = {0},
  year = {1998},
  month = {Aug},
  publisher = {American Physical Society},
  doi = {10.1103/PhysRevLett.81.938},
  url = {https://link.aps.org/doi/10.1103/PhysRevLett.81.938}
}

@article{bergeman2003atom,
  title = {Atom-Atom Scattering under Cylindrical Harmonic Confinement: Numerical and Analytic Studies of the Confinement Induced Resonance},
  author = {Bergeman, T. and Moore, M. G. and Olshanii, M.},
  journal = {Phys. Rev. Lett.},
  volume = {91},
  issue = {16},
  pages = {163201},
  numpages = {4},
  year = {2003},
  month = {Oct},
  publisher = {American Physical Society},
  doi = {10.1103/PhysRevLett.91.163201},
  url = {https://link.aps.org/doi/10.1103/PhysRevLett.91.163201}
}

@article{orso2024superfluid,
  title = {Superfluid fraction and {Leggett} bound in a density-modulated strongly interacting {Fermi} gas at zero temperature},
  author = {Orso, G. and Stringari, S.},
  journal = {Phys. Rev. A},
  volume = {109},
  issue = {2},
  pages = {023301},
  numpages = {9},
  year = {2024},
  month = {Feb},
  publisher = {American Physical Society},
  doi = {10.1103/PhysRevA.109.023301},
  url = {https://link.aps.org/doi/10.1103/PhysRevA.109.023301}
}

@article{yao2025paircorrelations,
  author  = {Yao, Ruixiao and Chi, Sungjae and Wang, Mingxuan and
             Fletcher, Richard J. and Zwierlein, Martin},
  title   = {Measuring Pair Correlations in {Bose} and {Fermi} Gases via
             Atom-Resolved Microscopy},
  journal = {Phys. Rev. Lett.},
  volume  = {134},
  number  = {18},
  pages   = {183402},
  year    = {2025},
  doi     = {10.1103/PhysRevLett.134.183402}
}

@article{regal2004resonance,
  author  = {Regal, C. A. and Greiner, M. and Jin, D. S.},
  title   = {Observation of Resonance Condensation of Fermionic Atom Pairs},
  journal = {Physical Review Letters},
  volume  = {92},
  number  = {4},
  pages   = {040403},
  year    = {2004},
  doi     = {10.1103/PhysRevLett.92.040403}
}

@article{dyke2021dynamics,
  author  = {Dyke, P. and Hogan, A. and Herrera, I. and
             Kuhn, C. C. N. and Hoinka, S. and Vale, C. J.},
  title   = {Dynamics of a {Fermi} Gas Quenched to Unitarity},
  journal = {Physical Review Letters},
  volume  = {127},
  number  = {10},
  pages   = {100405},
  year    = {2021},
  doi     = {10.1103/PhysRevLett.127.100405}
}

@article{zwettler2025cavity,
  author  = {Zwettler, Timo and Marijanovi{\'c}, Filip and
             B{\"u}hler, Tabea and Chattopadhyay, Sambuddha and
             Del Pace, Giulia and Skolc, Luka and Helson, Victor and
             Uchino, Shun and Demler, Eugene and Brantut, Jean-Philippe},
  title   = {Cavity-mediated charge and pair-density waves in a unitary {Fermi} gas},
  journal = {Nature Communications},
  volume  = {17},
  pages   = {496},
  year    = {2026},
  doi     = {10.1038/s41467-025-67184-8}
}

@article{buhler2025direct,
  title   = {Direct production of fermionic superfluids in a
             cavity-enhanced optical dipole trap},
  author  = {B{\"u}hler, Tabea N. C. and Zwettler, Timo and
             Bolognini, Gaia S. and Fabre, Aur{\'e}lien H. and
             Helson, Victor Y. and Del Pace, Giulia and
             Brantut, Jean-Philippe},
  journal = {SciPost Physics},
  volume  = {18},
  number  = {4},
  pages   = {133},
  year    = {2025},
  doi     = {10.21468/SciPostPhys.18.4.133}
}

@article{biagioni2024measurement,
	title = {Measurement of the superfluid fraction of a supersolid by {Josephson} effect},
	volume = {629},
	issn = {0028-0836, 1476-4687},
	url = {https://www.nature.com/articles/s41586-024-07361-9},
	doi = {10.1038/s41586-024-07361-9},
	number = {8013},
	urldate = {2026-09-07},
	journal = {Nature},
	author = {Biagioni, G. and Antolini, N. and Donelli, B. and Pezzè, L. and Smerzi, A. and Fattori, M. and Fioretti, A. and Gabbanini, C. and Inguscio, M. and Tanzi, L. and Modugno, G.},
	month = may,
	year = {2024},
	pages = {773--777}
}

\end{document}